\documentclass[12pt,letterpaper,a4paper]{article}
\usepackage[includeheadfoot,
marginratio={1:1,2:3},
width=525pt,
height=750pt,]{geometry}

\usepackage{amsmath}
\usepackage{amsfonts}
\usepackage{amssymb}
\usepackage{graphicx}
\usepackage{cancel}
\usepackage{empheq}
\usepackage{color}
\usepackage{hyperref}

\numberwithin{equation}{section}
\usepackage{float}
\restylefloat{table}
\restylefloat{figure}
\usepackage[utf8]{inputenc}
\usepackage{amsmath, amsthm, amssymb, amsfonts}
\usepackage[font=small, labelfont=bf]{caption}
\usepackage{amsmath, amsthm, amssymb, amsfonts}
\usepackage{multirow}
\usepackage{graphicx}
\usepackage{float}
\usepackage[numbers,sort&compress]{natbib}
\usepackage{enumerate}
\usepackage{amsmath}
\usepackage{amsfonts}
\usepackage{amssymb}
\usepackage{graphicx}
\usepackage{cancel}
\usepackage{empheq}
\usepackage{color}
\usepackage{hyperref}
\usepackage{float}
\restylefloat{table}
\restylefloat{figure}
\usepackage{longtable}
\usepackage{lipsum} 

\usepackage{amsfonts}
\usepackage{amssymb}
\usepackage{comment}
\usepackage{graphicx}
\usepackage{amsmath}
\usepackage{tabu}
\usepackage{dsfont}
\usepackage{float}
\usepackage{amsthm}
\usepackage{slashed}
\usepackage{setspace}
\usepackage[labelformat=simple]{subcaption}

\usepackage{pgfplots}
\usepackage{tikz}
\usepackage{tikz-cd}
\usetikzlibrary{shapes.misc,shadows,fit,decorations.pathmorphing,positioning,trees,decorations.markings,decorations.pathreplacing,calc,shapes,patterns,arrows,positioning,arrows.meta}
\usepackage{xcolor}
\usepackage{pgfplots}
\usepackage{pgfplotstable}
\usepackage{xcolor}
\usepackage[boxsize=0.5em,aligntableaux=center]{ytableau}
\usepackage{makecell}
\usepackage{diagbox}
\usepackage{environ}
\usepackage[utf8]{inputenc}
\usepackage{amsmath, amsthm, amssymb, amsfonts}
\usepackage{amsmath, amsthm, amssymb, amsfonts}
\usepackage{multirow}
\usepackage{graphicx}
\usepackage{float}
\usepackage[numbers,sort&compress]{natbib}
\usepackage{enumerate}
\usepackage{enumitem}
\usepackage[capitalize,sort]{cleveref}
\usepackage{hhline}
\usepackage{mathtools}
\usepackage{longtable}
\usepackage{makecell}
\usepackage{tabularx}
\usepackage{cellspace}
\usepackage{alphalph}
\usepackage{lscape}

\crefname{figure}{Figure}{Figures}
\crefname{table}{Table}{Tables}

\def\om{{\omega}}

\def\be{\begin{equation}}
\def\ee{\end{equation}}
\def\bea{\begin{eqnarray}}
\def\eea{\end{eqnarray}}
\def\bes{\begin{subequations}}
	\def\ees{\end{subequations}}

\newcommand{\bmat}{\left(\begin{array}}
	\newcommand{\emat}{\end{array}\right)}

\def\bZ{\mathbb{Z}}

\def\C{\mathbb{C}}

\def\a {\alpha}

\def\ov{\overline}

\def\RE{\text{Re}\,}
\def\ov{\overline}
\def\1{{\bf 1}}
\def\2{{\bf 2}}
\def\3{{\bf 3}}
\def\4{{\bf 4}}
\def\6{{\bf 6}}

\newcommand{\beq}{\begin{equation}}
\newcommand{\eeq}{\end{equation}}

\def\ov{\overline}

\allowdisplaybreaks[2]
\numberwithin{equation}{section}

\def\om{{\omega}}

\def\be{\begin{equation}}
\def\ee{\end{equation}}
\def\bea{\begin{eqnarray}}
\def\eea{\end{eqnarray}}
\def\bes{\begin{subequations}}
	\def\ees{\end{subequations}}

\usepackage{multicol}
\usepackage{float}
\usepackage{caption}
\def\mk {{\mathcal K}}

\def\r {{\rho}}

\def\tr {{\tilde{\rho}}}
\def\p {{\partial}}

\newcommand{\cK}{\mathcal{K}}
\newcommand{\cM}{\mathcal{M}}
\newcommand{\cN}{\mathcal{N}}

\newcommand{\cA}{\mathcal{A}}
\newcommand{\cB}{\mathcal{B}}

\newcommand{\la}{\langle}
\newcommand{\ra}{\rangle}

\allowdisplaybreaks[2]
\numberwithin{equation}{section}

\newcommand{\black}[1]{\textcolor{black}{#1}}

\begin{document}
		%
	
	\vspace{1.0cm}
	\begin{center}
		{\Large
			On the limitations of non-geometric fluxes to realize dS vacua}
		\vspace{0.4cm}
	\end{center}
	
	\vspace{0.35cm}
	\begin{center}
		David Prieto$^\diamond$, Joan Quirant$^{\star}$, and Pramod Shukla$^\dagger$ \footnote{Email: d.prietorodriguez@uu.nl, joanq@post.bgu.ac.il, pshukla@jcbose.ac.in}
	\end{center}

	\vspace{0.1cm}
	\begin{center}
{$^\diamond$ Institute for Theoretical Physics, Utrecht University,\\ Princetonplein 5, 3584 CC Utrecht, The Netherlands\\
\vskip0.5cm 
$^\star$ Department of Physics, Ben-Gurion University of the Negev, Be'er-Sheva 84105, Israel. \\
 \vskip0.5cm 
$^\dagger$ Department of Physical Sciences, Bose Institute,\\ 
Unified Academic Campus, EN 80, Sector V, Bidhannagar, Kolkata 700091, India.}
	\end{center}
	
	\vspace{1cm}
	
\abstract{

In this paper, we perform a systematic and analytical exploration of de Sitter conditions in type IIA compactifications with (non-)geometric fluxes along with the standard NS-NS and RR $p$-form fluxes. Exploiting the fact that the  F-term scalar potential can be written as a bilinear form, we start  by studying the most generic case. We find four conditions that the scalar fields and fluxes must satisfy to achieve de Sitter vacua. Particularizing to different configurations,  we recover and extend previous results in the literature. We then impose an Ansatz in which the F-terms are proportional to the respective K\"ahler derivatives. In this set-up we are able to derive additional constraints and to classify  the  possible dS no-go scenarios in terms of eight  axionic fluxes. Individually considering that any of these fluxes can be vanishing or non-vanishing leads to a total of 256 flux configurations. We find that 227 of these 256 possibilities result in a dS no-go scenario. The remaining 29 flux configurations, a priori, do not lead to  dS no-go cases and would deserve further investigation.

}

\clearpage
	
\tableofcontents


\section{Introduction}
\label{sec_intro}

Exploring de Sitter (dS) solutions in the landscape of superstring vacua has been a central and challenging goal in string phenomenology. Continuous studies have resulted in a series of no-go scenarios in the last two decades \cite{Maldacena:2000mw, Ooguri:2006in, DeWolfe:2005uu,Hertzberg:2007ke, Hertzberg:2007wc, Haque:2008jz, Flauger:2008ad, Caviezel:2008tf, deCarlos:2009fq, Caviezel:2009tu, Danielsson:2009ff, Danielsson:2010bc, Wrase:2010ew,  Shiu:2011zt, Dasgupta:2014pma, Danielsson:2018ztv, Bernardo:2020lar, Shukla:2019akv, Shukla:2019dqd}, setting important constraints on the possibility to obtain superstring vacua compatible with cosmological observations. More recently, in the spirit of the swampland program -see   \cite{Palti:2019pca, vanBeest:2021lhn, Agmon:2022thq} for a review- several conjectures questioning the existence of dS vacua in string theory have been formulated \cite{Obied:2018sgi, Garg:2018reu, Ooguri:2018wrx}.\footnote{Let us also mention that there are several proposals claiming for consistent dS realization, and we refer the interested readers to the reviews  \cite{Cicoli:2018kdo,Cicoli:2023opf}.} Building on the fact that most of the initially realized dS no-go results were based on  type IIA orientifold models \cite{DeWolfe:2005uu,Hertzberg:2007ke, Hertzberg:2007wc, Haque:2008jz, Flauger:2008ad, Caviezel:2008tf, deCarlos:2009fq, Caviezel:2009tu, Danielsson:2009ff, Danielsson:2010bc, Wrase:2010ew,Shiu:2011zt}, we aim to perform a detailed analytic study of the F-term potential in the landscape of Type IIA  flux vacua. To expand over the previous results, we will consider  (non-)geometric fluxes in addition to the usual form fluxes ($H_3$ and $F_p$).

The introduction and study of non-geometric fluxes have been motivated from various perspectives. They appear in the context of  gauged supergravities \cite{Derendinger:2005ph,Dall'Agata:2009gv,Aldazabal:2011yz}, in Double Field Theory (DFT) \cite{Aldazabal:2011nj,Andriot:2012wx,Blumenhagen:2013hva} and in superstring compactifications with generalized fluxes \cite{Derendinger:2004jn,Villadoro:2005cu,Aldazabal:2006up,Ihl:2006pp,Ihl:2007ah,Blumenhagen:2013hva,Marchesano:2020uqz}. In addition, the analysis of non-geometric flux compactifications has also led to some interesting utilizations of the symplectic geometries to formulate the four-dimensional effective scalar potentials in a concise form \cite{Shukla:2015hpa, Blumenhagen:2015lta, Gao:2017gxk, Shukla:2016hyy, Shukla:2019wfo,Leontaris:2023mmm}. These multidisciplinary interconnections have been very helpful for exploring the landscape of (generalized) flux vacua. A useful review of the developments of non-geometric flux compactifications can be found in \cite{Plauschinn:2018wbo}.

The initial motivation for studying the non-geometric flux compactifications lies in the fact that (non-)geometric fluxes naturally  arise from a successive application of  T-duality on the NS-NS three-form flux $H_3$ of  Type II supergravity theories \cite{Shelton:2005cf}. It turns out that such generalized fluxes  induce tree-level contributions to the scalar potential depending on the different moduli. As a consequence, this can lead to the possibility of dynamically stabilizing all the moduli in the absence of any (non-)perturbative corrections.\footnote{Notice that in type IIA with only RR and NSNS standard fluxes all moduli can be stabilized in AdS vacua \cite{DeWolfe:2005uu,Camara:2005dc}.} In this regard, incorporating such (non-)geometric fluxes certainly enriches the possibilities from a model-building perspective. 

However, it is still not fully known how many different kinds of fluxes can be consistently turned on simultaneously in a given construction \cite{Shelton:2005cf, Wecht:2007wu, Ihl:2007ah, Robbins:2007yv, Lanza:2019xxg,Lombardo:2016swq, Lombardo:2017yme}. Even the Bianchi identities, the simplest known constraints for the flux choice, are not fully settled for generic Calabi Yau orientifold compactifications, as can be observed from the fact that there are two inequivalent formulations for these relations \cite{Ihl:2007ah,Robbins:2007yv,Shukla:2016xdy,Gao:2018ayp}.

In the context of moduli stabilization and the attempts to construct de-Sitter (dS) vacua, the models based on type II orientifolds with (non-)geometric fluxes have attracted a considerable amount of attention in recent years \cite{Aldazabal:2006up, Ihl:2006pp, Ihl:2007ah, Aldazabal:2008zza, deCarlos:2009qm,  Danielsson:2012by, Blaback:2013ht, Blumenhagen:2015xpa, Blaback:2015zra,Marchesano:2019hfb,Marchesano:2021gyv,Shukla:2022srx,Damian:2023ote}. Understanding the effective four-dimensional (non-geometric) scalar potentials and the possible constraints are among the central  issues. On these same lines, exploring the ten-dimensional origin of four-dimensional effective potentials, which is known as the {\it dimensional oxidation} process, has also been  one of the  main topics of interest \cite{Blumenhagen:2013hva, Gao:2015nra,Shukla:2015bca,Shukla:2015hpa,Shukla:2016hyy,Gao:2017gxk,Leontaris:2023lfc,Leontaris:2023mmm}.

 In this paper, we consider the  generic bilinear structure of the non-geometric scalar potential \cite{Marchesano:2020uqz}. \black{As commented in the original reference, this form of the potential is derived assuming that possible  corrections are suppressed  and do not spoil the axionic shift symmetries of the K\"ahler  potentials\footnote{\black{$\alpha'$ corrections respect this symmetry -at least at first order-. For simplicity, we will assume we can ignore them.} } \cite{Grimm:2019bey,Grimm:2019ixq}. Taking this as a working assumption,\footnote{\black{Of course, one should check that this is true once a particular vacuum is chosen.}}}  we present a detailed analysis regarding  the constraints on the fluxes to realize dS extrema, exploring under what conditions dS no-go can be derived.
 In the first part of the paper, we will follow a completely general approach. We will see that just including more and more fluxes does not guarantee that one can find dS extrema. The main results of this first part are expression \eqref{eq: the conditions} and tables \ref{tab: no go Rmu=0} and \ref{tab:nogen}.
 
 After the general discussion, we will improve our analytical control by looking at a particular Ansatz. More specifically, we will impose a proportionality relation between the $F$-terms and the K\"ahler derivatives when both are evaluated at the vacuum. This is the same scenario already studied in the presence of only RR and NSNS fluxes in \cite{Marchesano:2019hfb} and with geometric fluxes in \cite{Marchesano:2020uqz}. We will see that, under these conditions, the scalar potential can be compactly expressed using a set of eight so-called {\it axionic fluxes}. Then, the simple choice of imposing an axionic flux to be zero or non-zero generates 256 flux configurations. Of these, $227$ cases lead to dS no-go scenarios, whereas the other 29  choices remain undecided.

The paper is organized as follows. In section \ref{sec_IIAorientifold} we begin with a brief review on the generic type IIA non-geometric flux compactifications where we provide the necessary ingredients relevant for the current work. Section \ref{sec_systematics} presents a detailed analysis of the generic conditions which can lead to dS no-go results. In Section \ref{sec_Ansatz} we impose the Ansatz that $F$-terms are proportional to the respective K\"ahler derivatives and analyze the subsequent simplified expressions for the scalar potential and its various derivatives. Section \ref{sec_dS} provides a detailed classification of the dS no-go scenarios in the Ansatz. Finally, we summarize the findings in Section \ref{sec_conclusions}, and present all the other details and necessary intermediate steps in appendices \ref{appa}, \ref{app_auxiliadS},  and \ref{app_inter}. Two detailed lists of all the dS no-go cases, and those for which a dS no-go could not be obtained, are displayed in appendix \ref{app_no-go-list}.

	
\section{Preliminaries}
\label{sec_IIAorientifold}

Type IIA compactifications on Calabi-Yau orientifolds with fluxes are a fascinating corner of the string landscape, since they provide a framework where one can stabilize all the moduli in a weakly coupled, scale separated AdS regime with the classical potential generated by the $p$-form fluxes alone \cite{DeWolfe:2005uu,Camara:2005dc}. One can even consider a more extensive set of NS fluxes related to each other by T-dualities, as discussed in \cite{Shelton:2005cf}. Putting all these fluxes into the game yields a richer scalar potential,  studied in \cite{Aldazabal:2006up,Shelton:2006fd,Micu:2007rd,Ihl:2007ah,Wecht:2007wu,Robbins:2007yv} and more recently in \cite{Marchesano:2020uqz}. In this section, we will review the scalar potential generated when all the $p$-form, geometric and non-geometric fluxes are turned on over a CY orientifold. As shown in \cite{Marchesano:2020uqz}, using some previous work \cite{Gao:2017gxk}, such potential can be rewritten in a very compact way thanks to the bilinear formalism developed in \cite{Bielleman:2015ina,Herraez:2018vae}. This bilinear structure  has turned out to be a potent tool to study the vacua structure, as demonstrated in \cite{Escobar:2018tiu,Escobar:2018rna,Marchesano:2019hfb, Marchesano:2020uqz}. In the following, we will restrict ourselves to recalling the main ingredients we will need from \cite{Marchesano:2020uqz}, referring the reader to this reference for a more detailed discussion.


\subsection{Type IIA orientifolds with general fluxes}

Let us consider (massive) type IIA string theory compactified on an orientifold of $X_4 \times X_6$ with $X_6$ a compact Calabi-Yau three-fold. We take the standard orientifold quotient by $\Omega_p (-)^{F_L} {\cal I}$ \cite{Blumenhagen:2005mu,Blumenhagen:2006ci,Marchesano:2007de,Ibanez:2012zz},\footnote{Here $\Omega_p$ the worldsheet parity reversal operator and  ${F_L}$ spacetime fermion number for the left movers.}  with ${\mathcal I}$ an involution of the Calabi-Yau metric acting on the K\"ahler 2-form $J$ and the holomorphic 3-form $\Omega$ as ${\cal I}(J) = - J$ and ${\cal I} (\Omega) = e^{2i\theta} \ov \Omega$, respectively. Under the action of $\mathcal{I}$, the harmonic forms split into even and odd  $H^p\left(X_6\right)=H^p_+\oplus H^p_-$. We can introduce a basis for each of these subspaces:
\begin{table}[h]
\begin{center}
\begin{tabular}{| c || c | c| c | c | c | c |} \hline
   \rule[-0.3cm]{0cm}{0.9cm} cohomology group &  $\ H^{(1,1)}_+\ $ & 
   $\ H^{(1,1)}_-\ $ & $\ H^{(2,2)}_+\ $ & $\ H^{(2,2)}_-\ $ & $\ H^{3}_+\ $ & $\ H^{3}_-\ $
   \\ \hline
   \rule[-0.3cm]{0cm}{0.8cm} dimension &  $h^{(1,1)}_+$  & $h^{(1,1)}_- $  
                                       &  $h^{(1,1)}_-$  & $h^{(1,1)}_+$ 
                                       &  $h^{(2,1)}+1$  &  $h^{(2,1)}+1$ 
   \\ \hline
   \rule[-0.3cm]{0cm}{0.8cm} basis     & $\omega_\alpha$ & $\omega_a$
                                       & $\tilde \omega^a$ & $\tilde \omega^\alpha$
   & $ \alpha_\mu$ & $\beta^\mu$ \\ \hline
\end{tabular}
\caption{Extracted from \cite{Grimm:2004ua} and adapted to our conventions.  We are working in units where $\ell_s\equiv 2\pi\sqrt{\alpha'}=1$. For simplicity, with respect to \emph{standard} notation, we have gathered the $\mathcal{I}$-odd three-forms $(\alpha_I, \beta^ \Lambda)$ into $(\alpha_I, \beta^ \Lambda)\equiv \alpha_\mu$ and the    $\mathcal{I}$-even three-forms $(\beta^ J,\alpha_\Sigma)$ into $(\beta^ J,\alpha_\Sigma)\equiv \beta^\mu$}
\label{base}
\end{center}
\end{table}

In the absence of background fluxes and neglecting worldsheet and D-brane instanton effects, dimensional reduction to 4d leads to a $\mathcal{N}=1$ supergravity theory whose massless field content can be organized as follows \cite{Grimm:2004ua}:
\begin{itemize}
\item There are  $h_-^{1,1}$ complex fields, the complexified  K\"ahler moduli $T^a$. They come from the complexification of the K\"ahler 2-form  $J=e^{\frac{\phi}{2}}t^a\omega_a$ and the $B=b^a\omega_a$ field, defined by $J_c \equiv B + i\, e^{\frac{\phi}{2}} J = \left( b^a + i \, t^a\right) \om_a $. Here $J$ is in the Einstein frame, $\phi$ is the 10d dilaton. Integrating on the correspondent cycles the complex scalar fields in $4d$ are 
\begin{align}
	T^a&=b^a+i t^a\, , &	a\in\{1,\, \dots,\, h_-^{1,1}\}\, .
	\end{align} 
The metric appearing in  the kinetic terms is derived from the K\"ahler potential 
	\be
	K_K \,  = \,  -{\rm log} \left(\frac{4}{3} \cK \right) \, ,
	\label{KK}
	\ee
	where $\cK = \cK_{abc} t^at^bt^c = 6 {\rm Vol}_{X_6} = \frac{3}{4} {\cal G}_T$ is homogeneous of degree three on the $t^a$ and $\cK_{abc}$ are the CY triple intersection numbers, $\cK_{abc}=- \int_{X_6} \om_a \wedge \om_b \wedge \om_c$
\item There are  $1+h^{2,1}$ complex fields,  so-called complex structure moduli, coming from the integration of the complex 3-form $\Omega$, the axio-dilaton and the RR 3-form potential $C_3$. Defining $\Omega_c \equiv C_3 + i \, \RE ({\cal C} \Omega)$ where ${\cal C} \equiv  e^{- \phi- i \theta} e^{\frac{1}{2}(K_{cs} - K_K)}$ and $K_{cs} = - \log \left(-i\, \int_{X_6} \Omega \wedge  \ov \Omega \right)$ the complex structure moduli are defined as
\begin{align}
U^\mu&=\xi^\mu+i u^\mu=\int_{X_6}\Omega_c\wedge \beta^\mu\, ,	&	 \mu&\in\{0,\,\dots\, h^{2,1}\}\, .
\end{align}
Their kinetic terms are given in terms of the following piece of the K\"ahler potential
\begin{align}
\label{KQ}
K_Q=4\log\left(\frac{e^\phi}{\sqrt{\rm Vol_{X_6}}}\right)\equiv -\log\left(e^{-4D}\right)\, ,
\end{align}
where $D$ is the four-dimensional dilaton. The function $\mathcal{G}_Q=e^{-K_Q/2}$ is a homogeneous function of degree two in $u^\mu$. These moduli are redefined in the presence of D6-brane moduli, and so is the K\"ahler potential \eqref{KQ} \cite{Grimm:2011dx,Kerstan:2011dy,Carta:2016ynn,Herraez:2018vae}. For simplicity, we will not consider compactifications with D6-brane moduli in the following.
\item In addition to the spectrum of the chiral scalar fields, vector multiplets can also arise from the dimensional reduction of the 3-form $C_3$ if the CY has $h_+^{1,1}\neq 0$. For simplicity, we will only consider compactifications such that $h_+^{1,1}=0$. This assumption also prevents us from considering the D-term flux potential, which does not contribute in this case.
\end{itemize}

\subsubsection*{The flux superpotential}

On top of this background, one can add RR, NSNS, geometric and non-geometric  fluxes. Following the conventions of \cite{Marchesano:2019hfb} we will define the flux quanta of the RR fields as 
\begin{equation}
m \, = \, G_0\, ,  \quad  m^a\, =\,  \int_{X_6} \bar{G}_2 \wedge \tilde \omega^a\, , \quad  e_a\, =\, -  \int_{X_6} \bar{G}_4 \wedge \omega_a \, , \quad e_0 \, =\, -  \int_{X_6} \bar{G}_6 \, .
\label{RRfluxes}
\end{equation}
The flux superpotential including RR and also  geometric and non-geometric NS fluxes is described in terms of a twisted differential operator \cite{Shelton:2006fd}
	\be
	\label{eq:twistedD}
	{\cal D} = d - H \wedge  +\  f \triangleleft  +\ Q \triangleright  +\  R\, \bullet \, ,
	\ee
where $H$ is the NS three-form flux, $f$ encodes the geometric fluxes, $Q$ that of globally-non-geometric fluxes and $R$ is the locally-non-geometric fluxes, see e.g. \cite{Wecht:2007wu,Plauschinn:2018wbo} for more details. The action of various fluxes appearing in ${\cal D}$ is such that for an arbitrary $p$-form $A_p$, the pieces $H\wedge A_p$, $f \triangleleft A_p$, $Q \triangleright A_p$ and $R \bullet A_p$ denote a $(p+3)$, $(p+1)$, $(p-1)$ and $(p-3)$-form respectively. Following the conventions in  \cite{Ihl:2007ah}, the action of the different NS fluxes on the basis of harmonic $p$-form is
	\begin{align}
\label{eq:fluxActions0}
H \wedge {\bf 1} &= -h_\mu \beta^\mu \, , &  H \wedge \alpha_\mu &=   h_\mu \Phi_6 \nonumber\, , \nonumber\\
f \triangleleft \om_a &= -f_{a \mu}\, \beta^\mu \, , & f \triangleleft \omega_\alpha &= \hat{f}_{\alpha}{}^\mu\, \alpha_\mu\, , \nonumber\\
f \triangleleft \alpha_\mu &= -f_{a \mu} \, \tilde\om^a\, ,& f \triangleleft \beta^\mu &= -\,f_{\alpha}{}^\mu \, \tilde{\omega}^\alpha \,,\nonumber\\
& & \\
Q \triangleright \tilde\om^a &= -Q^{a}{}_{\mu}\, \beta^\mu\, , & Q \triangleright  \tilde{\omega}^\alpha &=  Q^{\alpha \mu}\, \alpha_\mu\, , \nonumber\\
Q \triangleright \alpha_\mu &= \, Q^{a}{}_{\mu} \, \om_a\, , & Q \triangleright \beta^\mu &=  Q^{\alpha\, \mu} \, \omega_\alpha\,, \nonumber\\
R \bullet {\Phi_6} &= R_\mu\, \beta^\mu \, ,& R \bullet \alpha_\mu &= R_\mu   {\bf 1} \,, \nonumber
\end{align}
where $\Phi_6$ is the normalised volume form,  $\int_{X_6} \Phi_6 = 1$, and we also have that $H\wedge \beta^\mu  = R \bullet \beta^\mu = 0$. The NS flux quanta are $h_\mu, f_{a\, \mu},f_{\a}{}^\mu, Q^a{}_\mu, Q^{\a\, \mu},  R_\mu \in \bZ$. This specifies the action of the twisted differential operator \eqref{eq:twistedD} on each $p$-form.

In addition, the internal RR fluxes can be gathered in a single polyform
	\be
	\bar{\bf{G}} = G_0 + \bar{G}_2 + \bar{G}_4 + \bar{G}_6 \,.
	\ee
	Given these definitions, the flux-generated superpotential  is $W = W_{\rm RR} + W_{\rm NS}$ \cite{Shelton:2006fd,Aldazabal:2006up}
	\be
	\label{eq:Wrrnsns}
	W_{\rm RR}=-  \int_{X_6} \bar{\bf{G}} \wedge e^{J_c}\, , \qquad  W_{\rm NS} = \int_{X_6} \Omega_c \wedge {\cal D}\left( e^{-J_c} \right)\,.
	\ee
Skipping the intermediate steps and referring again to \cite{Marchesano:2019hfb} for a complete derivation, these superpotentials have the following expressions \cite{Shelton:2006fd,Aldazabal:2006up,Micu:2007rd,Wecht:2007wu,Ihl:2007ah}
	\bea
	\label{eq:Wgen}
	W_{\rm RR} &= &e_0 +  e_aT^a + \frac{1}{2}\, {\cal K}_{abc}  m^a T^b T^c  + \frac{m}{6}\, {\cal K}_{abc}\, T^a T^bT^c \, , \\
	W_{\rm NS} &= & U^\mu \Bigl[ h_\mu + f_{a\mu} T^a + \frac{1}{2} {\cal K}_{abc} \, T^b \, T^c \, Q^a{}_\mu + \frac{1}{6}\, {\cal K}_{abc} T^a T^b T^c \, R_\mu \Bigr] \, ,
	\label{eq:WgenNS}
	\eea
where $e_0, e_a, m^a, m, h_\mu,  f_{a\mu}, Q^a{}_\mu,  R_\mu$ are all integers.

\subsection{The F-term flux potential}
	
	Under the assumption that background fluxes do not affect the K\"ahler potential pieces \eqref{KK} and \eqref{KQ},\footnote{The validity of this assumption should not be taken for granted and will depend on the particular class of vacua. The results in \cite{Junghans:2020acz,Buratti:2020kda,Marchesano:2020qvg} suggest that it is valid in the presence of only $p$-form fluxes $F_{\rm RR}$, $H$. However,  \cite{Font:2019uva} gives an example of compactification with metric fluxes in which fluxes heavily correct the naive KK scale, and so should be the K\"ahler potential.} one can easily compute the F-term flux potential for closed string moduli via the standard supergravity expression
	\be
	\label{eq:VFgen}
	V_F =  e^K \left(K^{{\cal A}\ov{\cal B}}\, D_{\cal A} W \, \ov{D}_{\ov{\cal B}} \ov{W} - 3 \, |W|^2\right),
	\ee
	where the index ${\cal A} = \{a, \mu\}$ runs over all  moduli. In \cite{Marchesano:2019hfb}, following what was done in \cite{Bielleman:2015ina,Herraez:2018vae}, it was shown that this F-term potential displays a bilinear structure of the form
	\begin{equation}\label{VF}
	V_F  = {\rho}_\cA \, Z^{\cA\cB} \, {\rho}_\cB\, , 
	\end{equation}  
where the matrix entries $Z^{\cA\cB}$ only depend on the saxions $\{t^a, n^\mu \}$, while the ${\rho}_\cA$ only depend on the flux quanta and the axions $\{b^a, \xi^\mu\}$. Indeed, one can easily rewrite the results in \cite{Gao:2017gxk} to fit the above expression, obtaining the following result. The set of axion polynomials with flux-quanta coefficients are
	\begin{equation}
	\rho_\cA=\{\rho_0,\rho_a,\tilde{\rho}^a,\tilde{\rho},\rho_\mu,\rho_{a\mu},\tilde{\rho}^a_\mu ,\tilde{\rho}_\mu \}\, ,
	\label{rhos}
	\end{equation}
	and are defined as
	\bes
	\label{RRrhos}
	\begin{align}
	\rho_0&=e_0+e_ab^a+\frac{1}{2}\mathcal{K}_{abc}m^ab^bb^c+\frac{m}{6}\mathcal{K}_{abc}b^ab^bb^c+\rho_\mu\xi^\mu\, , \label{eq: rho0}\\
	\rho_a&=e_a+\mathcal{K}_{abc}m^bb^c+\frac{m}{2}\mathcal{K}_{abc}b^bb^c+\rho_{a\mu}\xi^\mu \, ,  \label{eq: rho_a}\\
	\tilde{\rho}^a&=m^a+m b^a + \tilde{\rho}^a_\mu\xi^\mu \, ,  \label{eq: rho^a}\\
	\tilde{\rho}&=m+\tilde{\rho}_\mu\xi^\mu \, ,   \label{eq: rhom}
	\end{align}   
	\ees 
	and
	\bes
	\label{NSrhos}
	\begin{align}    
	\rho_\mu&=h_\mu+f_{a\mu}b^a+\frac{1}{2}\mathcal{K}_{abc}b^bb^cQ_\mu^a+\frac{1}{6}\mathcal{K}_{abc}b^ab^bb^cR_\mu \, , \\
	\rho_{a\mu}&=f_{a\mu}+\mathcal{K}_{abc}b^bQ^c_\mu+\frac{1}{2}\mathcal{K}_{abc}b^bb^cR_\mu \, ,  \label{eq: rho_ak} \\
	\tilde{\rho}^a_\mu &=Q^a_\mu+b^aR_\mu \, , \\
	\tilde{\rho}_\mu &=R_\mu \, .
	\end{align}
	\ees
	The polynomials \eqref{NSrhos} are primarily new with respect to the Calabi--Yau case with $p$-form fluxes, as they highly depend on the presence of geometric and non-geometric fluxes. As in \cite{Herraez:2018vae}, both \eqref{RRrhos} and \eqref{NSrhos} have the interpretation of invariants under the discrete shift symmetries of the combined superpotential $W = W_{\rm RR} + W_{\rm NS}$.  
	
For the bilinear form $Z$, one finds the following expression
	\be
	\label{eq:Z-matrix}
	Z^{{\cal A}{\cal B}} =  e^K \, \begin{bmatrix}
		{\bf G}  \quad & \, \, {\cal O} \\
		{\cal O}^{\, t} \quad  & \, \, {\bf C}
	\end{bmatrix}\, ,
	\ee
	where
	\begin{equation}
	{\bf G} =\left(\begin{array}{cccc}
	4 & 0 & 0 & 0 \\
	0 & g^{ab} & 0 & 0\\
	0 & 0 & \frac{4\mathcal{K}^2}{9}g_{ab} & 0 \\
	0 & 0 & 0 & \frac{\mathcal{K}^2}{9}
	\end{array}
	\right)\, , \quad 
	\mathcal{O} =\left(\begin{array}{cccc}
	0 & 0 & 0 & -\frac{2\mathcal{K}}{3}u^\nu \\
	0 & 0 &  \frac{2\mathcal{K}}{3}u^\nu \delta^a_b & 0\\
	0 & -\frac{2\mathcal{K}}{3}u^\nu\delta^b_a & 0 & 0 \\
	\frac{2\mathcal{K}}{3}u^\nu & 0 & 0 & 0
	\end{array}
	\right)\, ,
	\label{eq:GOmatrix}
	\end{equation}
	\begin{equation}
	{\bf C} =\left(\begin{array}{cccc}
	c^{\mu\nu} & 0 & -\tilde{c}^{\mu\nu}\frac{\mathcal{K}_b}{2} & 0 \\
	0 & \tilde{c}^{\mu\nu}t^at^b+ g^{ab}u^\mu u^\nu  &  0 & -\tilde{c}^{\mu\nu}t^a\frac{\mathcal{K}}{6}\\
	-\tilde{c}^{\mu\nu}\frac{\mathcal{K}_a}{2} & 0 & \frac{1}{4}\tilde{c}^{\mu\nu}\mathcal{K}_a\mathcal{K}_b+\frac{4\mathcal{K}^2}{9}g_{ab}u^\mu u^\nu  & 0 \\
	0 & -\tilde{c}^{\mu\nu}t^b\frac{\mathcal{K}}{6} & 0 & \frac{\mathcal{K}^2}{36}c^{\mu\nu}
	\end{array}
	\right)\, .
	\end{equation}
	Here $K = K_K + K_Q$, $g_{ab} = \frac{1}{4} \partial_{t^a} \partial_{t^b} K_K\equiv \frac{1}{4}\p_a\p_b K_K$, and $c_{\mu\nu} = \frac{1}{4} \partial_{u^\mu}\partial_{u^\nu} K_Q\equiv \frac{1}{4}\p_\mu\p_\nu K_Q$, while upper indices denote their inverses. We have defined  $\mk_{a}=\mk_{abc}t^bt^c$ and $\tilde{c}^{\mu\nu}=c^{\mu\nu}-4u^\mu u^\nu $. 
	
Putting all this together, the final expression for the F-term potential is explicitly given in expressions \eqref{eq:VgenAnsatz1}-\eqref{eq:VgenAnsatz2}, which generalizes the results of \cite{Herraez:2018vae} -which only considered RR and NSNS fluxes- and can be easily connected to other known formulations of (non-)geometric potentials in the type IIA literature, e.g. \cite{Villadoro:2005cu,Flauger:2008ad,Blumenhagen:2013hva,Shukla:2019akv}.
	
\subsection{Bianchi identities and the tadpole constraints} 
	
Before coming to the study of flux vacua, it is important to look for the constraints arising from the Bianchi identities. One needs to impose them on the fluxes to get the genuine scalar potential. In our setup, these constraints can be obtained by imposing that the twisted differential ${\cal D}$ in \eqref{eq:twistedD} satisfies the nilpotency condition ${\cal D}^2 = 0$ when applied on the $p$-form basis of the different
cohomology groups \cite{Ihl:2007ah,Robbins:2007yv}. Using the flux actions (\ref{eq:fluxActions0}) in the nilpotency condition, one easily obtains the following constraints: 
    \bea
    \label{eq:bianchids2}
    & & h_\mu\, \hat{f}_\alpha{}^\mu = 0\, , \quad h_\mu\, \hat{Q}^{\alpha \mu} = 0\, , \quad f_{a\mu}\, \hat{f}_\alpha{}^\mu = 0\, , \quad f_{a\mu}\, \hat{Q}^{\alpha \mu} =0\, , \nonumber\\
    & & R_\mu\, \hat{Q}^{\alpha \mu} = 0\, , \quad R_\mu \, \hat{f}_\alpha{}^\mu = 0\, , \quad Q^a{}_\mu \, \hat{Q}^{\alpha \mu} = 0\, , \quad \hat{f}_\alpha{}^{\mu} \,Q^a{}_\mu=0\, ,\\
    & & \hat{f}_\alpha{}^{[\mu}\, \hat{Q}^{\alpha \nu]} = 0\, , \quad h_{[\mu} \, R_{\nu]} = f_{a[\mu}\, Q^a{}_{\nu]}\, .\nonumber
    \eea
Let us mention that, for our current work, we will focus on the orientifold constructions for which $h^{1,1}_+ = 0$ and therefore all the fluxes, moduli and axions with  $\alpha$ index  will be absent. This means that the only conditions which survive the orientifold project are given by
\bea
\label{eq:bianchids3}
& & h_{[\mu} \, R_{\nu]} = f_{a[\mu}\, Q^a{}_{\nu]} \quad \Longleftrightarrow \quad \r_{[\mu} \, \tr_{\nu]} = \r_{a[\mu}\, \tr^a{}_{\nu]},
\eea
where the last equation follows from the interesting fact that the set of constraints arising from the Bianchi identities continues to hold even after all the fluxes are promoted to their respective axionic-fluxes \cite{Shukla:2016xdy,Gao:2017gxk,Shukla:2019wfo}. However, as we have mentioned earlier, the two known formulations of the Bianchi identities, namely the {\it Standard formulation} and the {\it Cohomology formulation}, turn out to result in two inequivalent sets of constraints \cite{Ihl:2007ah,Robbins:2007yv,Shukla:2016xdy,Gao:2018ayp}. In fact, the constraints arising from the Cohomology formulation are contained in the Standard formulation, which, in addition has some more constraints, refereed as ``missing" Bianchi identities in \cite{Gao:2018ayp}. So, given such state-of-the-art, even if one manages to realize a suitable (dS) solution in a cohomology approach, the possibility of having additional constraints may not be generically ruled out! 

Complementing the flux constraints arising from the NSNS Bianchi identities, there are tadpole cancellation conditions that one must consider while constructing concrete models. These RR Bianchi identities are slightly non-trivial compared to those involving the NSNS flux. They can be encoded in the following,
\bea
\label{eq:tadpole}
& & {\cal D} \bar{\bf{G}} = \sum_{D6/D6'} q_{D6/D6'} + \sum_{O6} q_{O6},
\eea
where $q_{D6/D6'}$ and $q_{O6}$ denote the three-forms corresponding to the respective charges of the $D6$-brane, its image $D6'$-brane and the $O6$-plane which can be consistently allowed by the choice of orientifold involution. For our case, the left side of eqn.~(\ref{eq:tadpole}) turns out to be a flux-invariant given by
\bea
\label{eq:tadpole-flux-invariants}
& & {\cal D} \bar{\bf{G}} = \left(m\, h_\nu - m^a\,f_{a\nu}+ e_a\,{Q}^a_\nu-e_0\,R_\nu\right)\beta^\nu = \left(\tilde{\rho}\rho_\nu - \tilde{\rho}^a\rho_{a\nu}+\rho_a\tilde{\rho}^a_\nu-\rho_0\tilde{\rho}_\nu\right)\beta^\nu.
\eea
However, the introduction of appropriate local sources can be argued to cancel the tadpole charges, which can be determined once the local brane settings and $O$-planes are concretely specified in the model of interest. A recent update about (the challenges in) incorporating these constraints in the non-geometric models can be found in \cite{Plauschinn:2020ram}.

\subsection{Flux invariants and moduli fixing}
\label{app_invariants}

The bilinear formalism of the scalar potential and the use of gauge-invariant flux-axion polynomials provide a powerful tool to study the moduli stabilization processes systematically. Since the flux polynomials depend on the axions $b^a$, moduli fixing amounts to solve a system of algebraic equations in the saxions $t^a$ and the flux polynomials $\rho_\mathcal{A}$.  However, this change of framework comes at the cost of decoupling the number of variables from the number of moduli, since the size of the set $\rho_\mathcal{A}$ scales with the dimension of the flux lattice and not with the dimension of the moduli space. As a result, in most cases,  the number of polynomials will exceed the rank of the system of equations of motion. This is not a problem to achieve a well-determined solution because the set $\rho_\mathcal{A}$ is not composed of fully independent variables. There are many constraints arising from the way these objects have been defined. Such constraints are expressed in terms of a set of invariant combinations of flux-axion polynomials, which are invariant under shifts of the axions. The main invariants relevant to the construction considered in this paper were identified in \cite{Marchesano:2020uqz}. At the linear level, there is of course $\tilde{\rho}_\mu=R_\mu$. At the quadratic level, we have
\be
\tilde{\rho}\rho_\mu-\tilde{\rho}^a\rho_{a\mu} + \rho_a \tilde{\rho}^a_\mu -\rho_0\tilde{\rho}_\mu\, ,
   \label{invngr}
\ee
and
\be
\tilde{\rho}_{[\mu}^a \tilde{\rho}_{\nu]}\, , \qquad \r_{[\mu} \, \tr_{\nu]} - \r_{a[\mu}\, \tr^a{}_{\nu]}\, , \qquad    \rho_{a(\mu}\tilde{\rho}_{\nu)}-\mathcal{K}_{abc}\tilde{\rho}^b_\mu\tilde{\rho}^c_\nu\, ,
\label{finalNGinv}
\ee
where in the above $(\ )$ and $[\ ]$ stand for symmetrization and anti-symmetrization of indices, respectively. We again emphasize that the expressions above can be written exclusively in terms of the flux quanta and are therefore moduli independent. 

Let us also mention that there appears to be a close correlation between the flux invariants, the Bianchi identities and the tadpole contributions. For example, the flux invariant (\ref{invngr}) appears in the $V_{loc}$ piece in the scalar potential (\ref{eq:VgenAnsatz2}). Further, as mentioned in eq.~(\ref{eq:bianchids3}), the following NSNS Bianchi identity holds and happens to be the only one (in the cohomology formulation) which survives the orientifold projection with $h^{1,1}_+ = 0$ \cite{Gao:2017gxk},
\bea
& & h_{[\mu} \, R_{\nu]} = f_{a[\mu}\, Q^a{}_{\nu]} \quad \Longleftrightarrow \quad \r_{[\mu} \, \tr_{\nu]} = \r_{a[\mu}\, \tr^a{}_{\nu]}.
\eea
As mentioned in (\ref{finalNGinv}), it turns out that $\r_{[\mu} \, \tr_{\nu]} - \r_{a[\mu}\, \tr^a{}_{\nu]} = h_{[\mu} \, R_{\nu]} - f_{a[\mu}\, Q^a{}_{\nu]}$ is also an invariant flux combination. Moreover, it has been shown in \cite{Gao:2018ayp} that there are additional, the so-called missing Bianchi identities, in explicit toroidal orientifold models based on $\mathbb T^6/(\mathbb Z_2 \times \mathbb Z_2)$ and $\mathbb T^6/\mathbb Z_4$ orbifolds. Some of these are given as below,
\bea
& &  f_{a(\mu}\,{R}_{\nu)} = \mathcal{K}_{abc}{Q}^b_\mu\,{Q}^c_\nu\,, \qquad 3 h_{(\mu}\, R_{\nu)} = f_{a (\mu} \, Q^a_{\nu)}.
\eea
Since the set of  Bianchi identities which hold for the usual fluxes remains to hold (as the whole set of constraints) when those usual fluxes are promoted to their respective axionic-fluxes \cite{Shukla:2016xdy,Gao:2017gxk,Shukla:2019wfo}, such missing Bianchi identities can be written as follows,
\bea
& &  \rho_{a(\mu}\tilde{\rho}_{\nu)} = \mathcal{K}_{abc}\tilde{\rho}^b_\mu\tilde{\rho}^c_\nu\,, \qquad 3 \r_{(\mu}\, \tr_{\nu)} = \r_{a (\mu} \, \tr^a_{\nu)}.
\eea
And therefore, for such cases, $\rho_{a(\mu}\tilde{\rho}_{\nu)} - \mathcal{K}_{abc}\tilde{\rho}^b_\mu\tilde{\rho}^c_\nu$ and $3 \r_{(\mu}\, \tr_{\nu)} - \r_{a (\mu} \, \tr^a_{\nu)}$ are invariant fluxes as it has been observed in \cite{Marchesano:2020uqz}. This shows a very interesting connection between the set of Bianchi identities and the set of flux invariants. Following the study of Bianchi identities in type IIA context in \cite{Gao:2017gxk}, one would expect several additional ``missing" flux invariants corresponding to the ``missing" Bianchi identities in a given concrete construction. In more generic orientifold constructions with $h^{1,1}_+ \neq 0$ and therefore fluxes supporting the $D$-terms, one would expect to have even more flux invariants, which generate obstructions to achieve full moduli stabilization \cite{Plauschinn:2020ram}.

The existence of these invariants constrains the dimension of the orbit of possible values for the set $\rho_{\mathcal{A}}$. The surviving degrees of freedom,  which by construction are never larger than the number of axionic fields, are later fixed by the equations of motion.  Depending on the Hodge numbers of the underlying Calabi-Yau orientifold, even when $h^{1,1}_+ \neq 0$, the constraints arising from the invariants for generic flux configurations can be too strong to allow complete moduli stabilization. Such restrictions were already observed through the lenses of the Bianchi identities and the tadpole conditions in \cite{Plauschinn:2020ram}.

The rank of the choices of flux quanta also plays a crucial role, since new invariants may arise while others become trivial when lowering this value. For instance, if the non-geometric fluxes are set to zero (i.e. $Q^a_\mu=R_\mu=0$), we have the following  invariants at the linear level
\be 
 \tilde{\rho}= m\, ,\qquad \quad    \rho_{a\mu}=   f_{a\mu}\, ,
 \label{invmetl}
\ee
while at the quadratic level we have
\be
 \tilde{\rho}\rho_\mu-\tilde{\rho}^a\rho_{a\mu} =  \left(mh_\mu - m^a f_{a\mu}\right)  \, , \qquad  c^{a} \left(\tilde{\rho}\rho_{a}-\frac{1}{2}\mathcal{K}_{\bar{abc}}\tilde{\rho}^b\tilde{\rho}^c\right)\, .
 \label{invmetq}
\ee
Here the $c^a \in \mathbb{Z}$ are such that $c^a \rho_{a\mu} = 0$ $\forall \mu$. This means that there are $h^{1,1}_- - r_f$ of this last class of invariants, with $r_f$ the rank of $h^{1,1}\times (1+h^{2,1})$ matrix $f_{a\mu}$. Considering all these invariants, one can see that $\rho_\cA$ takes values in a $(1 + h^{1,1}_- + r_f)$-dimensional orbit.



\section{Analytics of the non-geometric flux vacua}
\label{sec_systematics}

Assuming that the choice of orientifold action is such that $h_+^{1,1}=0$ , which translates into the fact that there are no D-terms, the non-geometric scalar potential arising from the $F$-term contributions can  be clubbed into the following three pieces,
\bea
\label{eq:VgenAnsatz1}
& & \hskip-2cm  V \equiv V_F = e^K \left(V_{RR} + V_{NS} + V_{loc} \right),
\eea
where
\bea
\label{eq:VgenAnsatz2}
&& V_{RR} = e^K \, \biggl[4\rho_0^2+g^{ab}\rho_a\rho_b+\frac{4\mathcal{K}^2}{9}g_{ab}\tilde{\rho}^a\tilde{\rho}^b+\frac{\mathcal{K}^2}{9}\tilde{\rho}^2 \biggr], \\
& &  V_{NS} = e^K \, \biggl[c^{\mu\nu}\rho_\mu\rho_\nu+\left(\tilde{c}^{\mu\nu}t^at^b+g^{ab}u^\mu u^\nu \right)\rho_{a\mu}\rho_{b\nu} +\left(\tilde{c}^{\mu\nu}\frac{\mathcal{K}_a}{2}\frac{\mathcal{K}_b}{2}+\frac{4\mathcal{K}^2}{9}g_ {ab}u^\mu u^\nu \right)\tilde{\rho}^a_\mu\tilde{\rho}^b_\nu \nonumber\\
& & \hskip2cm +\frac{\mathcal{K}^2}{36}c^{\mu\nu}\tilde{\rho}_\mu\tilde{\rho}_\nu -\tilde{c}^{\mu\nu}\mathcal{K}_a\rho_\mu\tilde{\rho}^a_\nu-\frac{\mathcal{K}}{3}\tilde{c}^{\mu\nu}t^a\rho_{a\mu}\tilde{\rho}_\nu \biggr],\nonumber\\
& & V_{loc} =  \frac{4}{3} e^K \mathcal{K}u^\nu\, \biggl[ \tilde{\rho}\rho_\nu - \tilde{\rho}^a\rho_{a\nu}+\rho_a\tilde{\rho}^a_\nu-\rho_0\tilde{\rho}_\nu   \biggr]. \nonumber
\eea
For $V_{RR}$ and $V_{NS}$, we have considered this classification  promoting the usual scalar potential pieces for the standard RR and NS-NS fluxes to the corresponding axion polynomials. On the other hand, $V_{loc}$ can be related to the tadpole equations, which we choose to cancel without introducing any D6 branes. Let us also point out that $V_{RR}$ is a positive semi-definite quantity, whereas the sign of the $V_{NS}$ and the $V_{loc}$ contributions is a priori not fixed.

The generic superpotential can be expressed in terms of the axionic flux combinations after using eqs.~(\ref{RRrhos})-(\ref{NSrhos}) into eqs.~(\ref{eq:Wgen})-(\ref{eq:WgenNS}),
\bea
\label{eq:simpW}
& & W = \biggl[\left(\rho_0 - \frac{1}{2}\, {\cal K}_a \, \tilde\rho^a\right) - u^\mu \left(\rho_{a\mu}\, t^a - \frac{{\cal K}}{6} \, \tilde\rho_\mu\right)\biggr]\\
& & \qquad \qquad \qquad + \, i \, \biggl[\left(\rho_a\, t^a - \frac{{\cal K}}{6}\,\, \tilde\rho\right) + u^\mu \left(\rho_{\mu}\, - \frac{1}{2}\, {\cal K}_a\, \tilde\rho^a_\mu\right)\biggr]. \nonumber
\eea
Subsequently, the $F$-terms arising from the (non-)geometric flux superpotential pieces given in eqs.~\eqref{eq:Wgen}-\eqref{eq:WgenNS} simplify to the following form,
\begin{align}
F_a =&\left[\rho_a-\mk_{ab}\tr^b_\mu u^\mu-\frac{3}{2}\frac{\mk_a}{\mk}\left(t^b\rho_b+u^\mu\rho_\mu-\frac{1}{2}\mathcal{K}_b\tilde{\rho}^b_\mu u^\mu+\frac{1}{6}\mk\tr\right)\right]\nonumber\\ +&i\left[\mk_{ab}\tr^b+\rho_{a\mu}u^\mu+\frac{3}{2}\frac{\mk_a}{\mk}\left(\rho_0-t^au^\mu\rho_{a\mu}-\frac{1}{2}\mk_b\tr^b-\frac{1}{6}\mk \tilde{\rho}_\mu u^\mu\right)\right]\, \label{eq: F-Ta}\, ,\\
F_\mu =&\left[\rho_\mu-\frac{1}{2}\mk_a\tr^a_\mu+\frac{\p_\mu K}{2}\left(t^a\rho_a+u^\nu\rho_\nu-\frac{1}{2}\mathcal{K}_b\tilde{\rho}^b_\nu u^\nu-\frac{1}{6}\mk\tr\right)\right]\nonumber \\ +&i\left(t^a\rho_{a\mu}-\frac{1}{6}\mk \tilde{\rho}_\mu-\frac{\p_\mu K}{2}\left(\rho_0-t^au^\nu\rho_{a\nu}-\frac{1}{2}\mk_b\tr^b+\frac{1}{6}\mk \tilde{\rho}_\nu u^\nu\right)\right)\, .
\label{eq: F-Umu}
\end{align}

\subsection{SUSY vacua conditions}
Using the $F$-terms in eqs.~(\ref{eq: F-Ta})-(\ref{eq: F-Umu}), we impose the $F$-flatness conditions:
\bea
\label{eq:F-flatness}
& & F_{a} = 0 = F_{\mu},
\eea
which in particular leads to the following two complex constraints,
\begin{align}
t^a\, F_{a}=&\left[-\frac{1}{2}\rho_a\, t^a- \frac{1}{4}\mk_{a}\tr^a_\mu u^\mu - \frac{3}{2}u^\mu\rho_\mu-\frac{\mk}{4}\tilde{\rho}\right]\nonumber\\ +&i\left[\frac{3}{2}\rho_0+\frac{1}{4}\mk_{a}\tr^a - \frac{1}{2}\rho_{a\mu}\, t^a \, u^\mu-\frac{\mk}{4}\tr_\mu u^\mu\right] = 0\,,\label{eq: F-Ta1}
\end{align}
\begin{align}
u^\mu\, F_{\mu}=&\left[-\rho_\mu \, u^\mu +\frac{1}{2}\,\mk_a\tr^a_\mu \, u^\mu -2\, t^a\rho_a  +\frac{\mk}{3}\tr\right]\nonumber \\ 
+ & i\left[2 \rho_0 -\rho_{ a\mu}t^a \, u^\mu  -\mk_a \tr^a +\frac{\mk}{6} \tilde\rho_\mu \, u^\mu \right] = 0\, .\label{eq: F-Umu1}
\end{align}
These four real conditions can be solved for different axion flux polynomials taken as different variables. One such solution of (\ref{eq: F-Ta1})-(\ref{eq: F-Umu1}) can be given as under,
\bea
\label{eq:susy-simp-Ansatz1}
& & \rho_\mu \, u^\mu = -\frac{3\,\rho_a\, t^a}{4}\, - \frac{\mk\,\tr}{24}\,, \qquad \qquad \rho_{a\mu}\, t^a  \, u^\mu = \frac{9 \r_0}{4} -\,\frac{5\, \mk_a \, \tilde\rho^a}{8}\,,\\
& & \tilde\rho^a_\mu\, \mk_a\, u^\mu = \frac{5\, \rho_a\, t^a}{2} - \frac{3\, \mk\,\tr}{4}, \qquad \qquad \frac{\mk}{6} \, \tilde\rho_\mu\, u^\mu = \frac{\r_0}{4}+ \frac{3\, \mk_a \, \tilde\rho^a}{8}  \,. \nonumber
\eea
Note that the above solution is useful when we want to get rid of the NS-NS $\rho$ polynomials for determining the vacuum, as it helps in expressing them in terms of the generalized RR sector  polynomials. After imposing these four $F$-flatness conditions, the superpotential (\ref{eq:simpW}) simplifies into the following form,
\bea
\label{eq:simpWsusy1}
& & W_{SUSY} = \biggl[\left(-\rho_0 +\textcolor{black}{ \frac{1}{2}}\, {\cal K}_a \, \tilde\rho^a\right) - \, i \, \left(\rho_a\, t^a - \frac{{\cal K}}{6}\,\, \tilde\rho\right) \biggr]. 
\eea
Subsequently, the AdS minimum is given as,
\bea
& & \hskip-1.2cm \la V_{SUSY} \ra = - 3\, e^K \, |W_{SUSY}|^2 = - 3\, e^K \, \biggl[\left(\rho_0 - \textcolor{black}{\frac{1}{2}}\, {\cal K}_a \, \tilde\rho^a\right)^2 + \, \left(\rho_a\, t^a - \frac{{\cal K}}{6}\,\, \tilde\rho\right)^2 \,\biggr],
\eea
yielding a negative semidefinite scalar potential expressed as a sum of quadratic combinations of flux-axion polynomials. Now, in order to have the chance to obtain de Sitter vacua, we will explore the non-SUSY landscape.

\subsection{Non-SUSY vacua conditions}
In this subsection we perform an analytical exploration of the generic (non-SUSY) flux vacua.\footnote{\black{Strictly speaking, we should always talk about the  \emph{extrema} of the potential rather than about the vacua, since we still need to check the absence of flat directions and the perturbative stability through the Hessian. Nevertheless, we will use both words indistinguishably, bearing in mind that for a given extremum one should study the second derivatives of the potential to see if it is a vacuum.}} These vacua can be globally studied from the extremization of the scalar potential given in eqs.~(\ref{eq:VgenAnsatz1})-(\ref{eq:VgenAnsatz2}). In this regard, we present the relevant expressions of the various scalar potential derivatives. We will not calculate or discuss here the Hessian associated with these vacua, keeping in mind that it has to be computed to check its perturbative stability. Our main aim is to find swampland-like inequalities by considering the derivatives of the scalar potential and their various suitable combinations, as we will elaborate later on.

\vspace*{.5cm}
\noindent
\textbf{Axionic derivatives:}\\
\noindent
The derivatives of the scalar potential corresponding to the $C_3$ axions $\xi^\sigma$ are 
\bea
& & \hskip-1.5cm \frac{\partial V}{\partial \xi^\sigma} = e^K\, \biggl[8\rho_0 \, \rho_\sigma \, + 2 \, g^{ab}\rho_a\rho_{b\sigma} \, +\frac{8\mathcal{K}^2}{9}g_{ab}\tilde{\rho}^a\tilde{\rho}^b_\sigma \, +\frac{2\mathcal{K}^2}{9}\tilde{\rho_\sigma}\tilde{\rho} \, \nonumber\\
& & -\frac{4\mathcal{K}}{3}\rho_\sigma\tilde{\rho}_\nu u^\nu \, +\frac{4\mathcal{K}}{3}\rho_{a\sigma}\tilde{\rho}^a_\nu u^\nu \,  -\frac{4\mathcal{K}}{3}\tilde{\rho}^a_\sigma\rho_{a\nu}u^\nu \, +\frac{4\mathcal{K}}{3}\tilde{\rho}_\sigma\rho_\nu u^\nu \, \biggr].
\eea
Let us note that the last four terms  are trivial due to the fact that usual NS-NS Bianchi identities continue to be satisfied after promoting the fluxes into the axionic $\rho$-flux polynomials.  In fact, one has the following constraint from the generalized Bianchi identities \cite{Shukla:2016xdy,Gao:2017gxk,Shukla:2019wfo},
\bea
& & H_{[\underline\sigma} \, R_{\underline \mu]} = f_{a [\underline\sigma} \, Q^a_{\underline\mu]}, \qquad \Longrightarrow \qquad \rho_{[\underline\sigma} \, \tilde{\rho}_{\underline \mu]} = \rho_{a [\underline\sigma} \, \tilde{\rho}^a_{\underline\mu]}\,,
\eea
where the underlined indices are anti-symmetrized. Therefore, the $\xi^\sigma$ derivative can be recast as
\bea
\label{eq:der-ximu}
& & \hskip-1cm e^{-K}\, \frac{\partial V}{\partial \xi^\sigma} = 8\rho_o \, \rho_\sigma \, + 2\, g^{ab}\rho_a\rho_{b\sigma} \,  +\frac{8\mathcal{K}^2}{9}g_{ab}\tilde{\rho}^a\tilde{\rho}^b_\sigma \,  +\frac{2\mathcal{K}^2}{9}\tilde{\rho_\sigma}\tilde{\rho}.
\eea
Similarly, we get the following derivative of the scalar potential w.r.t. the $b^a$ axion,
\bea
\label{eq:der-ba}
& & \hskip-0.1cm e^{-K}\,\frac{\partial V}{\partial b^a} = 8\, \rho_0\, \rho_a\, + 2 \, g^{bc}\mathcal{K}_{abd} \, \tilde{\rho}^d \rho_c+\frac{8\mathcal{K}^2}{9}\, g_{ac} \, \tilde{\rho} \, \tilde{\rho}^c  + 2 \, c^{\mu\nu} \, \rho_{a\mu} \, \rho_\nu \nonumber\\
& & \hskip1cm + 2 \, \left(\tilde{c}^{\mu\nu}t^bt^c+g^{bc}u^\mu u^\nu \right)\mathcal{K}_{abd}\tilde{\rho}^d_\mu\rho_{c\nu} +  \, \left(\tilde{c}^{\mu\nu}\frac{\mathcal{K}_a \mathcal{K}_c}{2} + \frac{8\mathcal{K}^2}{9} g_{ac} \, u^\mu u^\nu \right)\tilde{\rho}_\mu\tilde{\rho}_\nu^c \nonumber\\
& & \hskip1cm - \, \tilde{c}^{\mu\nu} \left(\rho_{a\mu} \tilde{\rho}^b_\nu \, {\mathcal{K}_b} + \rho_\mu\tilde{\rho}_\nu {\mathcal{K}_a} \right) -\frac{1}{3}\, \tilde{c}^{\mu\nu} \mathcal{K}\mathcal{K}_{ac} \tilde{\rho}^c_\mu \tilde{\rho}_\nu\,.
\eea
\newpage
\textbf{Saxionic derivatives:}\\
\noindent
The two saxionic derivatives are presented below,
\begin{align}
 e^{-K} \, \frac{\partial V}{\partial u^\sigma}=&\left[e^{-K} V\partial_\sigma K-\frac{4\mathcal{K}}{3}\tilde{\rho}_\sigma\rho_0+\frac{4\mathcal{K}}{3}\tilde{\rho}^a_\sigma\rho_a-\frac{4\mathcal{K}}{3}\tilde{\rho}^a\rho_{a\sigma}+\frac{4\mathcal{K}}{3}\tilde{\rho} \rho_\sigma +\partial_\sigma c^{\mu\nu}\rho_\mu\rho_\nu \right.\nonumber\\
&-\mathcal{K}_b(\partial_\sigma c^{\mu\nu}\rho_\mu\tilde{\rho}_\nu^b-4\rho_\sigma\tilde{\rho}^b_\nu u^\nu -4\rho_\mu u^\mu \tilde{\rho}_\sigma^b)+t^at^b(\partial_\sigma c^{\mu\nu}\rho_{a\mu}\rho_{b\nu}-8\rho_{a\sigma}\rho_{b\nu}u^\nu )\nonumber\\
&+2g^{ab}\rho_{a\sigma}\rho_{b\nu}u^\nu -t^a\frac{\mathcal{K}}{3}(\partial_\sigma c^{\mu\nu}\rho_{a\mu}\tilde{\rho}_\nu-4\rho_{a\sigma}\tilde{\rho}_\nu u^\nu -4\rho_{a\mu}u^\mu \tilde{\rho}_\sigma)\nonumber\\
&\left. +\frac{\mathcal{K}_a\mathcal{K}_b}{4}(\partial_\sigma  c^{\mu\nu}\tilde{\rho}_\mu^a\tilde{\rho}_\nu^b-8\tilde{\rho}_\sigma^a\tilde{\rho}_\nu^bu^\nu) +\frac{8\mathcal{K}^2}{9}g_{ab}\tilde{\rho}^a_\sigma\tilde{\rho}^b_\nu u^\nu +\frac{\mathcal{K}^2}{ 36}\partial_\sigma c^{\mu\nu} \tilde{\rho}_\mu\tilde{\rho}_\nu \right],
\label{eq:der-umu}\\
\nonumber\\
e^{-K} \, \frac{\partial V}{\partial t^a}=&\left[e^{-K}V\partial_a K+\partial_a g^{bc}\rho_b\rho_c+\left(\black{\frac{8\mathcal{K}}{3}\mathcal{K}_a}g_{bc}+\frac{4\mathcal{K}^2}{9}\partial_a g_{bc}\right)\tilde{\rho}^b\tilde{\rho}^c+\black{\frac{2\mathcal{K}}{3} \mathcal{K}_a \tilde{\rho}^2}\right.\nonumber\\
&-\black{4\cK_a} \left(\rho_0 \tilde{\rho}_\nu - \rho_b \tilde{\rho}^b_\nu +\tilde{\rho}^b\rho_{b\nu} -\tilde{\rho}\rho_\nu \right) u^\nu -\black{2}\tilde{c}^{\mu\nu}\black{\cK_{ab}}\rho_\mu\tilde{\rho}^b_\nu +2\tilde{c}^{\mu\nu}t^c\rho_{a\mu}\rho_{c\nu} \nonumber\\
&+\partial_a g^{bc}\rho_{b\mu}u^\mu \rho_{c\nu}u^\nu -\tilde{c}^{\mu\nu}\left(\black{\cK_a}t^b\rho_{b\mu}\tilde{\rho}_\nu+\frac{\mathcal{K}}{\black{3}}\rho_{a\mu}\tilde{\rho}_\nu\right) +\tilde{c}^{\mu\nu}\black{\cK_{ab}\cK_c}\tilde{\rho}^b_\mu \tilde{\rho}^c_\nu \nonumber\\
&\left.+\black{\frac{8\mathcal{K}}{3}\cK_a}g_{bc}\tilde{\rho}_\mu^bu^\mu \tilde{\rho}^c_\nu u^\nu +\frac{4\mathcal{K}^2}{9}\partial_a g_{bc}\tilde{\rho}^b_\mu u^\mu \tilde{\rho}^c_\nu u^\nu  +\black{\frac{\mathcal{K}}{6}\cK_a}c^{\mu\nu}\tilde{\rho}_\mu\tilde{\rho}_\nu\right].
\label{eq:der-ta}
\end{align}
From the explicit expressions of the derivatives of the scalar potential, equations (\ref{eq:der-ximu})-(\ref{eq:der-ta}), one can construct the following four conditions as the initial step for exploring the non-SUSY landscape,
\bea
\label{eq:4extderV}
& & \hskip-1cm u^\sigma \, \frac{\partial V}{\partial \xi^\sigma} = 0, \qquad t^a \, \frac{\partial V}{\partial b^a} = 0, \qquad u^\mu\frac{\partial V}{\partial u^\mu}= 0, \qquad   t^a\frac{\partial V}{\partial t^a} = 0.
\eea
Note that the last two  quantities $\left(u^\mu\frac{\partial V}{\partial u^\mu}\right)$ and $\left(t^a\frac{\partial V}{\partial t^a}\right)$ can be expressed as a linear combination of the various scalar potential pieces, as we will exploit later. Moreover, it turns out that solving the last two (saxionic) extremization relations is equivalent to looking into the volume/dilaton $({\cal V}, D)$-plane as studied earlier, e.g. see \cite{Hertzberg:2007wc, Hertzberg:2007ke, Haque:2008jz, Flauger:2008ad}. More precisely, let us consider the first of these two relations, which is given by
\bea
\label{eq:derV1}
& & \hskip-1.3cm u^\mu\frac{\partial V}{\partial u^\mu} \equiv -4V+e^K\biggl[-\frac{4\mk}{3}u^\nu \rho_0\tilde{\rho}_\nu+\frac{4\mk}{3}u^\nu \rho_b\tilde{\rho}^b_\nu-\frac{4\mk}{3}u^\nu \tilde{\rho}^b\rho_{b\nu} +\frac{4\mk}{3}u^\nu \tilde{\rho}\rho_\nu \\
& & +2c^{\mu\nu}\rho_\mu\rho_\nu +2(\tilde{c}^{\mu\nu}t^bt^c+g^{bc}u^\mu u^\nu )\rho_{b\mu}\rho_{c\nu}-2\tilde{c}^{\mu\nu}\mathcal{K}_b\rho_\mu\tilde{\rho}^b_\nu-\frac{2\mk}{3}\tilde{c}^{\mu\nu}t^b\rho_{b\mu}\tilde{\rho}_\nu\nonumber\\
& & +\left(\frac{1}{2}\mathcal{K}_a\mathcal{K}_b\tilde{c}^{\mu\nu}+\frac{8\mathcal{K}^2}{9}g_{ab}u^\mu u^\nu \right)\tilde{\rho}^b_\mu\tilde{\rho}^c_\nu+\frac{\mathcal{K}^2}{18}c^{\mu\nu}\tilde{\rho}_\mu\tilde{\rho}_\nu \biggr] = 0. \nonumber
\eea
Now, using the decomposition of the potential given in eq.~(\ref{eq:VgenAnsatz1}), it can be observed that the above relation (\ref{eq:derV1}) can be compactly rewritten as
\bea
& & \hskip-1.3cm u^\mu\frac{\partial V}{\partial u^\mu}= - 4\, V_{RR} - 2 \, V_{NS} - 3 V_{loc}\,.
\eea
For the cases when $V_{loc} \ne 0$, using the fact that the above formula vanishes when evaluated at the vacuum, one can eliminate $V_{loc}$ -whose sign is undetermined- in \eqref{eq:VgenAnsatz1}. Subsequently, the potential at the vacuum can always be written as,
\bea
\label{eq:V-u-ext}
& & \la V \ra = \frac{\la V_{NS}\ra}{3}\,- \frac{\la V_{RR}\ra}{3}.
\eea
So, given the fact that the $V_{RR}$ contribution is always positive semidefinite (indeed \eqref{eq:VgenAnsatz2} directly implies $V_{RR}\geq 0$), whether one can realize a de-Sitter or not will be determined by the behaviour of the $V_{NS}$ piece when we impose the remaining three extremization conditions. 

We consider next the $t^a$ derivatives, which can result in the following useful relation,
\bea
\label{eq:derV2}
& & \hskip-1.1cm t^a\frac{\partial V}{\partial t^a}= -3V+e^K\biggl[2g^{bc}\rho_b\rho_c+\frac{16\mathcal{K}^2}{9}g_{bc}\tilde{\rho}^b \tilde{\rho}^c +\frac{2\mathcal{K}^2}{3}\tilde{\rho}^2\\ 
& & +2(\tilde{c}^{\mu\nu}t^bt^c+g^{bc}u^\mu u^\nu )\rho_{b\mu}\rho_{c\nu}+\left(\mathcal{K}_a\mathcal{K}_b\tilde{c}^{\mu\nu}+\frac{16\mathcal{K}^2}{9}g_{ab}u^\mu u^\nu \right)\tilde{\rho}^b_\mu\tilde{\rho}^c_\nu \nonumber\\
& & +\frac{\mathcal{K}^2}{6}c^{\mu\nu}\tilde{\rho}_\mu\tilde{\rho}_\nu-2\tilde{c}^{\mu\nu}\mathcal{K}_b\rho_\mu\tilde{\rho}^b_\nu -\frac{4\mk}{3}\tilde{c}^{\mu\nu}t^b\rho_{b\mu}\tilde{\rho}_\nu\nonumber\\
& & -4\mathcal{K}u^\nu \rho_0\tilde{\rho}_\nu+4\mathcal{K}u^\nu \rho_b\tilde{\rho}^b_\nu-4\mathcal{K}u^\nu \tilde{\rho}^b\rho_{b\nu} +4\mathcal{K}u^\nu \tilde{\rho}\rho_\nu \biggr].\nonumber
\eea
Note that the last line in the above expression is in fact ``$+3 V_{loc}$" which gets cancelled by the respective local piece in ``$-3V$" at the beginning of the equation. In this case, though, we could not derive an analogous expression to \eqref{eq:V-u-ext}.

In the following subsection we will mainly focus on equations \eqref{eq:derV1} and \eqref{eq:derV2}, which we will combine in order to determine necessary conditions for realizing a dS vacuum. In appendix \ref{app_derivatives} the reader interested can find  the explicit expressions of the four extremization conditions (\ref{eq:4extderV}) expressed  in terms of the  flux-axion polynomials.


\subsection{On generic dS no-go scenarios}
\label{s:ngeovacua}

As we have briefly commented in the introduction, finding dS vacua and exploring the possibility of dS no-go scenarios has been at the center of interest for decades. An important feature of the general analytical study of the vacuum equations is that it is possible to derive many No-Go conditions against de-Sitter realizations without explicitly solving the extremization conditions. Such has been the case for many type II constructions where the non-geometric fluxes are absent, e.g. see \cite{Hertzberg:2007wc, Hertzberg:2007ke, Haque:2008jz, Flauger:2008ad, Caviezel:2008tf, deCarlos:2009fq, Marchesano:2020uqz}. Some examples of non-geometric type II models that led to dS no-go results have been studied in \cite{Shukla:2019dqd, Shukla:2019akv}. In our framework this can be seen using \eqref{eq:derV1} in combination with \eqref{eq:derV2}. In the presence of all flux quanta, the same reasoning  can be generalized more systematically by using the following set of combinations of saxionic derivatives,
\bea
\label{eq:swamp-ineq}
& & \hskip-1cm u^\mu\frac{\partial V}{\partial u^\mu}+xt^a\frac{\partial V}{\partial t^a}= - \, (4+3x)V+4e^K \biggl[x\left(\frac{1}{2}g^{bc}\rho_b\rho_c+\frac{4\mathcal{K}^2}{9}g_{bc}\tilde{\rho}^b \tilde{\rho}^c +\frac{\mathcal{K}^2}{6}\tilde{\rho}^2\right)\\
& & +\frac{1}{2}c^{\mu\nu}\rho_\mu\rho_\nu +\left(\frac{1}{3}+x\right)\mathcal{K}u^\nu \left(\tilde{\rho}\rho_\nu - \tilde{\rho}^b\rho_{b\nu}+ \rho_b\tilde{\rho}^b_\nu  -\rho_0\tilde{\rho}_\nu\right) \nonumber\\
& & +\frac{1}{2}(1+x)(\tilde{c}^{\mu\nu}t^bt^c+g^{bc}u^\mu u^\nu )\rho_{b\mu}\rho_{c\nu} -\frac{1}{2}(1+x)\tilde{c}^{\mu\nu}\mathcal{K}_b\rho_\mu\tilde{\rho}^b_\nu\nonumber\\
& &+\frac{1}{8}(1+2x)\left(\cK_a\cK_b\tilde{c}^{\mu\nu}+\frac{16\mathcal{K}^2}{9}g_{ab}u^\mu u^\nu\right)\tilde{\rho}^b_\mu\tilde{\rho}^c_\nu -\frac{1}{6}(1+2x)\tilde{c}^{\mu\nu}\mathcal{K}t^b\rho_{b\mu}\tilde{\rho}_\nu \nonumber\\
& & +\frac{\mathcal{K}^2}{72}c^{\mu\nu}(1+3x)\tilde{\rho}_\mu\tilde{\rho}_\nu \biggr],\nonumber
\eea
which vanishes when evaluated at the vacuum. This relation naturally extends the results of \cite{Marchesano:2020uqz} beyond the geometric flux case and can be used to derive the general expression
\begin{align}
\label{eq:genx}
 3\la e^{-K} V \ra\equiv &\left(x+\frac{1}{3}\right)\left[-12\rho_0^2+\frac{c^{\mu\nu}\tilde{\rho}_\mu\tilde{\rho}_\nu}{12}\right]+\left(x-\frac{1}{3}\right)\left[\frac{\mk^2\tilde{\rho}^2}{3}-3c^{\mu\nu}\rho_\mu\rho_\nu\right]\\ &+(1-x)\left[(\tilde{c}^{\mu\nu}t^at^c+g^{ab}u^\mu u^\nu)\rho_{b\mu}\rho_{c\nu}-\frac{4\mk^2}{9}\tilde{\rho}^a\tilde{\rho}^ag_{ab}-\tilde{c}^{\mu\nu}\mathcal{K}_b\rho_\mu\tilde{\rho}^b_\nu\right]\nonumber\\&+(x+1)\left[\left(\tilde{c}^{\mu\nu}\frac{ \mathcal{K}_a}{2}\frac{\mathcal{K}_b}{2}+\frac{4\mk^2}{9}g_{ab}u^\mu u^\nu\right)\tilde{\rho}^a_\mu\tilde{\rho}^b_\nu-g^{ab}\rho_a\rho_b-\frac{\mk}{3}\tilde{c}^{\mu\nu}t^a\rho_{a\mu}\tilde{\rho}_\nu\right]\nonumber\,.
\end{align}
Notice that the elements of the set of polynomials $\rho_{\mathcal{A}}$ are not fully independent variables as the equations of motion constrain them.\footnote{As we discussed in section \ref{app_invariants}, even before imposing the equations of motion several constraints fix their relative values in terms of combinations of flux quanta.} Consequently, after demanding the vacuum conditions to be satisfied, there are many equivalent ways of writing the scalar potential in terms of these functions. The usefulness of relation \eqref{eq:genx} lies in its generality, since it has to be satisfied for any $x\in \mathbb{R}$.  By carefully choosing the values of $x$ in which the above formula is evaluated, one can arrive at several distinct simplified expressions for the scalar potential at its extremum points as a function of the flux-axion polynomials. We focus on those choices that cancel some of the terms in \eqref{eq:genx}, that is $x=\pm1$ and $x=\pm1/3$. The evaluation at those points is explicitly computed in appendix \ref{app_auxiliadS}. Each of them represents a boundary between regions in the flux-axion polynomial space  where different sets of  terms  can potentially contribute to a positive vacuum energy. Hence, analyzing them individually will provide a list of sufficient conditions to guarantee that the scalar potential is negative semi-definite. Such a list will not be exhaustive, but due to the method employed (trying to restrict the problem to lower dimensional subsectors of the space of $\rho_{\mathcal{A}}$) we expect these de Sitter no-go sufficient conditions to be the most restrictive non-trivial demands along each flux-axion direction, thus providing  valuable insight in the  search for de Sitter vacua.

Defining the following quantities
\begin{equation}
\label{eq: the definitions}
\begin{aligned}
    \mathcal{R}=& \left(\tilde{c}^{\mu\nu}t^at^b+g^{ab}u^\mu u^\nu \right)\rho_{a\mu}\rho_{b\nu}-\frac{4\mk^2}{9}\tilde{\rho}^a\tilde{\rho}^bg_{ab}\,, \qquad & \mathcal{S}=&\tilde{c}^{\mu\nu}\mathcal{K}_a\rho_\mu\tilde{\rho}^a_\nu\,,\\
    \tilde{\mathcal{R}}=& \left(\tilde{c}^{\mu\nu}\frac{\mathcal{K}_a}{2}\frac{\mathcal{K}_b}{2}+\frac{4\mathcal{K}^2}{9}g_ {ab}u^\mu u^\nu \right)\tilde{\rho}^a_\mu\tilde{\rho}^b_\nu-g^{ab}\rho_a\rho_b\,, \qquad & \tilde{\mathcal{S}}=& \mathcal{K}\tilde{c}^{\mu\nu}t^a\rho_{a\mu}\tilde{\rho}_\nu\,,
\end{aligned}
\end{equation}
we analyze how the different terms contribute to the sign of the potential in \eqref{eq:x=1}-\eqref{eq:x=-1/3}. The most general no-go conditions we can derive with this procedure are directly obtained from simply demanding that the expressions \eqref{eq:x=1}-\eqref{eq:x=-1/3} are negative. However, it is not very illustrative due to the large number of parameters involved. Therefore, we choose to soften the no-go claims, restricting ourselves to more basic conditions, in order to provide easy-to-check rules. Motivated by the observation below \eqref{eq:V-u-ext} -the fact that the sign of the potential in the vacuum can only be changed by the $V_{NS}$ piece- we consider  all the terms in \eqref{eq:x=1}-\eqref{eq:x=-1/3} that depend on the NSNS fluxes (i.e. $c^{\mu\nu}\rho_\mu\rho_\nu$, $c^{\mu\nu}\tilde{\rho}_\mu\tilde{\rho}_\nu$, $\mathcal{R},\tilde{\mathcal{R}},\mathcal{S},\tilde{\mathcal{S}}$) and include the additional RR contributions in the cases where they favour de Sitter vacua.\footnote{In particular, this means we will ignore the terms  $\mathcal{R}$ and $\tilde{\mathcal{R}}$ when they are purely  RR  -the cases in which $\rho_{a\mu}=0$ and $\tilde{\rho}^a_\nu=0$ respectively- since for these particular choices of fluxes $\mathcal{R}$ and $\tilde{\mathcal{R}}$ always favour AdS.} Doing so, we find the following four sufficient conditions to guarantee the absence of de Sitter vacua (note it is enough to satisfy one of them):
\begin{subequations}
\label{eq: the conditions}
    \begin{align}
        \label{eq:1}
        \frac{\cK^2\tilde{\rho}^2}{9}+\tilde{\mathcal{R}}-\frac{\tilde{\mathcal{S}}}{3}+\frac{\cK^2}{18}c^{\mu\nu}\tilde{\rho}_\mu\tilde{\rho}_\nu -c^{\mu\nu}\rho_\mu\rho_\nu\leq 0\, ,\\
            \label{eq:2}
        \mathcal{R}-\mathcal{S}+4\rho_0^2+2c^{\mu\nu}\rho_\mu\rho_\nu-\frac{\cK^2}{36}c^{\mu\nu} \tilde{\rho}_\mu\tilde{\rho}_\nu \leq 0\, ,\\\label{eq:3}
        \frac{\mathcal{R}}{2}-\frac{\mathcal{S}}{2}+\tilde{\mathcal{R}}-\frac{\tilde{\mathcal{S}}}{3}+\frac{\cK^2}{24}c^{\mu\nu}\tilde{\rho}_\mu\tilde{\rho}_\nu\leq0\, ,\\\label{eq:4}
        2\mathcal{R}-2\mathcal{S}+\tilde{\mathcal{R}}-\frac{\tilde{\mathcal{S}}}{3}+3c^{\mu\nu}\rho_\mu\rho_\nu\leq0\, .
    \end{align}
\end{subequations}
Though these four equations encode many possible no-go results against dS vacua, it is worth studying some particular choices of fluxes to shed some light on their implications.
\begin{itemize}
 \item 
 Previous results in the literature can be easily recovered when we omit the non-geometric fluxes. For instance, it is well known  that CY orientifold compactifications of massive type IIA with only RR and NSNS fluxes can never yield dS vacua \cite{Hertzberg:2007wc}. In our formalism, this  follows from \eqref{eq:3}, which reduces to $0\leq 0$ after setting $f_{a\mu} = Q^a_\mu = R_\mu = 0$ (and therefore $\rho_{a\mu}=0=\tilde{\rho}^a_\nu=\tilde{\rho}_\mu$). On the other hand, in the absence of non-geometric fluxes \eqref{eq:1} becomes  $\tilde{\rho}^2\leq 0$. This also proves that there cannot be dS if the Roman's parameter $\tilde{\rho}$ vanishes, even in the presence of  RR, NSNS and geometric fluxes, in agreement with what has been previously done \cite{Haque:2008jz,Caviezel:2008tf,Flauger:2008ad,Danielsson:2009ff,Danielsson:2010bc}. Finally, when we turn on simultaneously the Roman's mass and the metric fluxes, \eqref{eq:3} shows us that there are still important constraints to obtain de Sitter. In particular $\mathcal{R}$ must be positive, as was already observed in \cite{Marchesano:2020uqz}. All these results are summarized in rows 1 to 4 of table \ref{tab: no go Rmu=0}, labelled there by a \#.
    
\item 
The next step in the analysis is to introduce the non-geometric fluxes $Q_\mu^a$  while keeping $R_\mu$ (i.e.  $\tilde \rho^a_\mu\neq 0$ and $\tilde{\rho}_\mu=0$), so there is still a local notion of geometry in the internal space. The conditions that arise from \eqref{eq: the conditions} for the different flux configurations, summarized in rows 5 to 8 of table \ref{tab: no go Rmu=0}, become increasingly involved. In particular, the cross term $\mathcal{S}$ hinders the goal of making general and concise statements. It is worth mentioning, however, that when $\mathcal{R}$ and $\tilde{\mathcal{R}}$ are negative (as it happens naturally when considering the stability-motivated Ansatz employed in the following section) de Sitter no-go's are immediately recovered for configurations 5 and 7, while configurations 6 and 8 (generic case) do not rule out de Sitter vacua as long as the Roman's parameter is not trivial. 
\begin{table}[H]
\centering
\renewcommand{\arraystretch}{1.5}
\begin{tabular}{|c|c|c|c||c|c|}
\hline
$\tilde{\rho}_\mu$ & $\tilde{\rho}^a_\mu$ & $\rho_{a\mu}$  & $\rho_\mu$ & No-go condition & \# \\ \hline \hline
\multirow{8}{*}{0}  & \multirow{4}{*}{0}  & \multirow{2}{*}{0}  & 0  & \multirow{2}{*}{Always}  & 1  \\ \cline{4-4} \cline{6-6}
                    &                     &                     & -- &  & 2  \\ \cline{3-6} 
                    &                     & \multirow{2}{*}{--} & 0  & \multirow{2}{*}{$\mathcal{R}\leq 0$ \ or \ $\tilde{\rho}=0$} & 3  \\ \cline{4-4} \cline{6-6} 
                    &                     &                     & -- &  & 4  \\ \cline{2-6} 
                    & \multirow{4}{*}{--} & \multirow{2}{*}{0}  & 0  & $\tilde{\mathcal{R}}\leq 0$ or $\rho_0=0$ & 5  \\ \cline{4-6} 
                    &                     &                     & -- &  \begin{tabular}{@{}c@{}} $\cK^2\tilde{\rho}^2+9\tilde{\mathcal{R}}-9c^{\mu\nu}\rho_\mu\rho_\nu\leq0$\ or \ $2\tilde{\mathcal{R}}-\mathcal{S}\leq 0$ \ or \\ $4\rho_0^2+2c^{\mu\nu}\rho_\mu\rho_\nu-\mathcal{S}\leq 0$ \ or \ $\tilde{\mathcal{R}}+3c^{\mu\nu}\rho_\mu\rho_\nu -2\mathcal{S} \leq 0$ \end{tabular} & 6  \\ \cline{3-6} 
                    &                     & \multirow{2}{*}{--} & 0  & \begin{tabular}{@{}c@{}} $\cK^2\tilde{\rho}^2+9\tilde{\mathcal{R}}\leq 0$\ or \ $\mathcal{R}+2\tilde{\mathcal{R}}\leq 0$ \ or \\ $\mathcal{R}+4\rho_0^2\leq 0$ \ or \ $ 2\mathcal{R}+\tilde{\mathcal{R}} \leq 0$ \end{tabular} & 7  \\ \cline{4-6} 
                    &                     &                     & -- &  \begin{tabular}{@{}c@{}} $\cK^2\tilde{\rho}^2+9\tilde{\mathcal{R}}-9c^{\mu\nu}\rho_\mu \rho_\nu\leq0$\ or \ $2\tilde{\mathcal{R}}+\mathcal{R}-\mathcal{S}\leq 0$ \ or \\ $4\rho_0^2+2c^{\mu\nu}\rho_\mu\rho_\nu+\mathcal{R}-\mathcal{S}\leq 0$ \ or \ $\tilde{\mathcal{R}}+2\mathcal{R}+3c^{\mu\nu}\rho_\mu\rho_\nu -2\mathcal{S} \leq 0$ \end{tabular} & 8   \\ \hline
\end{tabular}
\caption{Summary of sufficient de Sitter no-go conditions for all the distinct basic flux-axion polynomial configurations with $R_\mu(\equiv \tilde\rho_\mu)=0$.}
\label{tab: no go Rmu=0}
\end{table}

\item 
Finally, we set $R_\mu\neq0$ which means $\tilde{\rho}_\mu\neq 0$. The most general no-go's we can derive with our simplified method are the relations \eqref{eq: the conditions}. For some particular configurations, these can be refined and shortened to prove the absence of de Sitter in a large region of the relevant flux-axion polynomial space. A selection of such simple cases is summarized in table \ref{tab:nogen}. Remarkably, as soon as we turn on the flux $\rho_\nu$ the no-go conditions become much more specific, which hints at the fact that the most general configurations are loosely constrained by \eqref{eq: the conditions}, leaving the door to find de Sitter still opened.\footnote{As an example, the de Sitter candidate found in \cite{deCarlos:2009fq} falls in the mirror type IIB version of the most general case, for which no subfamily of $\rho's$ is set to zero. The no-go employed there is a simplified version of \eqref{eq:3}. }
\begin{table}[H]
\centering
\renewcommand{\arraystretch}{1.5}
\begin{tabular}{|c|c|c|c||c|c|}
\hline
$\tilde{\rho}_\mu$ & $\tilde{\rho}^a_\mu$ & $\rho_{a\mu}$  & $\rho_\mu$ & No-go condition & \# \\ \hline \hline
\multirow{5}{*}{-}  & 0  & \multirow{2}{*}{0}  & \multirow{4}{*}{0}  & Always  & 9  \\ \cline{2-2} \cline{5-6} 
                    &     -                &                  &  &  $\tilde{\mathcal{R}}\leq 0$ \ or \ $\rho_0=0$ & 13  \\ \cline{2-3}\cline{5-6} 
                    &      0       &     \multirow{2}{*}{-}     &  & $6\mathcal{R}-\tilde{\mathcal{S}}\leq0$ \ or \ $4\rho_0^2+\mathcal{R}-\frac{\cK^2}{36}c^{\mu\nu}\tilde{\rho}_\mu\tilde{\rho}_\nu\leq 0$ \ or \ $\mathcal{R}\leq 0$   & 11  \\ \cline{2-2} \cline{5-6} 
                    &       -           &                    &  &  \begin{tabular}{@{}c@{}} $6\mathcal{R}+3\tilde{\mathcal{R}}-\tilde{\mathcal{S}}\leq0$ \ or \ $\frac{\cK^2\tilde{\rho}^2}{9}+\tilde{\mathcal{R}}-\frac{\tilde{\mathcal{S}}}{3}+\frac{\cK^2}{18}c^{\mu\nu}\tilde{\rho}_\mu\tilde{\rho}_\nu\leq0 $ \ or \\ $3\mathcal{R}+6\tilde{\mathcal{R}}-2\tilde{\mathcal{S}}+\frac{\cK^2}{4}c^{\mu\nu}\tilde{\rho}_\mu\tilde{\rho}_\nu\leq0$ \ or \  $4\rho_0^2+\mathcal{R}-\frac{\cK^2}{36}c^{\mu\nu}\tilde{\rho}_\mu\tilde{\rho}_\nu\leq 0$\end{tabular} & 15   \\ \cline{2-6} & 0 & 0 & - & $\cK^2\tilde{\rho}^2+\frac{\cK^2}{2}c^{\mu\nu}\tilde{\rho}_\mu\tilde{\rho}_\nu-9c^{\mu\nu}\rho_\mu\rho_\nu\leq 0$ \ or \ $2\rho_0^2+c^{\mu\nu}\rho_\mu\rho_\nu -\frac{\cK^2}{72}c^{\mu\nu}\tilde{\rho}_\mu\tilde{\rho}_\nu\leq 0$ & 10\\ \hline
\end{tabular}
\caption{Some configurations with $R_\mu(\equiv \tilde{\rho}_\mu)\neq 0$ and a selection of their most simple no-go's. Numbering of axion-flux configuration has been chosen to ease comparison with table \ref{tab:dS-no-go NS ansatz} in section \ref{sec_dS}.}
\label{tab:nogen}
\end{table}
\end{itemize}

\section{Imposing the Ansatz}
\label{sec_Ansatz}

In the previous section, we discussed the equations of motion of non-geometric flux compactifications and the nature of the resulting vacua. We now aim to particularize these techniques to a better-controlled framework where we can push the analytical characterization even further, thus refining the no-go results and granting a better understanding of the vacua structure. To do so, we recover the Ansatz studied in \cite{Marchesano:2020uqz}, which is given by the following proportionality relation between the F-terms and the derivatives of the K\"ahler potential of moduli space
\begin{align}
\label{solsfmax}\,
F_A =\left\{F_a, F_\mu \right\}=\left\{\a_K K_a,\a_Q K_\mu \right\}\, , \quad      \a_K, \a_Q \in \C\, .
\end{align}
This Ansatz, motivated through (perturbative) stability arguments \cite{GomezReino:2006dk}, was used in \cite{Marchesano:2020uqz} to perform a detailed study of a vast class of flux vacua. In that reference,  the central analysis was done by exclusively turning on RR, NSNS and geometric fluxes. In the present work, we want to go further by also including non-geometric fluxes. 

Translated to the F-terms computed in \eqref{eq: F-Ta}-\eqref{eq: F-Umu}, the Ansatz \eqref{solsfmax} requires 
\bea
\label{proprho}
& & \rho_a-\mk_{ab}\,\tr^b_\mu u^\mu = {\cal P}\, \partial_a K, \qquad \qquad \mk_{ab}\tr^b+\rho_{a\mu}u^\mu = {\cal Q}\, \partial_a K, \\
& & \rho_\mu- \frac{1}{2}\, \mk_a\tr^a_\mu = {\cal M} \, \partial_\mu K, \qquad \qquad \rho_{ a\mu}t^a-\frac{1}{6}\,\mk \, \tilde\rho_\mu = {\cal N} \, \partial_\mu K, \nonumber
\eea
where ${\mathcal P}$, ${\mathcal Q}$, $\cM$, $\cN$ are real functions of the moduli. The set of conditions (\ref{proprho})  can be manipulated to lead to the following four relations,
\begin{equation}
\begin{aligned}
    {\cal P} &= - \frac{1}{3} \left(\rho_a\, t^a-\mk_{a}\tr^a_\mu u^\mu \right) \, ,    &   {\cal Q} &= - \frac{1}{3}\left(\mk_{a}\tr^a+\rho_{a\mu}\, t^a \, u^\mu\right)\, ,\\  {\cal M} &= - \frac{1}{4}\left(\rho_\mu \, u^\mu- \frac{1}{2}\, \mk_a\tr^a_\mu  \, u^\mu\right)\, ,   &   {\cal N} &= -\frac{1}{4} \left(\rho_{ a\mu}\, t^a  \, u^\mu-\frac{\mk}{6}\tilde\rho_\mu  \, u^\mu \right)\, .
\end{aligned}
\label{eq:simp-Ansatz222}
\end{equation}
The plan  now is to  impose the Ansatz and to study the  scalar potential and its first derivatives, to see in which cases a dS extremum can be possible. We will not look at the second derivatives of the potential -the Hessian-, which we leave for future work.


\subsection{Simplified scalar potential and its derivatives}
Using the expressions ~(\ref{proprho})-(\ref{eq:simp-Ansatz222}), one can simplify the superpotential (\ref{eq:simpW}) and the $F$-terms (\ref{eq: F-Ta})-(\ref{eq: F-Umu})  to the following form
\bea
\label{eq:simp-W+Fterms}
& & W = \left(\rho_0 - \frac{1}{2}\, {\cal K}_a \, \tilde\rho^a + 4\, {\cal N} \right) + \, i \,\left(\rho_a\, t^a - \frac{{\cal K}}{6}\,\, \tilde\rho - 4\, {\cal M}\right), \\
& & F_{a} = \, \frac{1}{2}\,\p_a K \biggl[\left(t^a\rho_a +\frac{\mk}{6}\tilde{\rho} - 4\, {\cal M} + 2\, {\cal P}\right) - \, i \, \left(\rho_0 +\frac{3}{2}\,\mk_a\tr^a + 4\, {\cal Q} - 4 {\cal N} \right)\biggr], \nonumber\\
& & F_{\mu} = \, \frac{1}{2}\,\p_\mu K \biggl[\left( t^a\rho_a-\frac{\mk}{6}\tr - 2\, {\cal M}\right) - \, i\, \left(\rho_0 -\frac{1}{2}\mk_a \tr^a + 2\, {\cal N}\right) \biggr]. \nonumber
\eea
As it can be seen now, the $F$-terms are manifestly proportional to their respective K\"ahler derivatives. Combining these relations  with the standard identities: $(\p_\mu K)\, c^{\mu\nu} \,(\p_\mu K) = 16$ and $ u^\mu\, (\partial_\mu K) = -4$,  the expression of scalar potential can be greatly simplified. To illustrate how it works, here we note that six (out of 16) pieces of the scalar potential in \eqref{eq:VgenAnsatz1}, including the two cross-terms $\cal S$ and $\tilde{\cal S}$ introduced in \eqref{eq: the definitions} for which the signs are not a priory fixed, can be expressed as
\bea
\label{eq:general-Vpieces}
& & \hskip-1cm c^{\mu\nu}\rho_\mu\rho_\nu + \tilde{c}^{\mu\nu}t^at^b \rho_{a\mu}\rho_{b\nu} + \tilde{c}^{\mu\nu}\frac{\mathcal{K}_a}{2}\frac{\mathcal{K}_b}{2}\tilde{\rho}^a_\mu\tilde{\rho}^b_\nu + \frac{\mathcal{K}^2}{36}c^{\mu\nu}\tilde{\rho}_\mu\tilde{\rho}_\nu  -\tilde{c}^{\mu\nu}\mathcal{K}_a\rho_\mu\tilde{\rho}^a_\nu-\frac{\mathcal{K}}{3}\tilde{c}^{\mu\nu}t^a\rho_{a\mu}\tilde{\rho}_\nu \nonumber\\
& & \qquad \qquad = 4\, u^\mu\, u^\nu\, \rho_\mu  \, \rho_\nu + \frac{\mk^2}{9}\, u^\mu u^\nu\, \tilde\rho_\mu\,\tilde\rho_\nu -48\, ({\cal M}^2 + {\cal N}^2). 
\eea
Similarly, taking into account that $t^a\partial_a K=-3$ and $g^{ab}\cK_a\cK_b=4/3 \cK^2$, six more terms of the scalar potential \eqref{eq:VgenAnsatz1} can be written in the following manner,
\bea
& & \hskip-1cm g^{ab}\rho_a\rho_b+\frac{4\mathcal{K}^2}{9}g_{ab}\tilde{\rho}^a\tilde{\rho}^b + g^{ab}u^\mu u^\nu \rho_{a\mu}\rho_{b\nu} + \frac{4\mathcal{K}^2}{9}g_ {ab}u^\mu u^\nu \tilde{\rho}^a_\mu\tilde{\rho}^b_\nu + \frac{4\mathcal{K}}{3}u^\nu \rho_a\tilde{\rho}^a_\nu -\frac{4\mathcal{K}}{3}u^\nu \tilde{\rho}^a\rho_{a\nu} \nonumber\\
& &  = - 6\,({\cal P}^2+ {\cal Q}^2) + 2\,(t^a\, \rho_a)^2 + (\mk_a\, \tilde\rho^a_\mu\, u^\mu)^2 + (\mk_a\, \tilde\rho^a)^2 + 2\,(t^a\, \rho_{a \mu} u^\mu)^2. 
\eea
Merging all the above results and working with the remaining terms in $\rho_0$ and $\tilde{\rho}$, the full $F$-term scalar potential takes the following form
\bea
\label{eq:newV}
& & \hskip-1cm V = e^K \biggl[4\left(\rho_0+\mk_a\, \tilde\rho^a + 3 {\cal Q} - 4 {\cal N} \right)^2 + 4 \left(\frac{\mk}{6} \, \tilde\rho + \frac{1}{2}\, t^a \, \rho_a + \frac{3}{2} {\cal P} - 4 {\cal M}\right)^2 \\
& & \qquad \qquad  + 2\, (t^a\, \rho_a)^2 + (t^a\, \rho_a + 3 \, {\cal P})^2 + (\mk_a\, \tilde\rho^a)^2 + 2\,(\mk_a\, \tilde\rho^a + 3\, {\cal Q})^2 \nonumber\\
& & \qquad \qquad  - 48\,({\cal M}^2 + {\cal N}^2) - 6\,({\cal P}^2 + {\cal Q}^2) \biggr]. \nonumber
\eea
Notably, the introduction of the Ansatz (\ref{proprho}) allows us to express the scalar potential as a combination of pieces which are either positive semi-definite or negative semi-definite. This stands in stark contrast to the discussion in section \ref{sec_systematics}, where the overall sign of the terms of \eqref{eq:genx} involving $\mathcal{R}, \tilde{\mathcal{R}}, \mathcal{S}$ and $\tilde{\mathcal{S}}$ (defined in \eqref{eq: the definitions}) was left undetermined. Finally, in appendix \ref{app_miniansatz} we also write the four extremization conditions introduced in \eqref{eq:4extderV} in terms of the Ansatz.

To summarize the story so far, using the Ansatz \eqref{proprho} we have expressed the total $F$-term scalar potential as well as the four types of extremization conditions in terms of eight quantities, namely $\{{\cal M}, {\cal N}, {\cal P}, {\cal Q}\}$ and $\{\r_0, (\r_a t^a), (\mk_a \tr^a), \tr\}$. At first sight, it may naively look like an underdetermined system of four constraints with eight variables. However, let us note that these eight unknown quantities are not all independent, since there are several constraints due to the existence of moduli-invariant combinations of the flux-axion polynomials. We recall the previous discussion in section \ref{app_invariants} for more details on the flux invariants, and their correlations with the Bianchi identities. 


\subsection{Insights from a couple of examples}

In this subsection, we make use of the Ansatz \eqref{proprho}, the reformulated scalar potential (\ref{eq:newV}) and the four constraints arising from the derivatives of the potential (\ref{eq:etxV-gen-ansatz}), in order to study a subclass of simple but explicit models. Namely, we will focus on the SUSY vacuum as well as on the geometric scenario (where we set the non-geometric fluxes to zero). In the first of these two examples, we will limit ourselves to derive  the minimum of the potential  in terms of the Ansatz. Meanwhile, for the second case, we will connect our formalism to previous results in the literature \cite{Marchesano:2020uqz} and will have a first encounter with the no-go obtained in  section \ref{s:ngeovacua}.

\subsubsection*{SUSY vacua}

The supersymmetric vacua of our Ansatz could be derived as a subset of the solutions satisfying \eqref{eq:etxV-gen-ansatz}. However, it is far simpler to work directly with the $F$-flatness conditions $F_{T^a}=0=F_{U^{\mu}}$ using \eqref{eq:simp-W+Fterms}. We find
\bea
\label{eq:susy-sol1}
& t^a\rho_a +\frac{\mk}{6}\tilde{\rho} - 4\, {\cal M} + 2\, {\cal P} = 0, & \qquad \rho_0 +\frac{3}{2}\,\mk_a\tr^a + 4\, {\cal Q} - 4 {\cal N} = 0, \\
&  t^a\rho_a-\frac{\mk}{6}\tr - 2\, {\cal M} = 0, & \qquad \rho_0 -\frac{1}{2}\mk_a \tr^a + 2\, {\cal N} = 0. \nonumber
\eea
Subsequently, the space of solutions can be entirely described by the four parameters of our Ansatz. Given the relations above we can write
\bea
\label{eq:susy-sol2}
& & \hskip-1cm \r_0 = - \frac{{\cal N}}{2} - {\cal Q}, \qquad \qquad (\r_a t^a) = 3{\cal M} - {\cal P}, \\
& & \hskip-1cm (\mk_a \tr^a) = 3 {\cal N} - 2 {\cal Q}, \qquad \qquad \frac{\mk}{6} \tr= {\cal M} - {\cal P}, \nonumber
\eea
which, of course, is also a solution to the four extremization conditions given in~(\ref{eq:etxV-gen-ansatz}). Note that once a particular choice of flux quanta is made, the parameters $\mathcal{M},\mathcal{N},\mathcal{P},\mathcal{Q}$ will also be fixed by the constraints arising from the axion-invariant relations between the different elements of $\rho_A$. Such a process can be observed substituting the Ansatz on the expressions \eqref{invngr} and \eqref{finalNGinv}. Subsequently, the scalar potential at the SUSY minimum is given by 
\bea
& & \la V_{SUSY} \ra = - \, 12 \, e^K \left({\cal M}^2 + {\cal N}^2 \right).
\eea

\subsubsection*{Geometric model}
Let us now move away from the assumption of supersymmetry and consider some simplifications in the flux lattice. In the absence of non-geometric fluxes, one has $\tilde\rho^a_\mu = 0 = \tilde\rho_\mu$, and subsequently the set of 4 constraints in \eqref{proprho} takes the shorter form,
\bea
\label{eq:simp-Ansatz-geomONLY}
& & \rho_\mu \, u^\mu = - 4\, {\cal M}, \qquad \qquad \rho_{ a\mu}\, t^a  \, u^\mu = -4\, {\cal N},\\
& &  \rho_a\, t^a = -\, 3\,{\cal P}, \qquad \qquad  \mk_a \, \tilde\rho^a = - \, 3 \,{\cal Q} + 4\, {\cal N}. \nonumber
\eea
Given that only the last two of the above appear explicitly in our simplified formulation of the scalar potential and its derivatives, now we have a system of six unknowns, namely $\{{\cal M}, {\cal N}, {\cal P}, {\cal Q}, \r_0, \tr\}$ which are further constrained by the set of restrictions through the Bianchi identities and the flux invariants. Using the  expression (\ref{eq:simp-Ansatz-geomONLY}) the value of the potential in the vacuum can be written as
\bea
\label{eq:etxV-gen-ansatz-geomONLY}
& & \la V_{\rm geometric} \ra = e^K\, \left(4\,\rho_0^2+16{\cal M}^2 + 12{\cal P}^2 + 3 {\cal Q}^2 - 24 {\cal N} {\cal Q} - \frac{16}{3} {\cal K} \tilde\rho\,{\cal M} +\frac{{\cal K}^2}{9} \tr^2 \right),
\eea
under the conditions
\begin{subequations}
\label{eq:etxV-gen-ansatz-geom}
    \begin{align}
        &   {\cal N} \, {\cal P} - {\cal M} \, \rho_0 = 0, \\
&  8{\cal N} \left({\cal M} - {\cal P}\right) + 6\,{\cal P} \left({\cal Q} - \rho_0\right) +\frac{{\cal K}}{6}\,\tilde\rho\left(4{\cal N} - 3 {\cal Q} \right) = 0, \\
& \r_0^2+2{\cal M}^2 + 2{\cal N}^2 + 3 {\cal P}^2 + \frac{3}{4}{\cal Q}^2  - 5\, {\cal N}\, {\cal Q}-{\cal K} \, \tilde\rho \, {\cal M} + \frac{{\cal K}^2}{36} \, \tilde\rho^2 = 0, \label{eq: vacuum eq geometric umu} \\
& \r_0^2 + 4{\cal M}^2 - \frac{8}{3}{\cal N}^2 + {\cal P}^2 - \frac{1}{4}\,{\cal Q}^2+ \frac{2}{3}{\cal N}{\cal Q} - \frac{{\cal K}^2}{36} \, \tilde\rho^2 = 0, \label{eq: vacuum eq geometric ta}
    \end{align}
\end{subequations}
coming from the four extremization equations introduced in \eqref{eq:4extderV}.

Since this system corresponds to the geometric setup, it should reproduce some of the simple dS no-go scenarios discussed in the previous section. We know that one cannot have a dS extremum in geometric flux models without having non-trivial Romans mass terms and the geometric flux; for example, see \cite{Haque:2008jz,Caviezel:2008tf,Flauger:2008ad,Danielsson:2009ff,Danielsson:2010bc,Danielsson:2011au} and the recent works  \cite{Andriot:2022way, Andriot:2022yyj, Andriot:2022bnb}. Although in some of these studies dS extrema were obtained, those were found to be unstable. Under the conditions of our Ansatz (\ref{proprho}), such solutions are excluded and the dS no-go gets extended, which shows the effectiveness of the Ansatz to filter out unstable flux configurations. Indeed, given the special choice of Ansatz we have imposed, it is possible to combine \eqref{eq:etxV-gen-ansatz-geomONLY} with \eqref{eq: vacuum eq geometric umu} and \eqref{eq: vacuum eq geometric ta} to find the following negative semi-definite expression
\bea
\label{eq:geom-nogo2}
& & \la V \ra = - \frac{e^K}{3} \biggl[8 \rho_0^2 + 16 {\cal P}^2 + \frac{16}{3} \left(2{\cal N}- \frac{{\cal Q}}{4}\right)^2 + \frac{5{\cal Q}^2}{3}\biggr] \leq 0.
\eea
This shows that the no-go is restored  as long as the conditions (\ref{proprho}) are satisfied irrespective of whether the Romans mass term is zero or not! Such behaviour, which was already observed in \cite{Marchesano:2020uqz}, encompasses a particular realization of the no-go $\mathcal{R}\leq 0$ of configuration $4$ in table \ref{tab: no go Rmu=0}. As we will see in more detail in the next section, this no-go is more general and excludes a more significant set of flux configurations.

For completeness, let us conclude the discussion on the geometric flux setting by mentioning  that the four constraints in \eqref{eq:etxV-gen-ansatz-geom} have already been studied in  detail in \cite{Marchesano:2020uqz, Carrasco:2023hta} resulting in the identification of several classes of stable SUSY as well as non-SUSY $AdS$ vacua. In fact, one can connect the current analysis with those of \cite{Marchesano:2020uqz} by simply introducing six parameters $A, B, C, D, E, F$ in the following manner:
\bea
& & \r_0 = \mk\, A, \qquad {\cal P} = \mk\, B, \qquad {\cal Q} = -\frac{\mk (C -F)}{3}, \\
& & \tr = D, \qquad {\cal M} = \mk\, E, \qquad {\cal N} = \frac{\mk \, F}{4}. \nonumber
\eea
%

\section{dS no-go in the Ansatz}
\label{sec_dS}

Having introduced and familiarized ourselves  with the Ansatz in the previous section, we will now study how it behaves under the no-go's established in section \ref{sec_systematics}. As we will see, the Ansatz will improve the analytical control  and will allow us to refine the conditions required to go away from AdS vacua. We will start by discussing  the different  configurations of the NSNS fluxes, mirroring the general case of section \ref{s:ngeovacua}. Then, we will take advantage of the simplicity of the Ansatz and systematically study every possible configuration, including all the contributions (both RR and NSNS) in the scalar potential \eqref{eq:newV}.

\subsection{Reproducing and expanding the NSNS analysis}

In the context of the search of de Sitter vacua, one crucial outcome arising from the Ansatz \eqref{proprho} is given by the  expressions:
\begin{subequations}
\label{eq: R and S ansatz}
    \begin{align}
    \mathcal{R}&=-\frac{5}{3}(u^\mu t^a\rho_{a\mu})^2-\frac{1}{3}(\cK_a \tilde{\rho}^a)^2+\frac{\cK^2}{36}c^{\mu\nu}_{\rm P}\tilde{\rho}_\mu\tilde{\rho}_\nu\,,\\
    \tilde{\mathcal{R}}&=-\frac{5}{12}(u^\mu \cK_a \tilde{\rho}^a_\mu)^2-\frac{4}{3}(t^a\rho_a)^2+c^{\mu\nu}_{\rm P}\rho_\mu\rho_\nu\,,\\
    \mathcal{S}&=2c^{\mu\nu}_{\rm P}\rho_\mu\rho_\nu -3(\cK_a u^\mu \tilde{\rho}^a)( \rho_\mu u^\mu)\,,\\
    \tilde{\mathcal{S}}&=\frac{\cK^2}{6}c^{\mu\nu}_{\rm P}\tilde{\rho}_\mu\tilde{\rho}_\nu -3(u^\mu t^a\rho_{a\mu})(\tilde{\rho}_\mu u^\mu)\,,
    \end{align}
\end{subequations}
which can be derived from the following  relations, involving the primitive and non-primitive components of the moduli space metric,
\begin{equation}
    \begin{aligned}
        &  g^{ab} = g^{ab}_{\rm P} + g^{ab}_{\rm NP}, \qquad & & g^{ab}_{\rm P} \partial_b K = 0, \quad & & g^{ab}_{\rm NP} = \frac{4}{3}\, t^a\, t^b;\\
 & g_{ab} = g_{ab}^{\rm P} + g_{ab}^{\rm NP}, \qquad & & g_{ab}^{\rm P} t^b = 0, \quad & & \, g_{ab}^{\rm NP} = \frac{3}{4\cK^2}\, {\cal K}_a\, {\cal K}_b; \\
 & c^{\mu\nu} = c^{\mu\nu}_{\rm P} + c^{\mu\nu}_{\rm NP},   \qquad  & & c^{\mu\nu}_{\rm P} \partial_\nu K = 0, \quad & & c^{\mu\nu}_{\rm NP} = u^\mu u^\nu. 
    \end{aligned}
\end{equation}

Using these  results, one finds that the previous generic relation (\ref{eq:swamp-ineq}) simplifies to the following form: 
\begin{equation}
\label{eq: vev anstaz refined}
    \begin{aligned}
        3\langle e^{-K}V\rangle = & \left(x+\frac{1}{3}\right)\left[-12\rho_0^2+\frac{1}{12}(\tilde{\rho}_\mu u^\mu)^2\right]+\left(x-\frac{1}{3}\right)\left[\frac{\cK^2\tilde{\rho}^2}{3}-3(\rho_\mu u^\mu)^2\right]\\
        & + (1-x)\left[-\frac{5}{3}(\rho_{a\mu}t^au^\mu)^2-\frac{1}{3}(\cK_a \tilde{\rho}^a)^2+3(\tilde{\rho}^a_\mu \cK_a u^\mu)\rho_\mu u^\mu\right]\\
        &+(1+x)\left[-\frac{5}{12}(\tilde{\rho}^a_\mu u^\mu \cK_a)^2-\frac{4}{3}(t^a\rho_a)^2+(u^\mu t^a\rho_{a\mu})\tilde{\rho}_\nu u^\nu\right]\,.
    \end{aligned}
\end{equation}
where now all but two of the terms have fixed signs. 

From the relations  \eqref{eq: R and S ansatz} we see that some of the no-go conditions established in tables \ref{tab: no go Rmu=0} and \ref{tab:nogen} are automatically satisfied. For instance, in configurations $3,4,5,7$ of table \ref{tab: no go Rmu=0} we saw that a dS no-go was present if $\mathcal{R}\leq 0$, $\tilde{\mathcal{R}}\leq0$ or $\{\mathcal{R} \, ,\tilde{\mathcal{R}}\}\leq0$, depending on the case.\footnote{These configurations  correspond to the choices $\{\tilde\rho_\mu=0\, , \tilde\rho_\mu^a=0\, ,\rho_\mu=0\}\,\, , \{\tilde\rho_\mu=0\, , \tilde\rho_\mu^a=0\}\,, \{\tilde\rho_\mu=0\, , \rho_{a\mu}=0\, ,\rho_\mu=0\}$ and $ \{\tilde\rho_\mu=0\, ,\rho_\mu=0\}$ respectively.} We indeed verify that this is the behaviour displayed by the Ansatz, leaving only the two possibilities $6$ and $8$ of table \ref{tab: no go Rmu=0} for the search of dS vacua. Regarding these two cases, de Sitter is incompatible with our stability-inspired Ansatz whenever the Roman's mass vanishes.\footnote{This does not mean that $\tilde\rho\neq 0$ guarantees the existence of dS vacua, since using equation \eqref{eq: vev anstaz refined} one can obtain more involved constraints. We are only writing here the simplest conditions.} This discussion is summarized in rows 1-8 of table \ref{tab:dS-no-go NS ansatz}, matching there the labelling we used in table \ref{tab: no go Rmu=0}.

A similar analysis can be performed when the non-local non-geometric flux  $R_\mu$ is present. As it happened in the general case -recall table \ref{tab:nogen}- simple dS no-go conditions can be derived for some particular selection of fluxes, whereas for some others it is hard to extract easy-to-check equations from \eqref{eq: vev anstaz refined}. The main issue in the latter configurations is the lack of general control over terms depending on $\mathcal{S}$ and $\tilde{\mathcal{S}}$, since contrary to all others, they cannot be decomposed as  sums of quantities with definite signs. We summarize this discussion in rows 9-16 of table \ref{tab:dS-no-go NS ansatz}. Notice that we have broadened the analysis we made in table \ref{tab:nogen}, labelling the cases that appear in both tables with the same numbers.

\begin{table}[h!]
\centering
\renewcommand{\arraystretch}{1.5}
\begin{tabular}{|c|c|c|c||c|c|c|}
\hline
         $\tilde{\rho}_\mu$         &  $\tilde{\rho}^a_\mu$                 &   $\rho_{a\mu}$                & $\rho_\mu$ & de Sitter & No-go conditions &  \#\\ \hline \hline
\multirow{8}{*}{0} & \multirow{4}{*}{0} & \multirow{2}{*}{0} & 0 & No &  &  1\\ \cline{4-7} 
                  &                   &                   & - & No & &  2\\ \cline{3-7} 
                  &                   & \multirow{2}{*}{-} & 0 & No &  & 3 \\ \cline{4-7} 
                  &                   &                   & - & No &  &  4\\ \cline{2-7} 
                  & \multirow{4}{*}{-} & \multirow{2}{*}{0} & 0 & No &  & 5 \\ \cline{4-7} 
                  &                   &                   & - & Undecided & $\tilde{\rho}=0$ &  6\\ \cline{3-7} 
                  &                   & \multirow{2}{*}{-} & 0 & No &  & 7 \\ \cline{4-7} 
                  &                   &                   & - & Undecided & $\tilde{\rho}=0$ &  8\\ \hline
\multirow{8}{*}{-} & \multirow{4}{*}{0} & \multirow{2}{*}{0} & 0 & No &  & 9 \\ \cline{4-7} 
                  &                   &                   & - & Undecided &  &  10\\ \cline{3-7} 
                  &                   & \multirow{2}{*}{-} & 0 & Undecided & $\rho_0=0$ &  11\\ \cline{4-7} 
                  &                   &                   & - & Undecided &  &  12\\ \cline{2-7} 
                  & \multirow{4}{*}{-} & \multirow{2}{*}{0} & 0 & No &  &  13\\ \cline{4-7} 
                  &                   &                   & - & Undecided &  &  14\\ \cline{3-7} 
                  &                   & \multirow{2}{*}{-} & 0 & Undecided & $\rho_0=0$ &  15\\ \cline{4-7} 
                  &                   &                   & - & Undecided &  &  16\\ \hline
\end{tabular}
\caption{Viability of de Sitter with Ansatz \eqref{proprho} for the different NSNS axion-flux configurations. In the simplest configurations, de Sitter is ruled out entirely. In four configurations, we can provide some easy-to-check sufficient conditions that also lock the system away from dS. Finally, there are four other configurations, characterized by an increasing number of flux contributions,  where we are not able to make generic statements beyond \eqref{eq: vev anstaz refined}.  }
\label{tab:dS-no-go NS ansatz}
\end{table}


\subsection{A more comprehensive classification with all the axionic fluxes}
\label{sec:morecomp}

In this subsection, we comprehensively classify the dS no-go scenarios in the Ansatz introduced in \eqref{proprho}. This Ansatz, motivated by stability requirements, was characterized by four parameters constrained by the choice of flux quanta. Throughout the last two sections, we observed how it allowed us to cancel all contributions coming from the primitive sectors of the flux-axion polynomials and write the equations of motion and vacuum energy exclusively in terms of non-primitive components (see for instance \eqref{eq:newV}, \eqref{eq: vev anstaz refined} and \eqref{eq:etxV-gen-ansatz}). 

Until now, we have focused on the NSNS sector,  looking for some minimal requirements to achieve de Sitter, ignoring negative-definite terms in the scalar potential. We will now take full advantage of the powerful  simplifications generated by the Ansatz, systematically considering all possible flux contributions. To shorten the expressions, we will write the scalar potential and the equations of motion explicitly in terms of the non-primitive components of the flux-axion polynomials (eight in total, one per family) eliminating the dependence $\{{\cal M}, {\cal N}, {\cal P}, {\cal Q}\}$ \eqref{eq:simp-Ansatz222},
\bea
\label{eq:Nnew-8para}
& & \hskip-0.75cm V = e^K \biggl[4\, \gamma_1^2 + \frac{4}{3} \,\gamma_2^2 + \frac{4}{3}\,\gamma_3^2 + 4\, \gamma_4^2 + \lambda_1^2 -\frac{5}{3} \, \lambda_2^2 -\frac{5}{3}\, \lambda_3^2 + \lambda_4^2 + 6\, \lambda_1 \, \lambda_3  + 6 \, \lambda_2 \, \lambda_4 \\
& & +8\, \gamma_4\, \lambda_1 -\frac{8}{3} \gamma_3\, \lambda_2 + \frac{8}{3} \,\gamma_2\, \lambda_3 - 8\, \gamma_1\, \lambda_4 \biggr], \nonumber
\eea
where we have introduced four $\gamma_i$'s corresponding to the four generalized RR axionic fluxes and another four $\lambda_i$'s corresponding to the four generalized NS-NS axionic fluxes, defined as
\bea
& & \hskip-0.75cm \gamma_1 = \rho_0, \qquad \gamma_2 = \rho_a t^a, \qquad \gamma_3 = \frac{1}{2}{\cal K}_a \tilde\rho^a, \qquad \gamma_4 = \frac{{\cal K}}{6} \tilde\rho, \\
& & \hskip-0.75cm \lambda_1 = \rho_\mu u^\mu, \qquad \lambda_2 = \rho_{a\mu} t^a \, u^\mu, \qquad \lambda_3 = \frac{1}{2} \mathcal{K}_a\tilde{\rho}^a_\mu u^\mu, \qquad \lambda_4 = \frac{{\cal K}}{6}\,\tilde{\rho}_\mu u^\mu. \nonumber
\eea
We relegate to appendix \eqref{app_eomgammalambda} the usual extremization conditions \eqref{eq:4extderV}   written using this language.

Thus, we have eight axionic flux-dependent quantities $\{\gamma_i, \lambda_i\}$ which can be considered as parameters\footnote{Analogous to the original coefficients of the Ansatz, these parameters are not free once a particular point of the flux lattice is chosen. The invariant combinations of flux-axion polynomials constrain their values.} to seek solutions of the vacuum equations -see \eqref{eq:cond-8para}-.  Taking a fully analytic approach, we will now attempt to explore more dS no-go scenarios by considering special cases in a relatively more comprehensive manner. We will do this by switching off certain flux-axion polynomials at a time.

Let us pause the discussion here to explain what this means. In principle, if one wants to look for dS extrema by \emph{brute force} one needs to solve  the vacua equations and check that at that point  $V_{\rm solution}> 0$. For any of these solutions  $\{\gamma_i, \lambda_i\}$ will acquire some particular value  $\{\gamma_i, \lambda_i\}=\{\gamma_i, \lambda_i\}_{\rm solution}$, which can be equal or different from zero.  In other words, the $2^8$ possibilities of setting any of the eight quantities $\{\gamma_i, \lambda_i\}$ to zero or non-zero encode the full vacua landscape of the stability inspired Ansatz \eqref{solsfmax}. Notice that setting a flux polynomial to zero does not necessarily imply switching off a particular flux, rather, it should be interpreted as describing a possible branch of solutions.\footnote{We use the word ``possible" because it is not guaranteed that for a specific choice of values for $\{\gamma_i, \lambda_i\}$ we can find a solution to the full system of equations satisfying quantization requirements and tadpole constraints. However, the opposite is of course true, any vacua is described by some value (zero or non-zero) for  $\{\gamma_i, \lambda_i\}$ } This categorization allows us to perform a systematic search and see  whether any of these branches could contain a dS vacuum.  These are presented as follows:
\bea
\label{eq:Cis}
& {\bf {\mathbb S}8:}& \quad \{\gamma_1 = 0, \gamma_2 = 0, \gamma_3 = 0,\gamma_4 = 0, \lambda_1 =0 , \lambda_2 =0 , \lambda_3 =0 , \lambda_4 =0 \}\\
& {\bf {\mathbb S}7:}& \quad \{\gamma_1 \neq 0,\gamma_2 = 0, \gamma_3 = 0,\gamma_4 = 0, \lambda_1 =0 , \lambda_2 =0 , \lambda_3 =0 , \lambda_4 =0 \} \quad {\rm etc.}\nonumber\\
& {\bf {\mathbb S}6:}& \quad \{\gamma_1 \neq 0, \gamma_2 \neq 0, \gamma_3 = 0,\gamma_4 = 0, \lambda_1 =0 , \lambda_2 =0 , \lambda_3 =0 , \lambda_4 =0 \} \quad {\rm etc.}\nonumber\\
& {\bf {\mathbb S}5:}& \quad \{\gamma_1 \neq 0, \gamma_2 \neq 0, \gamma_3 \neq 0, \gamma_4 = 0, \lambda_1 =0 , \lambda_2 =0 , \lambda_3 =0 , \lambda_4 =0 \} \quad {\rm etc.}\nonumber\\
& {\bf {\mathbb S}4:}& \quad \{\gamma_1 \neq 0, \gamma_2 \neq 0, \gamma_3 \neq 0, \gamma_4 \neq 0, \lambda_1 =0 , \lambda_2 =0 , \lambda_3 =0 , \lambda_4 =0 \} \quad {\rm etc.}\nonumber\\
& {\bf {\mathbb S}3:}& \quad \{\gamma_1 \neq 0, \gamma_2 \neq 0, \gamma_3 \neq 0, \gamma_4 \neq 0, \lambda_1 \neq 0 , \lambda_2 =0 , \lambda_3 =0 , \lambda_4 =0 \} \quad {\rm etc.}\nonumber\\
& {\bf {\mathbb S}2:}& \quad \{\gamma_1 \neq 0, \gamma_2 \neq 0, \gamma_3 \neq 0, \gamma_4 \neq 0, \lambda_1 \neq 0 , \lambda_2 \neq 0, \lambda_3 =0 , \lambda_4 =0 \} \quad {\rm etc.}\nonumber\\
& {\bf {\mathbb S}1:}& \quad \{\gamma_1 \neq 0, \gamma_2 \neq 0, \gamma_3 \neq 0, \gamma_4 \neq 0, \lambda_1 \neq 0 , \lambda_2 \neq 0, \lambda_3 \neq 0 , \lambda_4 =0 \} \quad {\rm etc.}\nonumber\\
& {\bf {\mathbb S}0:}& \quad \{\gamma_1 \neq 0, \gamma_2 \neq 0, \gamma_3 \neq 0, \gamma_4 \neq 0, \lambda_1 \neq 0 , \lambda_2 \neq 0, \lambda_3 \neq 0 , \lambda_4 \neq 0 \}.\nonumber
\eea
For this classification, we denote the class ${\bf {\mathbb S}n}$ to be the one in which $n$ out of 8 parameters are set to zero, while the remaining $(8-n)$ parameters are non-zero. For the case of  ${\bf {\mathbb S}0}$ we mean that all the eight quantities are non-zero, however they can still be dependent, via the Bianchi identities and tadpole cancellation conditions. In total, there can be the following number of axionic flux configurations in each of the nine sub-classes summing to a total of 256:
\bea
\label{eq:CisNo}
& & {\bf {\mathbb S}8:} = 1(1), \quad {\bf {\mathbb S}7:} = 8(8), \quad {\bf {\mathbb S}6:} = 28(28), \quad {\bf {\mathbb S}5:} = 56(54), \quad {\bf {\mathbb S}4:} = 70(68), \\
& & {\bf {\mathbb S}3:} = 56(56), \quad {\bf {\mathbb S}2:} = 28(12), \quad {\bf {\mathbb S}1:} = 8(0), \quad {\bf {\mathbb S}0:} = 1(0), \nonumber
\eea
where the number in the respective brackets corresponds to the total number of cases which we could confirm to result in a no-go scenario for dS realization.

Demanding the extremization constraints (\ref{eq:cond-8para}) along with $V>0$, we interestingly find that 227 (of the 256) cases correspond to the set of dS no-go scenarios. All these 227 cases are listed in Table \ref{tab_dS-no-go-220} of the appendix \ref{app_no-go-list}. The new table contains and refines the results of table \ref{tab:dS-no-go NS ansatz}, providing valuable information regarding some of the undetermined cases. For instance, we now have no-go results for some of the configurations of row $6$ of table \ref{tab:dS-no-go NS ansatz} even when the no-go $\tilde{\rho}=0$ is evaded, such as case $199$ of table \ref{tab_dS-no-go-220}. Similarly, we have more control over the generic configuration $16$ of table \ref{tab:dS-no-go NS ansatz} accounting for the knowledge of the RR-sector, such as no-go $92$ of table \ref{tab_dS-no-go-220}. From these results, we observe that even when non-geometric fluxes are introduced, simple flux configurations (in which some of the flux quanta are turned to zero or tuned so they induce cancellations in the equations of motion) are generally not interesting for the search of de Sitter.

\subsection{An illustrative example}
There are still many configurations that have the potential to realize phenomenologically compelling vacua. Out of the 256 possible choices, for 29 of them we could not decide whether there is a dS no-go or not using our analytic approach. These 29 cases are listed in Table \ref{tab_36cases-without-nogo}, which notably include the most generic case in which none of the eight parameters are set to zero. Among them,  the seemingly simplest configurations are displayed in rows 1 and 2 of table  \ref{tab:possible}, which in the axion-polynomial language  read
\begin{subequations}
  \begin{align}
 \rho_0&=0\, ,    &     \rho_a t^a&=0\, , &  {\cal K}_a \tilde\rho^a&=0\, ,   &   \rho_{a\mu} t^a \, u^\mu&=0\, , &  {\cal K}\tilde{\rho}_\mu u^\mu&=0 \, , \\
\rho_a t^a&=0\, , &  {\cal K}_a \tilde\rho^a&=0\, ,   & \cal K \tilde\rho&=0\, , &   \rho_\mu u^\mu&=0\, ,  &   \mathcal{K}_a\tilde{\rho}^a_\mu u^\mu&=0\, ,
\end{align}  
\end{subequations}
and correspond to $\mathcal{N}=0=\mathcal{Q}$ and $\mathcal{P}=0=\mathcal{M}$ respectively in the Ansatz \eqref{proprho}. These two cases offer a relatively simple path to construct explicit vacua where dS is not openly excluded. Let us briefly illustrate how one could take advantage of the information learnt from row $1$ in table \ref{tab_36cases-without-nogo} to proceed with an analytical search of dS vacua.

First we use \eqref{eq:cond-8para} to fix some of the parameters we have not turned to zero.  The simplest solution compatible with row 1 of table \ref{tab_36cases-without-nogo} is easily written in terms of $\gamma_4$ 
\begin{equation}
    \begin{gathered}
        \gamma_1=\gamma_2=\gamma_3=\lambda_2=\lambda_4=0\, ,\\
        \lambda_1=-\frac{\gamma_4}{4}\,,\qquad \lambda_3=-\frac{9}{4}\gamma_4\,,\qquad  \gamma_4\geq 0\,.
    \end{gathered}
\end{equation}
Meanwhile the parameters of the Ansatz \eqref{proprho} are given by
\begin{equation}
    \mathcal{N}=\mathcal{Q}=0\,,\qquad \mathcal{P}=-\frac{3}{2}\gamma_4\,, \qquad \mathcal{M}=-\frac{9}{16}\gamma_4\, .
\end{equation}
Note that $\lambda_4=0$ allows us to turn off the non-local non-geometric fluxes ($R_\mu=0$). By making this choice, $\gamma_4$ and  all the other parameters of the Ansatz are determined by the value of the Romans mass. Putting together all the constraints of our Ansatz  in this particular case we have 
\begin{equation}
    \begin{aligned}
        \rho_a-\cK_{ab}\tilde{\rho}^b_\mu u^\mu &= -\frac{3}{2}\gamma_4\partial_a K\,, & \qquad \cK_b\tilde{\rho}^b &=0\,, \\
        \cK_{ab}\tilde{\rho}^b +\rho_{a\mu} u^\mu&=0\,, & \qquad \rho_0 &=0\,,\\
        \rho_\mu-\frac{1}{2}\cK_a \tilde{\rho}^a\mu &=-\frac{9}{16}\gamma_4 \partial_\mu K\,, & \qquad \rho_a t^a &=0\,,\\
        \rho_{a\mu} t^a &=0\,.
    \end{aligned}
\end{equation}
These relations can be employed to simplify the general equations of motion \eqref{eq:der-ximu}-\eqref{eq:der-ta} and the invariants of section \ref{app_invariants}, which are then used to fix the value of the moduli for a given choice of flux quanta. 

Using the table to look for de Sitter in the most generic configurations can be very involved, given the non-trivial relation between the flux quanta and the parameters of the Ansatz. However, it is worth noting that the connection becomes more straightforward for isotropic geometries leading to $STU$-like models (where one denotes the chiral coordinates as $T^a=T$, $U^0=S$  and $U^i=U$). A detailed numerical analysis may help one in exploring such cases. Given that our current work focuses on taking an analytic approach for studying a generic $F$-term type IIA scalar potential induced via the inclusion of (non-)geometric fluxes, we leave the numerical analysis of these undecided cases for future work.


\section{Conclusions}
\label{sec_conclusions}

In this paper, we have performed a systematic analysis of the de-Sitter no-go scenarios in the context of type IIA string theory, including (non-)geometric fluxes, along with the standard NS-NS and RR fluxes. To begin with, we have considered the generic $F$-term scalar potential, expressed in the bilinear formulation \cite{Bielleman:2015ina,Herraez:2018vae}. This splits the dependence between saxions, encoded in a bilinear form, and axions, which are combined with the flux quanta to build a set of flux-axion polynomials. We have used such formulation to derive the extremization conditions for the  scalar potential and  have discussed the solutions to these conditions. Moreover,  we have found that the generic extremization conditions lead to four sets of constraints, see expression \eqref{eq: the conditions}, which can be useful for deriving swampland-like inequalities in the form of de-Sitter  no-go equations.  Studying these relations we recover many previously known no-go results as well as some new ones which show that the simplest non-geometric flux configurations are generically not enough to enable de Sitter vacua. \black{Therefore,  any chance to find a de Sitter extremum involves carefully handling many different contributions simultaneously (the standard RR and NSNS, geometric and non-geometric fluxes), significantly increasing the difficulty of having a reliable analytical control. }

Given the complicated form of the generic scalar potential, some overall simplification is needed to perform any flux vacua analysis further in this generic setup. Two distinct paths could be taken. One could either perform a numerical search or make some simplifications in order to proceed with an analytic approach. We  have chosen the second option and imposed an Ansatz characterized by the $F$-term being proportional to the respective K\"ahler derivatives \cite{Marchesano:2020uqz}. This results in significant simplifications for the scalar potential and its derivatives, helping us to obtain a large number of dS no-go results by refining the cases found in the generic search. These no-go scenarios, classified on the basis of various axionic fluxes, are listed in Table (\ref{tab_dS-no-go-220}) and rule out many of the simplest flux configurations one can attempt to construct. Crucially, there remain 29 cases out of 256 for which no-go results could not be derived from our analytic approach. These cases, listed in Table (\ref{tab_36cases-without-nogo}), are still undecided. It is worth mentioning that finding a solution to the moduli equations of motion with a positive value of the scalar potential in any of these $29$ configurations does not guarantee the existence of a real de Sitter vacua. One still needs to make sure that the flux quantization and the tadpole conditions are satisfied and look for possible instabilities. Such conditions can be very restrictive and provide an additional challenge in many setups \cite{Plauschinn:2020ram}.

\black{We emphasize once again that the analysis we have performed throughout this paper is subjected to the assumption of corrections to the K\"ahler potential being small enough so they can be ignored, especially those breaking its axionic shift. Such an assumption should always be verified once a particular vacuum has been found, which is far from trivial, as noted in \cite{Banlaki:2018ayh,Junghans:2018gdb,Grimm:2019ixq}. Moving away from this regime is a compelling but challenging path  to expand on the current results. Other alternative ways to make further progress using  this same formalism include generalizing our Ansatz, introducing  $\alpha'$ corrections as done in \cite{Escobar:2018rna},  or accounting for the effect of D-terms in the scalar potential.\footnote{This last option would come at the cost of reducing the number of stabilized moduli, as discussed in \cite{Plauschinn:2020ram}.}} In any case, we hope the classification we have provided here will be helpful for future vacua searches. It can allow to take controlled cancellations on the vacuum equations without turning off the option to find de Sitter from the start. Complementary to this, it would be interesting to perform a numerical exploration of some of the simplest configurations of this flux landscape subset. They offer a compelling avenue for building explicit models of phenomenological interest.


\section*{Acknowledgments}
We would like to thank Fernando Marchesano for useful discussions, and collaboration at the initial stage of this project. We would like to thank Xin Gao, Thomas Grimm, George Leontaris, Erik Plauschinn and Filippo Revello  for useful discussions and comments. The work of D.P. is supported by the Dutch
Research Council (NWO) via a Vici grant. The work of J.Q. is supported by the Israel Science Foundation (grant No. 741/20) and by the German Research Foundation through a German-Israeli Project Cooperation (DIP) grant ``Holography and the Swampland''. P.S. would like to thank the {\it Department of Science and Technology (DST), India} for the kind support.


\newpage
\appendix
\setcounter{equation}{0}


\section{Derivatives of the potential}
\label{appa}
\subsection{General equations}
\label{app_derivatives}
In this appendix we explicitly give the four extremization conditions 
\begin{align}
\label{eq:excond}
    u^\sigma \, \frac{\partial V}{\partial \xi^\sigma} &= 0\equiv {\bf C1}\, ,  &   t^a \, \frac{\partial V}{\partial b^a} &= 0\equiv {\bf C2}\, ,&   u^\mu\frac{\partial V}{\partial u^\mu} &= 0\equiv {\bf C3}\, ,    &   t^a\frac{\partial V}{\partial t^a} &= 0\equiv {\bf C4}\, ,
\end{align}
in terms of the flux-axion polynomials introduced in \eqref{RRrhos} and \eqref{NSrhos}
\bea
\label{eq:4extderV-gen}
& {\bf C1:} & \quad  8\rho_0 \, \rho_\sigma u^\sigma \, + 2\, g^{ab}\rho_a\rho_{b\sigma} u^\sigma \,  +\frac{8\mathcal{K}^2}{9}g_{ab}\tilde{\rho}^a\tilde{\rho}^b_\sigma u^\sigma \,  +\frac{2\mathcal{K}^2}{9}\tilde{\rho_\sigma}\tilde{\rho} u^\sigma = 0, \\
& {\bf C2:} & \quad  8\, \rho_0\, \rho_a\,  t^a + 4\, \rho_a\,  t^a \, \mathcal{K}_{b} \, \tilde{\rho}^b -\frac{4\mk}{3}\, \rho_a\, \tilde{\rho}^a +\frac{2\mathcal{K}}{3}\,\tilde{\rho} \,\mathcal{K}_{a} \, \tilde{\rho}^a  + 2 \, c^{\mu\nu} \, \rho_{a\mu}\,  t^a \, \rho_\nu   \nonumber\\
& & - {\mathcal{K}} \, \tilde{c}^{\mu\nu} \rho_\mu\tilde{\rho}_\nu + \, {c}^{\mu\nu} \rho_{a\mu}\, t^a \tilde{\rho}^b_\nu \, \mathcal{K}_{b}  - \frac{4\mathcal{K}}{3}\, u^\mu u^\nu \tilde{\rho}^a_\mu \, \rho_{a\nu} +\frac{\mathcal{K}}{6}\, {c}^{\mu\nu} \mathcal{K}_{a}\, \tilde{\rho}^a_\mu \, \tilde{\rho}_\nu = 0, \nonumber\\
& {\bf C3:} & \quad 16\rho_0^2+ 4 g^{ab}\rho_a\rho_b +\frac{16\mathcal{K}^2}{9} g_{ab} \tilde{\rho}^a\tilde{\rho}^b +\frac{4\mathcal{K}^2}{9}\tilde{\rho}^2+ 2c^{\mu\nu}\rho_\mu\rho_\nu+2(\tilde{c}^{\mu\nu}t^at^b+g^{ab}u^\mu u^\nu )\rho_{a\mu}\rho_{b\nu}\nonumber\\
& & +\left(\frac{1}{2}\tilde{c}^{\mu\nu}\cK_a\cK_b+\frac{8\mathcal{K}^2}{9}g_ {ab}u^\mu u^\nu \right)\tilde{\rho}^a_\mu\tilde{\rho}^b_\nu+\frac{\mathcal{K}^2}{18}c^{\mu\nu}\tilde{\rho}_\mu\tilde{\rho}_\nu -2\tilde{c}^{\mu\nu}\mathcal{K}_a\rho_\mu\tilde{\rho}^a_\nu-\frac{2\mathcal{K}}{3}\tilde{c}^{\mu\nu}t^a\rho_{a\mu}\tilde{\rho}_\nu \nonumber\\
& & -{4\mathcal{K}} u^\nu \rho_0\tilde{\rho}_\nu+{4\mathcal{K}}u^\nu \rho_a\tilde{\rho}^a_\nu -{4\mathcal{K}}u^\nu \tilde{\rho}^a\rho_{a\nu}+{4\mathcal{K}}u^\nu \tilde{\rho}\rho_\nu = 0, \nonumber\\
& {\bf C4:} & \quad  12\rho_0^2+g^{ab}\rho_a\rho_b - \frac{4\mathcal{K}^2}{9}g_{ab}\tilde{\rho}^a\tilde{\rho}^b -\frac{\mathcal{K}^2}{3}\tilde{\rho}^2+ 3c^{\mu\nu}\rho_\mu\rho_\nu+(\tilde{c}^{\mu\nu}t^at^b+g^{ab}u^\mu u^\nu )\rho_{a\mu}\rho_{b\nu} \nonumber\\
& & -\left(\frac{1}{4}\tilde{c}^{\mu\nu}\cK_a\cK_b+\frac{4\mathcal{K}^2}{9}g_ {ab}u^\mu u^\nu \right)\tilde{\rho}^a_\mu\tilde{\rho}^b_\nu - \frac{\mathcal{K}^2}{12}c^{\mu\nu}\tilde{\rho}_\mu\tilde{\rho}_\nu -\tilde{c}^{\mu\nu}\mathcal{K}_a\rho_\mu\tilde{\rho}^a_\nu+\frac{\mathcal{K}}{3}\tilde{c}^{\mu\nu}t^a\rho_{a\mu}\tilde{\rho}_\nu = 0.\nonumber
\eea
All the possible extrema in the presence of generic (non-)geometric fluxes will be contained in the set of solutions of these four constraints.

\subsection{Imposing the Ansatz}
\label{app_miniansatz}
Under the Ansatz presented in section \ref{sec_Ansatz} the equations \eqref{eq:excond}, take the following form
\bea
\label{eq:etxV-gen-ansatz}
& {\bf C1:} & \qquad   4\, \r_0 \left(\frac{\r_at^a}{2} + \frac{3 {\cal P}}{2} - 4 {\cal M}\right) - \frac{2\mk}{3} \tr\left(\mk_a \tr^a + 3 {\cal Q} - 4 {\cal N}\right) \\
& & \quad - \left(\r_a t^a\right) \left(\mk_a \tr^a\right) - 4 {\cal Q} (\r_a t^a) + {\cal P} (\mk_a \tr^a) = 0, \nonumber\\
& {\bf C2:} & \qquad  2(\r_a t^a) \left(\r_0 - 4{\cal Q} + 5 {\cal N}\right) + (\mk_a \tr^a)\left(\frac{\mk}{6} \tr - 8 {\cal P} + 20 {\cal M} \right)\nonumber\\
& & \quad + 6 \left(5 {\cal N}{\cal P}+ 10 {\cal M} {\cal Q} - 12 {\cal M}{\cal N} - 4 {\cal P}{\cal Q} \right) - 2 (\mk_a \tr^a) (\r_a t^a) = 0, \nonumber\\
& {\bf C3:} & \qquad  \r_0^2 + \frac{3}{4}\left(\r_a t^a\right)^2 + \left(\mk_a \tr^a\right)^2 + \frac{\mk^2}{36} \, \tr^2 + 2{\cal M}^2 + 2{\cal N}^2 + \frac{3 {\cal P}^2}{2} + 6{\cal Q}^2 \nonumber\\
& & \quad - 6\, {\cal M} {\cal P} - 12 {\cal N} {\cal Q} - 6 \rho_0\, {\cal N} + \frac{9 \r_0 \,{\cal Q}}{2} - \mk {\cal M} \tr + \frac{3\mk}{8}\, {\cal P} \, \tr \nonumber\\
& & \quad + \left(\r_a t^a\right) \left(\frac{\mk \, \tr}{8} - 2{\cal M} + \frac{7 {\cal P}}{4}\right) + \left(\mk_a \tr^a\right) \left(\frac{3 \r_0}{2} - 4\, {\cal N} + \frac{19 {\cal Q}}{4}\right) = 0, \nonumber\\
& {\bf C4:} & \qquad  \r_0^2 + \frac{1}{3}\left(\r_a t^a\right)^2 - \frac{11}{12}\left(\mk_a \tr^a\right)^2 - \frac{\mk^2}{36} \, \tr^2 + 4{\cal M}^2 - 4{\cal N}^2 + 2{\cal P}^2 - 8{\cal Q}^2 \nonumber\\
& & \quad - 6{\cal M}{\cal P} + 12 {\cal N} {\cal Q} + 4 \left({\cal N} - \frac{4}{3} {\cal Q} \right)\left(\mk_a \tr^a\right) -2 \left({\cal M} - \frac{2 {\cal P}}{3}\right) \left(\r_a t^a\right) = 0.\nonumber
\eea

\subsection{Non-primitive decomposition}
\label{app_eomgammalambda}

Finally, we write again the conditions \eqref{eq:excond} using the $\{\gamma, \lambda\}$ language introduced in section \ref{sec:morecomp}
\bea
\label{eq:cond-8para}
& {\bf C1:}& \qquad 4 {\gamma_1} {\lambda_1}+\frac{4 {\gamma_2} {\lambda_2}}{3}+\frac{4 {\gamma_3} {\lambda_3}}{3}+4 {\gamma_4} {\lambda_4} =0, \\
& {\bf C2:}& \qquad  2 {\gamma_1} {\gamma_2}+\frac{4 {\gamma_2} {\gamma_3}}{3}+2 {\gamma_3} {\gamma_4}+\frac{{\lambda_1} {\lambda_2}}{2}+\frac{9 {\lambda_1} {\lambda_4}}{2}-\frac{{\lambda_2} {\lambda_3}}{6}+\frac{{\lambda_3} {\lambda_4}}{2}=0,\nonumber\\
& {\bf C3:}& \qquad  {\gamma_1}^2-\frac{3 {\gamma_1} {\lambda_4}}{2}+\frac{{\gamma_2}^2}{3}+\frac{{\gamma_2} {\lambda_3}}{2}+\frac{{\gamma_3}^2}{3}-\frac{{\gamma_3} {\lambda_2}}{2}+{\gamma_4}^2+\frac{3 {\gamma_4} {\lambda_1}}{2}+\frac{{\lambda_1}^2}{8}+\frac{3 {\lambda_1} {\lambda_3}}{4}\nonumber\\
& & \hskip0.5cm -\frac{5 {\lambda_2}^2}{24}+\frac{3 {\lambda_2} {\lambda_4}}{4}-\frac{5 {\lambda_3}^2}{24}+\frac{{\lambda_4}^2}{8}=0,\nonumber\\
& {\bf C4:}& \qquad  {\gamma_1}^2+\frac{{\gamma_2}^2}{9}-\frac{{\gamma_3}^2}{9}-{\gamma_4}^2+\frac{{\lambda_1}^2}{4}+\frac{{\lambda_1} {\lambda_3}}{2}-\frac{5 {\lambda_2}^2}{36}-\frac{{\lambda_2} {\lambda_4}}{2}+\frac{5 {\lambda_3}^2}{36}-\frac{{\lambda_4}^2}{4} = 0.\nonumber
\eea


\section{Auxiliary equations for dS no-go}
\label{app_auxiliadS}

In this appendix we present the explicit form of the expression \eqref{eq:genx} for the particular choices of values $x=\pm 1$ and $x=\pm1/3$. As explained in the main text, the equation \eqref{eq:genx}  is always satisfied in the vacuum, independently of the values of $x$. Nevertheless, some specific  selections of  $x$ can be more illuminating in the search for dS no-go.

From $x = 1$ we obtain the following constraint,
\bea
\label{eq:x=1}
& & \hskip-1cm \la V \ra = -\frac{4e^K}{3} \biggl[4\rho_0^2+ \frac{1}{2} g^{ab}\rho_a\rho_b- \frac{\mathcal{K}^2}{18}\tilde{\rho}^2  + \frac{1}{2} \,c^{\mu\nu}\rho_\mu\rho_\nu \\
& & - \frac{1}{2} \, \left(\tilde{c}^{\mu\nu}\frac{\mathcal{K}_a}{2}\frac{\mathcal{K}_b}{2}+\frac{4\mathcal{K}^2}{9}g_ {ab}u^\mu u^\nu \right)\tilde{\rho}^a_\mu\tilde{\rho}^b_\nu  +\frac{\mathcal{K}}{6}\tilde{c}^{\mu\nu}t^a\rho_{a\mu}\tilde{\rho}_\nu -\frac{\mathcal{K}^2}{36}c^{\mu\nu}\tilde{\rho}_\mu\tilde{\rho}_\nu \biggr]\,, \nonumber
\eea
whereas from $x=-1$ one has 
\bea
\label{eq:x=-1}
& & \hskip-1cm \la V \ra = -\, \frac{4e^K}{3} \biggl[-2\rho_0^2 +\frac{2\mathcal{K}^2}{9}g_{ab}\tilde{\rho}^a\tilde{\rho}^b+\frac{\mathcal{K}^2}{9}\tilde{\rho}^2  - c^{\mu\nu}\rho_\mu\rho_\nu \\
& & -\frac{1}{2}\left(\tilde{c}^{\mu\nu}t^at^b+g^{ab}u^\mu u^\nu \right)\rho_{a\mu}\rho_{b\nu} +\frac{1}{2}\tilde{c}^{\mu\nu}\mathcal{K}_a\rho_\mu\tilde{\rho}^a_\nu +\frac{\mathcal{K}^2}{72}c^{\mu\nu}\tilde{\rho}_\mu\tilde{\rho}_\nu \biggr]. \nonumber
\eea
Similarly the cases with $x =1/3$ and $x = -1/3$  result in the following respective relations,
\bea
\label{eq:x=1/3}
& & \hskip-0.5cm \la V \ra = -\, \frac{4e^K}{3} \biggl[2\rho_0^2+ \frac{1}{3} g^{ab}\rho_a\rho_b+\frac{2\mathcal{K}^2}{27}g_{ab}\tilde{\rho}^a\tilde{\rho}^b - \frac{1}{6} \, \left(\tilde{c}^{\mu\nu}t^at^b+g^{ab}u^\mu u^\nu \right)\rho_{a\mu}\rho_{b\nu}\\
& & + \frac{1}{6}\tilde{c}^{\mu\nu}\mathcal{K}_a\rho_\mu\tilde{\rho}^a_\nu - \frac{1}{3} \,\left(\tilde{c}^{\mu\nu}\frac{\mathcal{K}_a}{2}\frac{\mathcal{K}_b}{2}+\frac{4\mathcal{K}^2}{9}g_ {ab}u^\mu u^\nu \right)\tilde{\rho}^a_\mu\tilde{\rho}^b_\nu +\frac{\mathcal{K}}{9}\tilde{c}^{\mu\nu}t^a\rho_{a\mu}\tilde{\rho}_\nu -\frac{\mathcal{K}^2}{72}c^{\mu\nu}\tilde{\rho}_\mu\tilde{\rho}_\nu \biggr], \nonumber
\eea
and
\bea
\label{eq:x=-1/3}
& & \hskip-1.5cm \la V \ra = -\, \frac{4e^K}{3} \biggl[\frac{1}{6}g^{ab}\rho_a\rho_b+\frac{4\mathcal{K}^2}{27}g_{ab}\tilde{\rho}^a\tilde{\rho}^b+\frac{\mathcal{K}^2}{18}\tilde{\rho}^2  - \frac{1}{2} \,c^{\mu\nu}\rho_\mu\rho_\nu \\
& & - \frac{1}{3} \left(\tilde{c}^{\mu\nu}t^at^b+g^{ab}u^\mu u^\nu \right)\rho_{a\mu}\rho_{b\nu}+\frac{1}{3}\tilde{c}^{\mu\nu}\mathcal{K}_a\rho_\mu\tilde{\rho}^a_\nu \nonumber\\
& & -\frac{1}{6}\left(\tilde{c}^{\mu\nu}\frac{\mathcal{K}_a}{2}\frac{\mathcal{K}_b}{2}+\frac{4\mathcal{K}^2}{9}g_ {ab}u^\mu u^\nu \right)\tilde{\rho}^a_\mu\tilde{\rho}^b_\nu \,  +\frac{\mathcal{K}}{18}\tilde{c}^{\mu\nu}t^a\rho_{a\mu}\tilde{\rho}_\nu \biggr]\,. \nonumber
\eea


\section{Intermediate results for scalar potential simplifications using the Ansatz}
\label{app_inter}

In this section,  after imposing the Ansatz given in eqs.~(\ref{proprho}), we present some intermediate steps which can be relevant for simplifying the scalar potential pieces collected in eqs.~(\ref{eq:VgenAnsatz1})-(\ref{eq:VgenAnsatz2}). Some of those are presented as follows,
\bea
\label{eq:intermediate}
& & \hskip-1cm \left(\tilde{c}^{\mu\nu}t^at^b+g^{ab}u^\mu u^\nu \right)\rho_{a\mu}\rho_{b\nu} -\tilde{c}^{\mu\nu}\mathcal{K}_a\rho_\mu\tilde{\rho}^a_\nu = \frac{4\mathcal{K}^2}{9}g_{ab}\tilde{\rho}^a\tilde{\rho}^b - 2 \tilde{c}^{\mu\nu}\rho_\mu\rho_\nu + \frac{\mathcal{K}^2}{36} \tilde{c}^{\mu\nu}\tilde{\rho}_\mu\tilde{\rho}_\nu \nonumber\\
& & - 48 {\cal N}^2 + 12 {\cal Q}^2 + 24 {\cal M} \r_\mu u^\mu + (\mk_a\, \tr^a)^2 - \frac{4 \mk}{3} {\cal N} \tr_\mu u^\mu + 8 {\cal Q} (\mk_a\, \tr^a),\nonumber\\
& & \\
& & \hskip-1cm \left(\tilde{c}^{\mu\nu}\frac{\mathcal{K}_a}{2}\frac{\mathcal{K}_b}{2}+\frac{4\mathcal{K}^2}{9}g_ {ab}u^\mu u^\nu \right)\tilde{\rho}^a_\mu\tilde{\rho}^b_\nu -\tilde{c}^{\mu\nu}t^a\frac{\mathcal{K}}{3}\rho_{a\mu}\tilde{\rho}_\nu =g^{ab}\rho_a\rho_b + \tilde{c}^{\mu\nu}\rho_\mu\rho_\nu - \frac{\mathcal{K}^2}{18} \tilde{c}^{\mu\nu}\tilde{\rho}_\mu\tilde{\rho}_\nu \nonumber\\
& &  - 48 {\cal M}^2 + 3 {\cal P}^2 + 8 {\cal M} \r_\mu u^\mu - (t^a\, \r_a)^2 - 4 \mk {\cal N} \tr_\mu u^\mu + 2 {\cal P} (t^a\, \r_a).\nonumber
\eea
Note that the sum of the above two pieces helps in eliminating the presence of $c^{\mu\nu}$ metric in the $V_{NS}$ piece given in eq.~(\ref{eq:VgenAnsatz2}), which subsequently simplifies to the following form,
\bea
\label{eq:VNS-ansatz}
& & \hskip-1cm V_{NS}^\ast = e^K \biggl[g^{ab}\rho_a\rho_b + \frac{4\mathcal{K}^2}{9}g_{ab}\tilde{\rho}^a\tilde{\rho}^b + 4\, (\r_\mu u^\mu)^2 + \frac{\mathcal{K}^2}{9} (\tr_\mu u^\mu)^2 \\
& & - 48 {\cal N}^2 + 12 {\cal Q}^2 + 32 {\cal M} \r_\mu u^\mu + (\mk_a\, \tr^a)^2 - \frac{16 \mk}{3} {\cal N} \tr_\mu u^\mu + 8 {\cal Q} (\mk_a\, \tr^a)\nonumber\\
& &  - 48 {\cal M}^2 + 3 {\cal P}^2 - (t^a\, \r_a)^2 + 2 {\cal P} (t^a\, \r_a)\biggr]. \nonumber
\eea
The first two pieces in the above expression are precisely the generalized RR flux contributions from the generalized version of the $F_2$ and $F_4$ form fluxes in the presence of (non-)geometric fluxes. Similarly, the $V_{loc}$ piece in eq.~(\ref{eq:VgenAnsatz2}) simplifies to the following form,
\bea
\label{eq:Vloc-ansatz}
& & \hskip-1cm V_{loc}^\ast = - \, e^K \biggl[\frac{4\mk}{3} \r_0 (\tr_\mu u^\mu) - \frac{4\mk}{3} \tr (\r_\mu u^\mu) + 2 g^{ab}\rho_a\rho_b + \frac{8\mathcal{K}^2}{9}g_{ab}\tilde{\rho}^a\tilde{\rho}^b \\
& & + 4 (t^a\, \r_a)^2 + 2(\mk_a\, \tr^a)^2 - 4 {\cal P} (\r_a\, t^a) - 4 {\cal Q} (\mk_a\, \tr^a) \biggr], \nonumber
\eea
where we remark that none of the terms appearing in $V_{loc}^\ast$ involve $c^{\mu\nu}$ either and, as a result of this simplification, the scalar potential in eqs.~(\ref{eq:VgenAnsatz1})-(\ref{eq:VgenAnsatz2}) can be expressed without using any of the metrics $c^{\mu\nu}, g_{ab}$ and $g^{ab}$. This is because the pieces with $g_{ab}$ and its inverse appearing in the scalar potential precisely cancel out in the combination $V = V_{RR}^\ast + V_{NS}^\ast + V_{loc}^\ast$. Moreover, using eq.~(\ref{proprho}) the terms involving $(\r_\mu u^\mu)$ and $(\tr_\mu u^\mu)$ in both the eqs.~(\ref{eq:VNS-ansatz}) and (\ref{eq:Vloc-ansatz}) can further be expressed in terms of ${\cal M}, {\cal N}, {\cal P}, {\cal Q}$ and $(\r_a t^a)$ and $(\mk_a \tr^a)$. Hence, a direct implication of using the Ansatz in eq.(\ref{proprho}) is the fact that it helps in expressing the total $F$-term scalar potential in terms of eight parameters, namely $\{{\cal M}, {\cal N}, {\cal P}, {\cal Q}\}$ and $\{\r_0, (\r_a t^a), (\mk_a \tr^a), \tr\}$, and one can completely get rid of the metric. This is reflected in the collection (\ref{eq:newV}).

Finally, let us also mention that the two relations in eq.~(\ref{eq:intermediate}) help to eliminate the presence of all the involved metrics, namely $c^{\mu\nu}, g_{ab}$ and $g^{ab}$, within the four extremization conditions given in eq.~(\ref{eq:4extderV-gen}) and the same subsequently leads to the four relations collected in the eq.~(\ref{eq:etxV-gen-ansatz}).


\newpage
\section{List of possible dS no-go scenarios}
\label{app_no-go-list}

\begin{center}
\renewcommand{\arraystretch}{1.15}
\begin{longtable}{|c||c|c|}
\caption{List of 227 flux configurations with their respective vanishing flux-dependent parameters leading to dS no-go scenarios. The parameters not mentioned in a given configuration are considered to be non-zero.} \\
\hline 
Sr. No. & Vanishing Axionic Flux parameters & $\exists$ dS \\
    \hhline{|=|=|=|}
    \endfirsthead
    \hline 
Sr. No. & Vanishing Axionic Flux parameters & $\exists$ dS \\
    \hhline{|=|=|=|}
    \endhead
    \label{tab_dS-no-go-220}1 & $\left\{\gamma _4 = 0, \, \gamma _3 = 0, \, \gamma _2 = 0, \, \gamma _1 = 0, \, \lambda _1 = 0, \, \lambda _2 =
   0, \, \lambda _3 = 0, \, \lambda _4 = 0\right\}$ & \text{False} \\
\hline
 2 & $\left\{\gamma _4 = 0, \, \gamma _3 = 0, \, \gamma _2 = 0, \, \gamma _1 = 0, \, \lambda _1 = 0, \, \lambda _2 =
   0, \, \lambda _3 = 0\right\}$ & \text{False} \\
 3 & $\left\{\gamma _4 = 0, \, \gamma _3 = 0, \, \gamma _2 = 0, \, \gamma _1 = 0, \, \lambda _1 = 0, \, \lambda _2 =
   0, \, \lambda _4 = 0\right\}$ & \text{False} \\
 4 & $\left\{\gamma _4 = 0, \, \gamma _3 = 0, \, \gamma _2 = 0, \, \gamma _1 = 0, \, \lambda _1 = 0, \, \lambda _3 =
   0, \, \lambda _4 = 0\right\}$ & \text{False} \\
 5 & $\left\{\gamma _4 = 0, \, \gamma _3 = 0, \, \gamma _2 = 0, \, \gamma _1 = 0, \, \lambda _2 = 0, \, \lambda _3 =
   0, \, \lambda _4 = 0\right\}$ & \text{False} \\
 6 & $\left\{\gamma _4 = 0, \, \gamma _3 = 0, \, \gamma _2 = 0, \, \lambda _1 = 0, \, \lambda _2 = 0, \, \lambda _3 =
   0, \, \lambda _4 = 0\right\}$ & \text{False} \\
 7 & $\left\{\gamma _4 = 0, \, \gamma _3 = 0, \, \gamma _1 = 0, \, \lambda _1 = 0, \, \lambda _2 = 0, \, \lambda _3 =
   0, \, \lambda _4 = 0\right\}$ & \text{False} \\
 8 & $\left\{\gamma _4 = 0, \, \gamma _2 = 0, \, \gamma _1 = 0, \, \lambda _1 = 0, \, \lambda _2 = 0, \, \lambda _3 =
   0, \, \lambda _4 = 0\right\}$ & \text{False} \\
 9 & $\left\{\gamma _3 = 0, \, \gamma _2 = 0, \, \gamma _1 = 0, \, \lambda _1 = 0, \, \lambda _2 = 0, \, \lambda _3 =
   0, \, \lambda _4 = 0\right\}$ & \text{False} \\
\hline
 10 & $\left\{\gamma _4 = 0, \, \gamma _3 = 0, \, \gamma _2 = 0, \, \gamma _1 = 0, \, \lambda _1 = 0, \, \lambda _2 =
   0\right\}$ & \text{False} \\
 11 & $\left\{\gamma _4 = 0, \, \gamma _3 = 0, \, \gamma _2 = 0, \, \gamma _1 = 0, \, \lambda _1 = 0, \, \lambda _3 =
   0\right\}$ & \text{False} \\
 12 & $\left\{\gamma _4 = 0, \, \gamma _3 = 0, \, \gamma _2 = 0, \, \gamma _1 = 0, \, \lambda _1 = 0, \, \lambda _4 =
   0\right\}$ & \text{False} \\
 13 & $\left\{\gamma _4 = 0, \, \gamma _3 = 0, \, \gamma _2 = 0, \, \gamma _1 = 0, \, \lambda _2 = 0, \, \lambda _3 =
   0\right\}$ & \text{False} \\
 14 & $\left\{\gamma _4 = 0, \, \gamma _3 = 0, \, \gamma _2 = 0, \, \gamma _1 = 0, \, \lambda _2 = 0, \, \lambda _4 =
   0\right\}$ & \text{False} \\
 15 & $\left\{\gamma _4 = 0, \, \gamma _3 = 0, \, \gamma _2 = 0, \, \gamma _1 = 0, \, \lambda _3 = 0, \, \lambda _4 =
   0\right\}$ & \text{False} \\
 16 & $\left\{\gamma _4 = 0, \, \gamma _3 = 0, \, \gamma _2 = 0, \, \lambda _1 = 0, \, \lambda _2 = 0, \, \lambda _3 =
   0\right\}$ & \text{False} \\
 17 & $\left\{\gamma _4 = 0, \, \gamma _3 = 0, \, \gamma _2 = 0, \, \lambda _1 = 0, \, \lambda _2 = 0, \, \lambda _4 =
   0\right\}$ & \text{False} \\
 18 & $\left\{\gamma _4 = 0, \, \gamma _3 = 0, \, \gamma _2 = 0, \, \lambda _1 = 0, \, \lambda _3 = 0, \, \lambda _4 =
   0\right\}$ & \text{False} \\
 19 & $\left\{\gamma _4 = 0, \, \gamma _3 = 0, \, \gamma _2 = 0, \, \lambda _2 = 0, \, \lambda _3 = 0, \, \lambda _4 =
   0\right\}$ & \text{False} \\
 20 & $\left\{\gamma _4 = 0, \, \gamma _3 = 0, \, \gamma _1 = 0, \, \lambda _1 = 0, \, \lambda _2 = 0, \, \lambda _3 =
   0\right\}$ & \text{False} \\
 21 & $\left\{\gamma _4 = 0, \, \gamma _3 = 0, \, \gamma _1 = 0, \, \lambda _1 = 0, \, \lambda _2 = 0, \, \lambda _4 =
   0\right\}$ & \text{False} \\
 22 & $\left\{\gamma _4 = 0, \, \gamma _3 = 0, \, \gamma _1 = 0, \, \lambda _1 = 0, \, \lambda _3 = 0, \, \lambda _4 =
   0\right\}$ & \text{False} \\
 23 & $\left\{\gamma _4 = 0, \, \gamma _3 = 0, \, \gamma _1 = 0, \, \lambda _2 = 0, \, \lambda _3 = 0, \, \lambda _4 =
   0\right\}$ & \text{False} \\
 24 & $\left\{\gamma _4 = 0, \, \gamma _3 = 0, \, \lambda _1 = 0, \, \lambda _2 = 0, \, \lambda _3 = 0, \, \lambda
   _4 = 0\right\}$ & \text{False} \\
 25 & $\left\{\gamma _4 = 0, \, \gamma _2 = 0, \, \gamma _1 = 0, \, \lambda _1 = 0, \, \lambda _2 = 0, \, \lambda _3 =
   0\right\}$ & \text{False} \\
 26 & $\left\{\gamma _4 = 0, \, \gamma _2 = 0, \, \gamma _1 = 0, \, \lambda _1 = 0, \, \lambda _2 = 0, \, \lambda _4 =
   0\right\}$ & \text{False} \\
 27 & $\left\{\gamma _4 = 0, \, \gamma _2 = 0, \, \gamma _1 = 0, \, \lambda _1 = 0, \, \lambda _3 = 0, \, \lambda _4 =
   0\right\}$ & \text{False} \\
 28 & $\left\{\gamma _4 = 0, \, \gamma _2 = 0, \, \gamma _1 = 0, \, \lambda _2 = 0, \, \lambda _3 = 0, \, \lambda _4 =
   0\right\}$ & \text{False} \\
 29 & $\left\{\gamma _4 = 0, \, \gamma _2 = 0, \, \lambda _1 = 0, \, \lambda _2 = 0, \, \lambda _3 = 0, \, \lambda
   _4 = 0\right\}$ & \text{False} \\
 30 & $\left\{\gamma _4 = 0, \, \gamma _1 = 0, \, \lambda _1 = 0, \, \lambda _2 = 0, \, \lambda _3 = 0, \, \lambda
   _4 = 0\right\}$ & \text{False} \\
 31 & $\left\{\gamma _3 = 0, \, \gamma _2 = 0, \, \gamma _1 = 0, \, \lambda _1 = 0, \, \lambda _2 = 0, \, \lambda _3 =
   0\right\}$ & \text{False} \\
 32 & $\left\{\gamma _3 = 0, \, \gamma _2 = 0, \, \gamma _1 = 0, \, \lambda _1 = 0, \, \lambda _2 = 0, \, \lambda _4 =
   0\right\}$ & \text{False} \\
 33 & $\left\{\gamma _3 = 0, \, \gamma _2 = 0, \, \gamma _1 = 0, \, \lambda _1 = 0, \, \lambda _3 = 0, \, \lambda _4 =
   0\right\}$ & \text{False} \\
 34 & $\left\{\gamma _3 = 0, \, \gamma _2 = 0, \, \gamma _1 = 0, \, \lambda _2 = 0, \, \lambda _3 = 0, \, \lambda _4 =
   0\right\}$ & \text{False} \\
 35 & $\left\{\gamma _3 = 0, \, \gamma _2 = 0, \, \lambda _1 = 0, \, \lambda _2 = 0, \, \lambda _3 = 0, \, \lambda
   _4 = 0\right\}$ & \text{False} \\
 36 & $\left\{\gamma _3 = 0, \, \gamma _1 = 0, \, \lambda _1 = 0, \, \lambda _2 = 0, \, \lambda _3 = 0, \, \lambda
   _4 = 0\right\}$ & \text{False} \\
 37 & $\left\{\gamma _2 = 0, \, \gamma _1 = 0, \, \lambda _1 = 0, \, \lambda _2 = 0, \, \lambda _3 = 0, \, \lambda
   _4 = 0\right\}$ & \text{False} \\
\hline
 38 & $\left\{\gamma _4 = 0, \, \gamma _3 = 0, \, \gamma _2 = 0, \, \gamma _1 = 0, \, \lambda _1 = 0\right\}$ &
   \text{False} \\
 39 & $\left\{\gamma _4 = 0, \, \gamma _3 = 0, \, \gamma _2 = 0, \, \gamma _1 = 0, \, \lambda _2 = 0\right\}$ &
   \text{False} \\
 40 & $\left\{\gamma _4 = 0, \, \gamma _3 = 0, \, \gamma _2 = 0, \, \gamma _1 = 0, \, \lambda _3 = 0\right\}$ &
   \text{False} \\
 41 & $\left\{\gamma _4 = 0, \, \gamma _3 = 0, \, \gamma _2 = 0, \, \gamma _1 = 0, \, \lambda _4 = 0\right\}$ &
   \text{False} \\
 42 & $\left\{\gamma _4 = 0, \, \gamma _3 = 0, \, \gamma _2 = 0, \, \lambda _1 = 0, \, \lambda _2 = 0\right\}$ &
   \text{False} \\
 43 & $\left\{\gamma _4 = 0, \, \gamma _3 = 0, \, \gamma _2 = 0, \, \lambda _1 = 0, \, \lambda _4 = 0\right\}$ &
   \text{False} \\
 44 & $\left\{\gamma _4 = 0, \, \gamma _3 = 0, \, \gamma _2 = 0, \, \lambda _2 = 0, \, \lambda _3 = 0\right\}$ &
   \text{False} \\
 45 & $\left\{\gamma _4 = 0, \, \gamma _3 = 0, \, \gamma _2 = 0, \, \lambda _2 = 0, \, \lambda _4 = 0\right\}$ &
   \text{False} \\
 46 & $\left\{\gamma _4 = 0, \, \gamma _3 = 0, \, \gamma _2 = 0, \, \lambda _3 = 0, \, \lambda _4 = 0\right\}$ &
   \text{False} \\
 47 & $\left\{\gamma _4 = 0, \, \gamma _3 = 0, \, \gamma _1 = 0, \, \lambda _1 = 0, \, \lambda _2 = 0\right\}$ &
   \text{False} \\
 48 & $\left\{\gamma _4 = 0, \, \gamma _3 = 0, \, \gamma _1 = 0, \, \lambda _1 = 0, \, \lambda _3 = 0\right\}$ &
   \text{False} \\
 49 & $\left\{\gamma _4 = 0, \, \gamma _3 = 0, \, \gamma _1 = 0, \, \lambda _1 = 0, \, \lambda _4 = 0\right\}$ &
   \text{False} \\
 50 & $\left\{\gamma _4 = 0, \, \gamma _3 = 0, \, \gamma _1 = 0, \, \lambda _2 = 0, \, \lambda _3 = 0\right\}$ &
   \text{False} \\
 51 & $\left\{\gamma _4 = 0, \, \gamma _3 = 0, \, \gamma _1 = 0, \, \lambda _2 = 0, \, \lambda _4 = 0\right\}$ &
   \text{False} \\
 52 & $\left\{\gamma _4 = 0, \, \gamma _3 = 0, \, \gamma _1 = 0, \, \lambda _3 = 0, \, \lambda _4 = 0\right\}$ &
   \text{False} \\
 53 & $\left\{\gamma _4 = 0, \, \gamma _3 = 0, \, \lambda _1 = 0, \, \lambda _2 = 0, \, \lambda _3 = 0\right\}$ &
   \text{False} \\
 54 & $\left\{\gamma _4 = 0, \, \gamma _3 = 0, \, \lambda _1 = 0, \, \lambda _2 = 0, \, \lambda _4 = 0\right\}$ &
   \text{False} \\
 55 & $\left\{\gamma _4 = 0, \, \gamma _3 = 0, \, \lambda _1 = 0, \, \lambda _3 = 0, \, \lambda _4 = 0\right\}$ &
   \text{False} \\
 56 & $\left\{\gamma _4 = 0, \, \gamma _3 = 0, \, \lambda _2 = 0, \, \lambda _3 = 0, \, \lambda _4 = 0\right\}$ &
   \text{False} \\
 57 & $\left\{\gamma _4 = 0, \, \gamma _2 = 0, \, \gamma _1 = 0, \, \lambda _1 = 0, \, \lambda _2 = 0\right\}$ &
   \text{False} \\
 58 & $\left\{\gamma _4 = 0, \, \gamma _2 = 0, \, \gamma _1 = 0, \, \lambda _1 = 0, \, \lambda _3 = 0\right\}$ &
   \text{False} \\
 59 & $\left\{\gamma _4 = 0, \, \gamma _2 = 0, \, \gamma _1 = 0, \, \lambda _1 = 0, \, \lambda _4 = 0\right\}$ &
   \text{False} \\
 60 & $\left\{\gamma _4 = 0, \, \gamma _2 = 0, \, \gamma _1 = 0, \, \lambda _2 = 0, \, \lambda _3 = 0\right\}$ &
   \text{False} \\
 61 & $\left\{\gamma _4 = 0, \, \gamma _2 = 0, \, \gamma _1 = 0, \, \lambda _2 = 0, \, \lambda _4 = 0\right\}$ &
   \text{False} \\
 62 & $\left\{\gamma _4 = 0, \, \gamma _2 = 0, \, \gamma _1 = 0, \, \lambda _3 = 0, \, \lambda _4 = 0\right\}$ &
   \text{False} \\
 63 & $\left\{\gamma _4 = 0, \, \gamma _2 = 0, \, \lambda _1 = 0, \, \lambda _2 = 0, \, \lambda _3 = 0\right\}$ &
   \text{False} \\
 64 & $\left\{\gamma _4 = 0, \, \gamma _2 = 0, \, \lambda _1 = 0, \, \lambda _2 = 0, \, \lambda _4 = 0\right\}$ &
   \text{False} \\
 65 & $\left\{\gamma _4 = 0, \, \gamma _2 = 0, \, \lambda _1 = 0, \, \lambda _3 = 0, \, \lambda _4 = 0\right\}$ &
   \text{False} \\
 66 & $\left\{\gamma _4 = 0, \, \gamma _2 = 0, \, \lambda _2 = 0, \, \lambda _3 = 0, \, \lambda _4 = 0\right\}$ &
   \text{False} \\
 67 & $\left\{\gamma _4 = 0, \, \gamma _1 = 0, \, \lambda _1 = 0, \, \lambda _2 = 0, \, \lambda _3 = 0\right\}$ &
   \text{False} \\
 68 & $\left\{\gamma _4 = 0, \, \gamma _1 = 0, \, \lambda _1 = 0, \, \lambda _2 = 0, \, \lambda _4 = 0\right\}$ &
   \text{False} \\
 69 & $\left\{\gamma _4 = 0, \, \gamma _1 = 0, \, \lambda _1 = 0, \, \lambda _3 = 0, \, \lambda _4 = 0\right\}$ &
   \text{False} \\
 70 & $\left\{\gamma _4 = 0, \, \gamma _1 = 0, \, \lambda _2 = 0, \, \lambda _3 = 0, \, \lambda _4 = 0\right\}$ &
   \text{False} \\
 71 & $\left\{\gamma _4 = 0, \, \lambda _1 = 0, \, \lambda _2 = 0, \, \lambda _3 = 0, \, \lambda _4 = 0\right\}$ &
   \text{False} \\
 72 & $\left\{\gamma _3 = 0, \, \gamma _2 = 0, \, \gamma _1 = 0, \, \lambda _1 = 0, \, \lambda _2 = 0\right\}$ &
   \text{False} \\
 73 & $\left\{\gamma _3 = 0, \, \gamma _2 = 0, \, \gamma _1 = 0, \, \lambda _1 = 0, \, \lambda _3 = 0\right\}$ &
   \text{False} \\
 74 & $\left\{\gamma _3 = 0, \, \gamma _2 = 0, \, \gamma _1 = 0, \, \lambda _1 = 0, \, \lambda _4 = 0\right\}$ &
   \text{False} \\
 75 & $\left\{\gamma _3 = 0, \, \gamma _2 = 0, \, \gamma _1 = 0, \, \lambda _2 = 0, \, \lambda _3 = 0\right\}$ &
   \text{False} \\
 76 & $\left\{\gamma _3 = 0, \, \gamma _2 = 0, \, \gamma _1 = 0, \, \lambda _3 = 0, \, \lambda _4 = 0\right\}$ &
   \text{False} \\
 77 & $\left\{\gamma _3 = 0, \, \gamma _2 = 0, \, \lambda _1 = 0, \, \lambda _2 = 0, \, \lambda _3 = 0\right\}$ &
   \text{False} \\
 78 & $\left\{\gamma _3 = 0, \, \gamma _2 = 0, \, \lambda _1 = 0, \, \lambda _2 = 0, \, \lambda _4 = 0\right\}$ &
   \text{False} \\
 79 & $\left\{\gamma _3 = 0, \, \gamma _2 = 0, \, \lambda _1 = 0, \, \lambda _3 = 0, \, \lambda _4 = 0\right\}$ &
   \text{False} \\
 80 & $\left\{\gamma _3 = 0, \, \gamma _2 = 0, \, \lambda _2 = 0, \, \lambda _3 = 0, \, \lambda _4 = 0\right\}$ &
   \text{False} \\
 81 & $\left\{\gamma _3 = 0, \, \gamma _1 = 0, \, \lambda _1 = 0, \, \lambda _2 = 0, \, \lambda _3 = 0\right\}$ &
   \text{False} \\
 82 & $\left\{\gamma _3 = 0, \, \gamma _1 = 0, \, \lambda _1 = 0, \, \lambda _2 = 0, \, \lambda _4 = 0\right\}$ &
   \text{False} \\
 83 & $\left\{\gamma _3 = 0, \, \gamma _1 = 0, \, \lambda _1 = 0, \, \lambda _3 = 0, \, \lambda _4 = 0\right\}$ &
   \text{False} \\
 84 & $\left\{\gamma _3 = 0, \, \gamma _1 = 0, \, \lambda _2 = 0, \, \lambda _3 = 0, \, \lambda _4 = 0\right\}$ &
   \text{False} \\
 85 & $\left\{\gamma _3 = 0, \, \lambda _1 = 0, \, \lambda _2 = 0, \, \lambda _3 = 0, \, \lambda _4 = 0\right\}$ &
   \text{False} \\
 86 & $\left\{\gamma _2 = 0, \, \gamma _1 = 0, \, \lambda _1 = 0, \, \lambda _2 = 0, \, \lambda _3 = 0\right\}$ &
   \text{False} \\
 87 & $\left\{\gamma _2 = 0, \, \gamma _1 = 0, \, \lambda _1 = 0, \, \lambda _2 = 0, \, \lambda _4 = 0\right\}$ &
   \text{False} \\
 88 & $\left\{\gamma _2 = 0, \, \gamma _1 = 0, \, \lambda _1 = 0, \, \lambda _3 = 0, \, \lambda _4 = 0\right\}$ &
   \text{False} \\
 89 & $\left\{\gamma _2 = 0, \, \gamma _1 = 0, \, \lambda _2 = 0, \, \lambda _3 = 0, \, \lambda _4 = 0\right\}$ &
   \text{False} \\
 90 & $\left\{\gamma _2 = 0, \, \lambda _1 = 0, \, \lambda _2 = 0, \, \lambda _3 = 0, \, \lambda _4 = 0\right\}$ &
   \text{False} \\
 91 & $\left\{\gamma _1 = 0, \, \lambda _1 = 0, \, \lambda _2 = 0, \, \lambda _3 = 0, \, \lambda _4 = 0\right\}$ &
   \text{False} \\
\hline
 92 & $\left\{\gamma _4 = 0, \, \gamma _3 = 0, \, \gamma _2 = 0, \, \gamma _1 = 0\right\}$ & \text{False} \\
 93 & $\left\{\gamma _4 = 0, \, \gamma _3 = 0, \, \gamma _2 = 0, \, \lambda _2 = 0\right\}$ & \text{False} \\
 94 & $\left\{\gamma _4 = 0, \, \gamma _3 = 0, \, \gamma _2 = 0, \, \lambda _4 = 0\right\}$ & \text{False} \\
 95 & $\left\{\gamma _4 = 0, \, \gamma _3 = 0, \, \gamma _1 = 0, \, \lambda _1 = 0\right\}$ & \text{False} \\
 96 & $\left\{\gamma _4 = 0, \, \gamma _3 = 0, \, \gamma _1 = 0, \, \lambda _2 = 0\right\}$ & \text{False} \\
 97 & $\left\{\gamma _4 = 0, \, \gamma _3 = 0, \, \gamma _1 = 0, \, \lambda _3 = 0\right\}$ & \text{False} \\
 98 & $\left\{\gamma _4 = 0, \, \gamma _3 = 0, \, \gamma _1 = 0, \, \lambda _4 = 0\right\}$ & \text{False} \\
 99 & $\left\{\gamma _4 = 0, \, \gamma _3 = 0, \, \lambda _1 = 0, \, \lambda _2 = 0\right\}$ & \text{False} \\
 100 & $\left\{\gamma _4 = 0, \, \gamma _3 = 0, \, \lambda _1 = 0, \, \lambda _4 = 0\right\}$ & \text{False} \\
 101 & $\left\{\gamma _4 = 0, \, \gamma _3 = 0, \, \lambda _2 = 0, \, \lambda _3 = 0\right\}$ & \text{False} \\
 102 & $\left\{\gamma _4 = 0, \, \gamma _3 = 0, \, \lambda _2 = 0, \, \lambda _4 = 0\right\}$ & \text{False} \\
 103 & $\left\{\gamma _4 = 0, \, \gamma _3 = 0, \, \lambda _3 = 0, \, \lambda _4 = 0\right\}$ & \text{False} \\
 104 & $\left\{\gamma _4 = 0, \, \gamma _2 = 0, \, \gamma _1 = 0, \, \lambda _1 = 0\right\}$ & \text{False} \\
 105 & $\left\{\gamma _4 = 0, \, \gamma _2 = 0, \, \gamma _1 = 0, \, \lambda _2 = 0\right\}$ & \text{False} \\
 106 & $\left\{\gamma _4 = 0, \, \gamma _2 = 0, \, \gamma _1 = 0, \, \lambda _3 = 0\right\}$ & \text{False} \\
 107 & $\left\{\gamma _4 = 0, \, \gamma _2 = 0, \, \gamma _1 = 0, \, \lambda _4 = 0\right\}$ & \text{False} \\
 108 & $\left\{\gamma _4 = 0, \, \gamma _2 = 0, \, \lambda _1 = 0, \, \lambda _2 = 0\right\}$ & \text{False} \\
 109 & $\left\{\gamma _4 = 0, \, \gamma _2 = 0, \, \lambda _1 = 0, \, \lambda _4 = 0\right\}$ & \text{False} \\
 110 & $\left\{\gamma _4 = 0, \, \gamma _2 = 0, \, \lambda _2 = 0, \, \lambda _3 = 0\right\}$ & \text{False} \\
 111 & $\left\{\gamma _4 = 0, \, \gamma _2 = 0, \, \lambda _2 = 0, \, \lambda _4 = 0\right\}$ & \text{False} \\
 112 & $\left\{\gamma _4 = 0, \, \gamma _2 = 0, \, \lambda _3 = 0, \, \lambda _4 = 0\right\}$ & \text{False} \\
 113 & $\left\{\gamma _4 = 0, \, \gamma _1 = 0, \, \lambda _1 = 0, \, \lambda _2 = 0\right\}$ & \text{False} \\
 114 & $\left\{\gamma _4 = 0, \, \gamma _1 = 0, \, \lambda _1 = 0, \, \lambda _3 = 0\right\}$ & \text{False} \\
 115 & $\left\{\gamma _4 = 0, \, \gamma _1 = 0, \, \lambda _1 = 0, \, \lambda _4 = 0\right\}$ & \text{False} \\
 116 & $\left\{\gamma _4 = 0, \, \gamma _1 = 0, \, \lambda _2 = 0, \, \lambda _3 = 0\right\}$ & \text{False} \\
 117 & $\left\{\gamma _4 = 0, \, \gamma _1 = 0, \, \lambda _2 = 0, \, \lambda _4 = 0\right\}$ & \text{False} \\
 118 & $\left\{\gamma _4 = 0, \, \gamma _1 = 0, \, \lambda _3 = 0, \, \lambda _4 = 0\right\}$ & \text{False} \\
 119 & $\left\{\gamma _4 = 0, \, \lambda _1 = 0, \, \lambda _2 = 0, \, \lambda _3 = 0\right\}$ & \text{False} \\
 120 & $\left\{\gamma _4 = 0, \, \lambda _1 = 0, \, \lambda _2 = 0, \, \lambda _4 = 0\right\}$ & \text{False} \\
 121 & $\left\{\gamma _4 = 0, \, \lambda _1 = 0, \, \lambda _3 = 0, \, \lambda _4 = 0\right\}$ & \text{False} \\
 122 & $\left\{\gamma _4 = 0, \, \lambda _2 = 0, \, \lambda _3 = 0, \, \lambda _4 = 0\right\}$ & \text{False} \\
 123 & $\left\{\gamma _3 = 0, \, \gamma _2 = 0, \, \gamma _1 = 0, \, \lambda _1 = 0\right\}$ & \text{False} \\
 124 & $\left\{\gamma _3 = 0, \, \gamma _2 = 0, \, \gamma _1 = 0, \, \lambda _3 = 0\right\}$ & \text{False} \\
 125 & $\left\{\gamma _3 = 0, \, \gamma _2 = 0, \, \lambda _1 = 0, \, \lambda _2 = 0\right\}$ & \text{False} \\
 126 & $\left\{\gamma _3 = 0, \, \gamma _2 = 0, \, \lambda _1 = 0, \, \lambda _4 = 0\right\}$ & \text{False} \\
 127 & $\left\{\gamma _3 = 0, \, \gamma _2 = 0, \, \lambda _2 = 0, \, \lambda _3 = 0\right\}$ & \text{False} \\
 128 & $\left\{\gamma _3 = 0, \, \gamma _2 = 0, \, \lambda _3 = 0, \, \lambda _4 = 0\right\}$ & \text{False} \\
 129 & $\left\{\gamma _3 = 0, \, \gamma _1 = 0, \, \lambda _1 = 0, \, \lambda _2 = 0\right\}$ & \text{False} \\
 130 & $\left\{\gamma _3 = 0, \, \gamma _1 = 0, \, \lambda _1 = 0, \, \lambda _3 = 0\right\}$ & \text{False} \\
 131 & $\left\{\gamma _3 = 0, \, \gamma _1 = 0, \, \lambda _1 = 0, \, \lambda _4 = 0\right\}$ & \text{False} \\
 132 & $\left\{\gamma _3 = 0, \, \gamma _1 = 0, \, \lambda _2 = 0, \, \lambda _3 = 0\right\}$ & \text{False} \\
 133 & $\left\{\gamma _3 = 0, \, \gamma _1 = 0, \, \lambda _3 = 0, \, \lambda _4 = 0\right\}$ & \text{False} \\
 134 & $\left\{\gamma _3 = 0, \, \lambda _1 = 0, \, \lambda _2 = 0, \, \lambda _3 = 0\right\}$ & \text{False} \\
 135 & $\left\{\gamma _3 = 0, \, \lambda _1 = 0, \, \lambda _2 = 0, \, \lambda _4 = 0\right\}$ & \text{False} \\
 136 & $\left\{\gamma _3 = 0, \, \lambda _1 = 0, \, \lambda _3 = 0, \, \lambda _4 = 0\right\}$ & \text{False} \\
 137 & $\left\{\gamma _3 = 0, \, \lambda _2 = 0, \, \lambda _3 = 0, \, \lambda _4 = 0\right\}$ & \text{False} \\
 138 & $\left\{\gamma _2 = 0, \, \gamma _1 = 0, \, \lambda _1 = 0, \, \lambda _2 = 0\right\}$ & \text{False} \\
 139 & $\left\{\gamma _2 = 0, \, \gamma _1 = 0, \, \lambda _1 = 0, \, \lambda _3 = 0\right\}$ & \text{False} \\
 140 & $\left\{\gamma _2 = 0, \, \gamma _1 = 0, \, \lambda _1 = 0, \, \lambda _4 = 0\right\}$ & \text{False} \\
 141 & $\left\{\gamma _2 = 0, \, \gamma _1 = 0, \, \lambda _2 = 0, \, \lambda _3 = 0\right\}$ & \text{False} \\
 142 & $\left\{\gamma _2 = 0, \, \gamma _1 = 0, \, \lambda _3 = 0, \, \lambda _4 = 0\right\}$ & \text{False} \\
 143 & $\left\{\gamma _2 = 0, \, \lambda _1 = 0, \, \lambda _2 = 0, \, \lambda _3 = 0\right\}$ & \text{False} \\
 144 & $\left\{\gamma _2 = 0, \, \lambda _1 = 0, \, \lambda _2 = 0, \, \lambda _4 = 0\right\}$ & \text{False} \\
 145 & $\left\{\gamma _2 = 0, \, \lambda _1 = 0, \, \lambda _3 = 0, \, \lambda _4 = 0\right\}$ & \text{False} \\
 146 & $\left\{\gamma _2 = 0, \, \lambda _2 = 0, \, \lambda _3 = 0, \, \lambda _4 = 0\right\}$ & \text{False} \\
 147 & $\left\{\gamma _1 = 0, \, \lambda _1 = 0, \, \lambda _2 = 0, \, \lambda _3 = 0\right\}$ & \text{False} \\
 148 & $\left\{\gamma _1 = 0, \, \lambda _1 = 0, \, \lambda _2 = 0, \, \lambda _4 = 0\right\}$ & \text{False} \\
 149 & $\left\{\gamma _1 = 0, \, \lambda _1 = 0, \, \lambda _3 = 0, \, \lambda _4 = 0\right\}$ & \text{False} \\
 150 & $\left\{\gamma _1 = 0, \, \lambda _2 = 0, \, \lambda _3 = 0, \, \lambda _4 = 0\right\}$ & \text{False} \\
 151 & $\left\{\lambda _1 = 0, \, \lambda _2 = 0, \, \lambda _3 = 0, \, \lambda _4 = 0\right\}$ & \text{False} \\
 152 & $\left\{\gamma_2 = 0, \, \,  \gamma_1 = 0, \, \,  \lambda_2 = 0, \, \,  \lambda_4 = 0\right\}$ & \text{False} \\
 153 & $\left\{\gamma_3 = 0, \, \,  \gamma_2 = 0, \, \,  \lambda_2 = 0, \, \,  \lambda_4 = 0\right\}$ & \text{False} \\
 154 & $\left\{\gamma_3 = 0, \, \,  \gamma_2 = 0, \, \,  \lambda_1 = 0, \, \,  \lambda_3 = 0\right\}$ & \text{False} \\
 155 & $\left\{\gamma_3 = 0, \, \,  \gamma_2 = 0, \, \,  \gamma_1 = 0, \, \,  \lambda_2 = 0\right\}$ & \text{False} \\
 156 & $\left\{\gamma_3 = 0, \, \,  \gamma_2 = 0, \, \,  \gamma_1 = 0, \, \,  \lambda_4 = 0\right\}$ & \text{False} \\
 157 & $\left\{\gamma_4 = 0, \, \,  \gamma_3 = 0, \, \,  \lambda_1 = 0, \, \,  \lambda_3 = 0\right\}$ & \text{False} \\
 158 & $\left\{\gamma_4 = 0, \, \,  \gamma_3 = 0, \, \,  \gamma_2 = 0, \, \,  \lambda_3 = 0\right\}$ & \text{False} \\
 159 & $\left\{\gamma_4 = 0, \, \,  \gamma_3 = 0, \, \,  \gamma_2 = 0, \, \,  \lambda_1 = 0\right\}$ & \text{False} \\
 \hline
 160 & $\left\{\gamma _4 = 0, \, \gamma _3 = 0, \, \gamma _1 = 0\right\}$ & \text{False} \\
 161 & $\left\{\gamma _4 = 0, \, \gamma _3 = 0, \, \lambda _2 = 0\right\}$ & \text{False} \\
 162 & $\left\{\gamma _4 = 0, \, \gamma _3 = 0, \, \lambda _4 = 0\right\}$ & \text{False} \\
 163 & $\left\{\gamma _4 = 0, \, \gamma _2 = 0, \, \gamma _1 = 0\right\}$ & \text{False} \\
 164 & $\left\{\gamma _4 = 0, \, \gamma _2 = 0, \, \lambda _2 = 0\right\}$ & \text{False} \\
 165 & $\left\{\gamma _4 = 0, \, \gamma _2 = 0, \, \lambda _4 = 0\right\}$ & \text{False} \\
 166 & $\left\{\gamma _4 = 0, \, \gamma _1 = 0, \, \lambda _1 = 0\right\}$ & \text{False} \\
 167 & $\left\{\gamma _4 = 0, \, \gamma _1 = 0, \, \lambda _2 = 0\right\}$ & \text{False} \\
 168 & $\left\{\gamma _4 = 0, \, \gamma _1 = 0, \, \lambda _3 = 0\right\}$ & \text{False} \\
 169 & $\left\{\gamma _4 = 0, \, \gamma _1 = 0, \, \lambda _4 = 0\right\}$ & \text{False} \\
 170 & $\left\{\gamma _4 = 0, \, \lambda _1 = 0, \, \lambda _2 = 0\right\}$ & \text{False} \\
 171 & $\left\{\gamma _4 = 0, \, \lambda _1 = 0, \, \lambda _4 = 0\right\}$ & \text{False} \\
 172 & $\left\{\gamma _4 = 0, \, \lambda _2 = 0, \, \lambda _3 = 0\right\}$ & \text{False} \\
 173 & $\left\{\gamma _4 = 0, \, \lambda _2 = 0, \, \lambda _4 = 0\right\}$ & \text{False} \\
 174 & $\left\{\gamma _4 = 0, \, \lambda _3 = 0, \, \lambda _4 = 0\right\}$ & \text{False} \\
 175 & $\left\{\gamma _3 = 0, \, \gamma _1 = 0, \, \lambda _1 = 0\right\}$ & \text{False} \\
 176 & $\left\{\gamma _3 = 0, \, \gamma _1 = 0, \, \lambda _3 = 0\right\}$ & \text{False} \\
 177 & $\left\{\gamma _3 = 0, \, \lambda _1 = 0, \, \lambda _2 = 0\right\}$ & \text{False} \\
 178 & $\left\{\gamma _3 = 0, \, \lambda _1 = 0, \, \lambda _4 = 0\right\}$ & \text{False} \\
 179 & $\left\{\gamma _3 = 0, \, \lambda _2 = 0, \, \lambda _3 = 0\right\}$ & \text{False} \\
 180 & $\left\{\gamma _3 = 0, \, \lambda _3 = 0, \, \lambda _4 = 0\right\}$ & \text{False} \\
 181 & $\left\{\gamma _2 = 0, \, \gamma _1 = 0, \, \lambda _1 = 0\right\}$ & \text{False} \\
 182 & $\left\{\gamma _2 = 0, \, \gamma _1 = 0, \, \lambda _3 = 0\right\}$ & \text{False} \\
 183 & $\left\{\gamma _2 = 0, \, \lambda _1 = 0, \, \lambda _2 = 0\right\}$ & \text{False} \\
 184 & $\left\{\gamma _2 = 0, \, \lambda _1 = 0, \, \lambda _4 = 0\right\}$ & \text{False} \\
 185 & $\left\{\gamma _2 = 0, \, \lambda _2 = 0, \, \lambda _3 = 0\right\}$ & \text{False} \\
 186 & $\left\{\gamma _2 = 0, \, \lambda _3 = 0, \, \lambda _4 = 0\right\}$ & \text{False} \\
 187 & $\left\{\gamma _1 = 0, \, \lambda _1 = 0, \, \lambda _2 = 0\right\}$ & \text{False} \\
 188 & $\left\{\gamma _1 = 0, \, \lambda _1 = 0, \, \lambda _3 = 0\right\}$ & \text{False} \\
 189 & $\left\{\gamma _1 = 0, \, \lambda _1 = 0, \, \lambda _4 = 0\right\}$ & \text{False} \\
 190 & $\left\{\gamma _1 = 0, \, \lambda _2 = 0, \, \lambda _3 = 0\right\}$ & \text{False} \\
 191 & $\left\{\gamma _1 = 0, \, \lambda _3 = 0, \, \lambda _4 = 0\right\}$ & \text{False} \\
 192 & $\left\{\lambda _1 = 0, \, \lambda _2 = 0, \, \lambda _4 = 0\right\}$ & \text{False} \\
 193 & $\left\{\lambda _1 = 0, \, \lambda _3 = 0, \, \lambda _4 = 0\right\}$ & \text{False} \\
 194 & $\left\{\gamma_1 = 0, \, \,  \lambda_2 = 0, \, \,  \lambda_4 = 0\right\}$ & \text{False} \\
 195 & $\left\{\gamma_2 = 0, \, \,  \lambda_2 = 0, \, \,  \lambda_4 = 0\right\}$ & \text{False} \\
 196 & $\left\{\gamma_2 = 0, \, \,  \lambda_1 = 0, \, \,  \lambda_3 = 0\right\}$ & \text{False} \\
 197 & $\left\{\gamma_2 = 0, \, \,  \gamma_1 = 0, \, \,  \lambda_2 = 0\right\}$ & \text{False} \\
 198 & $\left\{\gamma_2 = 0, \, \,  \gamma_1 = 0, \, \,  \lambda_4 = 0\right\}$ & \text{False} \\
 199 & $\left\{\gamma_3 = 0, \, \,  \lambda_2 = 0, \, \,  \lambda_4 = 0\right\}$ & \text{False} \\
 200 & $\left\{\gamma_3 = 0, \, \,  \lambda_1 = 0, \, \,  \lambda_3 = 0\right\}$ & \text{False} \\
 201 & $\left\{\gamma_3 = 0, \, \,  \gamma_1 = 0, \, \,  \lambda_2 = 0\right\}$ & \text{False} \\
 202 & $\left\{\gamma_3 = 0, \, \,  \gamma_1 = 0, \, \,  \lambda_4 = 0\right\}$ & \text{False} \\
 203 & $\left\{\gamma_3 = 0, \, \,  \gamma_2 = 0, \, \,  \lambda_2 = 0\right\}$ & \text{False} \\
 204 & $\left\{\gamma_3 = 0, \, \,  \gamma_2 = 0, \, \,  \lambda_4 = 0\right\}$ & \text{False} \\
 205 & $\left\{\gamma_3 = 0, \, \,  \gamma_2 = 0, \, \,  \lambda_3 = 0\right\}$ & \text{False} \\
 206 & $\left\{\gamma_3 = 0, \, \,  \gamma_2 = 0, \, \,  \lambda_1 = 0\right\}$ & \text{False} \\
 207 & $\left\{\gamma_3 = 0, \, \,  \gamma_2 = 0, \, \,  \gamma_1 = 0\right\}$ & \text{False} \\
 208 & $\left\{\gamma_4 = 0, \, \,  \lambda_1 = 0, \, \,  \lambda_3 = 0\right\}$ & \text{False} \\
 209 & $\left\{\gamma_4 = 0, \, \,  \gamma_2 = 0, \, \,  \lambda_3 = 0\right\}$ & \text{False} \\
 210 & $\left\{\gamma_4 = 0, \, \,  \gamma_2 = 0, \, \,  \lambda_1 = 0\right\}$ & \text{False} \\
 211 & $\left\{\gamma_4 = 0, \, \,  \gamma_3 = 0, \, \,  \lambda_3 = 0\right\}$ & \text{False} \\
 212 & $\left\{\gamma_4 = 0, \, \,  \gamma_3 = 0, \, \,  \lambda_1 = 0\right\}$ & \text{False} \\
 213 & $\left\{\gamma_4 = 0, \, \,  \gamma_3 = 0, \, \,  \gamma_2 = 0\right\}$ & \text{False} \\
 {214} & $\left\{\lambda _2 = 0, \, \lambda _3 = 0, \, \lambda _4 = 0\right\}$ & \text{False} \\
 {215} & $\left\{\lambda _1 = 0, \, \lambda _2 = 0, \, \lambda _3 = 0\right\}$ & \text{False} \\
 \hline
{216} & $\left\{\lambda _1 = 0, \, \lambda _2 = 0\right\}$ & \text{False} \\
{217} & $\left\{\lambda_1 = 0, \, \,  \lambda_3 = 0\right\}$ & \text{False} \\
{218} & $\left\{\lambda_1 = 0, \, \,  \lambda_4 = 0\right\}$ & \text{False} \\
{219} & $\left\{\lambda_2 = 0, \, \,  \lambda_3 = 0\right\}$ & \text{False} \\
{220} & $\left\{\lambda_2 = 0, \, \,  \lambda_4 = 0\right\}$ & \text{False} \\
{221} & $\left\{\lambda _3 = 0, \, \lambda _4 = 0\right\}$ & \text{False} \\
{222} & $\left\{\gamma _4 = 0, \, \lambda _2 = 0\right\}$ & \text{False} \\
{223} & $\left\{\gamma _4 = 0, \, \lambda _4 = 0\right\}$ & \text{False} \\
{224} & $\left\{\gamma_3 = 0, \, \,  \lambda_2 = 0\right\}$ & \text{False} \\
{225} & $\left\{\gamma_2 = 0, \, \,  \lambda_3 = 0\right\}$ & \text{False} \\
{226} & $\left\{\gamma_1 = 0, \, \,  \lambda_1 = 0\right\}$ & \text{False} \\
{227} & $\left\{\gamma_1 = 0, \, \,  \lambda_3 = 0\right\}$ & \text{False} \\
\hline
\end{longtable}
\end{center}

\begin{center}
\renewcommand{\arraystretch}{1.15}
\label{tab:possible}
\begin{longtable}{|c||c|}
\caption{List of 29 flux configurations with their respective vanishing flux dependent parameters for which dS no-go case could not be found. The parameters not mentioned in a given configuration are considered to be non-zero.} \\
\hline 
Sr. No. & Vanishing Axionic Flux parameters \\
    \hhline{|=|=|}
    \endfirsthead
    \hline 
Sr. No. & Vanishing Axionic Flux parameters \\
    \hhline{|=|=|}
    \endhead
    \label{tab_36cases-without-nogo}1 & $\left\{\gamma_1 = 0, \, \,  \gamma_2 = 0, \, \,  \gamma_3 = 0, \, \,  \lambda_2 = 0, \, \,  \lambda_4 = 0\right\}$ \\
 2 & $\left\{\gamma_2 = 0, \, \,  \gamma_3 = 0, \, \,  \gamma_4 = 0, \, \,  \lambda_1 = 0, \, \,  \lambda_3 = 0\right\}$ \\
 \hline
  3 & $\left\{\gamma_4 = 0, \, \,  \gamma_2 = 0, \, \,  \lambda_1 = 0, \, \,  \lambda_3 = 0\right\}$ \\
 4 & $\left\{\gamma_3 = 0, \, \,  \gamma_1 = 0, \, \,  \lambda_2 = 0, \, \,  \lambda_4 = 0\right\}$ \\
 \hline
 \hline
 5 & $\left\{\gamma_1 = 0, \, \,  \lambda_2 = 0\right\}$ \\
 6 & $\left\{\gamma_1 = 0, \, \,  \lambda_4 = 0\right\}$ \\
 7 & $\left\{\gamma_2 = 0, \, \,  \lambda_4 = 0\right\}$ \\
 8 & $\left\{\gamma_2 = 0, \, \,  \lambda_1 = 0\right\}$ \\
 9 & $\left\{\gamma_2 = 0, \, \,  \gamma_1 = 0\right\}$ \\
 10 & $\left\{\gamma_2 = 0, \, \,  \lambda_2 = 0\right\}$ \\
 11 & $\left\{\gamma_3 = 0, \, \,  \lambda_4 = 0\right\}$ \\
 12 & $\left\{\gamma_3 = 0, \, \,  \lambda_3 = 0\right\}$ \\
 13 & $\left\{\gamma_3 = 0, \, \,  \lambda_1 = 0\right\}$ \\
 14 & $\left\{\gamma_3 = 0, \, \,  \gamma_1 = 0\right\}$ \\
 15 & $\left\{\gamma_3 = 0, \, \,  \gamma_2 = 0\right\}$ \\
 16 & $\left\{\gamma_4 = 0, \, \,  \lambda_3 = 0\right\}$ \\
 17 & $\left\{\gamma_4 = 0, \, \,  \lambda_1 = 0\right\}$ \\
 18 & $\left\{\gamma_4 = 0, \, \,  \gamma_1 = 0\right\}$ \\
 19 & $\left\{\gamma_4 = 0, \, \,  \gamma_2 = 0\right\}$ \\
 20 & $\left\{\gamma_4 = 0, \, \,  \gamma_3 = 0\right\}$ \\
 \hline
 21 & $\left\{\lambda_2 = 0\right\}$ \\
 22 & $\left\{\lambda_4 = 0\right\}$ \\
 23 & $\left\{\lambda_3 = 0\right\}$ \\
 24 & $\left\{\lambda_1 = 0\right\}$ \\
 25 & $\left\{\gamma_1 = 0\right\}$ \\
 26 & $\left\{\gamma_2 = 0\right\}$ \\
 27 & $\left\{\gamma_3 = 0\right\}$ \\
 28 & $\left\{\gamma_4 = 0\right\}$ \\
 \hline
 29 & None of the $\lambda_i$'s and $\gamma_i$'s are set to zero. \\
 \hline
\end{longtable}
\end{center}

\bibliographystyle{JHEP}
\bibliography{reference}


\end{document}